\documentclass[a4paper, 11pt, oneside]{Thesis}  
\graphicspath{{./}}  

\usepackage[square, numbers, comma, sort&compress]{natbib}  
\usepackage{verbatim}  
\usepackage{vector}  
\hypersetup{urlcolor=blue, colorlinks=true}  
\include{Qcircuit}

\begin{document}
\frontmatter	  

\title  {Quantum Information Processing with Continuous Variables and Atomic Ensembles}
\authors  {\texorpdfstring
            {\href{php07mz@sheffield.ac.uk}{Marcin Zwierz}}
            {Marcin Zwierz}
            }
\addresses  {\groupname\\\deptname\\\univname}  
\date       {January 2011}
\subject    {}

\maketitle

\setstretch{1.3}  

\fancyhead{}  
\rhead{\thepage}  
\lhead{}  

\pagestyle{fancy}  

\Declaration{

\addtocontents{toc}{\vspace{1em}}  

I, MARCIN ZWIERZ, declare that the work presented in this thesis, expect where otherwise state, is based on my own research and has not been submitted previously for a degree in this or any other university. Parts of the work reported in this thesis have been published as follows:

\textbf{\LARGE{Publications}}
\begin{enumerate}
\item M. Zwierz, C. A. P\'{e}rez-Delgado, and P. Kok. General optimality of the Heisenberg limit for quantum metrology. \textit{Phys. Rev. Lett.} \textbf{105}, 180402 (2010)
\item M. Zwierz, C. A. P\'{e}rez-Delgado, and P. Kok. Unifying parameter estimation and the Deutsch-Jozsa algorithm for continuous variables. \textit{Phys. Rev. A} \textbf{82}, 042320 (2010)
\item M. Zwierz and P. Kok. Applications of atomic ensembles in distributed quantum computing. \textit{International Journal of Quantum Information} \textbf{8}, 181-218 (2010)
\item M. Zwierz and P. Kok. High-efficiency cluster-state generation with atomic ensembles via the dipole-blockade mechanism. \textit{Phys. Rev. A} \textbf{79}, 022304 (2009)
\end{enumerate}


Signed:\\
\rule[1em]{25em}{0.5pt}  

Date:\\
\rule[1em]{25em}{0.5pt}  
}
\clearpage  





\addtotoc{Abstract}  
\abstract{
\addtocontents{toc}{\vspace{1em}}  

Quantum information theory promises many advances in science and technology. This thesis presents three different results in quantum information theory.

The first result addresses the theoretical foundations of quantum metrology. It is now well known that quantum-enhanced metrology promises improved sensitivity in parameter estimation over classical measurement procedures. The Heisenberg limit is considered to be the ultimate limit in quantum metrology imposed by the laws of quantum mechanics. It sets a lower bound on how precisely a physical quantity can be measured given a certain amount of resources in any possible measurement. Recently, however, several measurement procedures have been proposed in which the Heisenberg limit seemed to be surpassed. This led to an extensive debate over the question how the sensitivity scales with the physical resources such as the average photon number and the computational resources such as the number of queries that are used in estimation procedures. Here, we reconcile the physical definition of the relevant resources used in parameter estimation with the information-theoretical scaling in terms of the query complexity of a quantum network. This leads to a novel and ultimate Heisenberg limit that applies to all conceivable measurement procedures. Our approach to quantum metrology not only resolves the mentioned paradoxical situations, but also strengths the connection between physics and computer science.

A clear connection between physics and computer science is also present in other results. The second result reveals a close relationship between quantum metrology and the Deutsch-Jozsa algorithm over continuous-variable quantum systems. The Deutsch-Jozsa algorithm, being one of the first quantum algorithms, embodies the remarkable computational capabilities offered by quantum information processing. Here, we develop a general procedure, characterized by two parameters, that unifies parameter estimation and the Deutsch-Jozsa algorithm. Depending on which parameter we keep constant, the procedure implements either the parameter estimation protocol or the Deutsch-Jozsa algorithm. The procedure estimates a value of an unknown parameter with Heisenberg-limited precision or solves the Deutsch-Jozsa problem in a single run without the use of any entanglement.

The third result illustrates how physical principles that govern interaction of light and matter can be efficiently employed to create a computational resource for a (one-way) quantum computer. More specifically, we demonstrate theoretically a scheme based on atomic ensembles and the dipole blockade mechanism for generation of the so-called cluster states \emph{in a single step}. The entangling protocol requires nearly identical single-photon sources, one ultra-cold ensemble per physical qubit, and regular photo detectors. This procedure is significantly more efficient than any known robust probabilistic entangling operation.
}

\clearpage  

\setstretch{1.3}  

\acknowledgements{
\addtocontents{toc}{\vspace{1em}}  

I am truly grateful to my supervisors Pieter Kok and Stefan Weigert for all their patient help, guidance, encouragement and support. This gratitude is also warmly extended to my collaborator Carlos P\'{e}rez-Delgado for all his valuable suggestions.

Many special thanks to the people who made my time in Sheffield particularly enjoyable: Frank Bello, Andrew Carter, Alexander Chalcraft, Christopher Duffy, Entesar Ganash, Domnic Hosler, Carlos P\'{e}rez-Delgado, Mark Pogson, Nusrat Rafique, Andrew Ramsay, Samantha Walker and friends from the Department of Geography.

I would also like to express my deepest gratitude to my wife Agnieszka and my parents for their love and endless support.

Finally, I would like to thank the White Rose Foundation for funding my programme of study.
}
\clearpage  

\pagestyle{fancy}  

\lhead{\textsc{Contents}}  
\tableofcontents  

\lhead{\textsc{List of Figures}}  
\listoffigures  


\setstretch{1.5}  
\clearpage  
\lhead{\textsc{Abbreviations}}  
\listofsymbols{ll}  
{
\textbf{CX} & \textbf{C}ontrolled-\textbf{X} \\
\textbf{CZ} & \textbf{C}ontrolled-\textbf{Z} \\
\textbf{EPR} & \textbf{E}instein-\textbf{P}odolsky-\textbf{R}osen \\
\textbf{POVM} & \textbf{P}ositive \textbf{O}perator \textbf{V}alued \textbf{M}easure \\
\textbf{SQL} & \textbf{S}tandard \textbf{Q}uantum \textbf{L}imit \\
\textbf{CVs} & \textbf{C}ontinuous \textbf{V}ariable\textbf{s} \\
\textbf{MOT} & \textbf{M}agneto-\textbf{O}ptical \textbf{T}rapping \\
\textbf{BEC} & \textbf{B}ose-\textbf{E}instein \textbf{C}ondensate \\
\textbf{GLM} & \textbf{G}iovannetti \textbf{L}loyd \textbf{M}accone \\
\textbf{BFCG} & \textbf{B}oixo \textbf{F}lammia \textbf{C}aves \textbf{G}eremia \\
\textbf{RB} & \textbf{R}oy \textbf{B}raunstein \\
\textbf{DJ} & \textbf{D}eutsch-\textbf{J}ozsa \\
\textbf{EIT} &  \textbf{E}lectromagnetically \textbf{I}nduced \textbf{T}ransparency\\
\textbf{STIRAP} &  \textbf{STI}mulated \textbf{R}aman \textbf{A}diabatic \textbf{P}assage\\
\textbf{DLCZ} & \textbf{D}uan \textbf{L}ukin \textbf{C}irac \textbf{Z}oller \\
\textbf{DQC} & \textbf{D}istributed \textbf{Q}uantum \textbf{C}omputing  \\
\textbf{GHZ} &  \textbf{G}reenberger–-\textbf{H}orne–-\textbf{Z}eilinger\\
\textbf{HOM} &  \textbf{H}ong-\textbf{O}u-\textbf{M}andel\\
\textbf{SPDC} &  \textbf{S}pontaneous \textbf{P}arametric \textbf{D}own-\textbf{C}onversion\\
}




\setstretch{1.5}  

\pagestyle{empty}  
\dedicatory{To Agnieszka and my parents}

\addtocontents{toc}{\vspace{2em}}  

\mainmatter	  
\pagestyle{fancy}  



\part{Introduction}

\chapter{Quantum Information Processing}
\label{Chapter1}
\lhead{\textsc{Chapter 1. Quantum Information Processing}}

\section{Introduction}
Quantum information theory is a novel branch of science that exploits the remarkable features of quantum mechanics to store, manipulate and transfer information in ways that are unattainable to any classical device. It is arguably one of the most exciting branches of science that promises a huge impact on many other disciplines. Quantum information theory lays at the intersection of theoretical and experimental physics, and computer science. Thus, the impact it may have on these disciplines is quite clear. Surprisingly, quantum effects also seem to have a large significance in some phenomena in biology such as the light-harvesting complexes that are capable to efficiently transmit a single quantum of light on a relatively long distance or the avian compass that birds use to navigate in the magnetic field of Earth. Therefore, a fundamentally deeper understanding of some biological systems may be impossible without an ``insight" from the field of quantum information. Also, molecular chemistry can be greatly influenced by quantum information science, if we are able to build a quantum simulator that would allow us to study the behaviour of complex molecules. Quantum computation over discrete or continuous-variable quantum systems is main field of quantum information theory. The quantum phenomena can also be harnessed to perform measurements of physical quantities with a precision inaccessible to any classical device.

The organisation of Chapter~\ref{Chapter1} reflects the order at which different subjects are introduced in the remaining chapters. In Sec.~\ref{sec:qc}, we recall basic notions of quantum computation such as a qubit and a quantum gate. In Sec.~\ref{dqc}, we introduce a more practical and less abstract form of quantum computation, namely distributed quantum computation. Distributed quantum computation is closely related to quantum communication. This relation is so close that people perceive them as two faces of the same coin, that is, if you can establish quantum network and transfer information between its nodes, you can perform quantum computation. In the same section, we present the measurement-based model of quantum computation that can be implemented in a distributed manner. In Sec.~\ref{qmet}, we review the basic foundations of quantum metrology - an important discipline of quantum information theory that is concerned with high-precision measurements. In Sec.~\ref{cv}, we introduce an alternative to quantum computation based on discrete quantum systems (qubits), namely continuous-variable quantum computation. In this section, we review basic properties of continuous quantum systems and present some basic continuous-variable quantum gates. Finally, in Sec.~\ref{atomic}, we introduce the concept of an atomic ensemble, a physical system that can be used in distributed quantum computation.

\section{Quantum computation}\label{sec:qc}
The construction of a quantum computer is an important goal of modern science, which requires an effort from both experimental and theoretical physicists, and quantum computer scientists. A quantum computer is a computing device whose operation is based on the principles of quantum mechanics. The quantum computer exploits the non-classical and counterintuitive phenomena of quantum mechanics such as superposition, entanglement, quantum interference and quantum measurement to perform some computations more efficiently than any classical computer \cite{nielsen}. The basic unit of information for a quantum computer is called a quantum bit or qubit. A qubit is an abstraction of a two-dimensional quantum system that consists of two addressable quantum states, so-called basis states $|0 \rangle$ and $|1 \rangle$, that is, the computational basis. A qubit is represented as a vector that lives in a two-dimensional Hilbert space. The $|0 \rangle$ and $|1 \rangle$ states are analogous to the 0 and 1 of a classical bit. In contrast with classical bits, qubits can exist in any superposition of basis states such as $|\psi \rangle = \alpha |0 \rangle + \beta |1 \rangle$, where $\alpha$ and $\beta$ are complex numbers called amplitudes that obey $|\alpha|^{2} + |\beta|^{2} = 1$. This is the so-called superposition principle. A qubit can exist in a superposition of both basis states, until we try to observe it by performing a measurement. By means of a measurement, we find a qubit in one of the basis states with a probability given by a square of the amplitudes: $|\alpha|^{2}$ for the $|0 \rangle$ state and $|\beta|^{2}$ for the $|1 \rangle$ state. For the state $|\psi \rangle = |+ \rangle = \frac{1}{\sqrt{2}} (|0 \rangle + |1 \rangle)$, the qubit has an equal probability: 50\%, of being in the $|0 \rangle$ or $|1 \rangle$ state. Therefore, if we repeat the measurement in the computational basis many times, on average half of the outcomes will yield a classical value of either 0 or 1 and the state of the qubit will be collapsed to the basis state $|0 \rangle$ or $|1 \rangle$, respectively. The superposition principle applies not only to a single qubit but to many qubits as well. In the case of two qubits $|\phi \rangle = \alpha |0 \rangle + \beta |1 \rangle$ and $|\varphi \rangle = \gamma |0 \rangle + \delta |1 \rangle$, the state of a composite system is given by the tensor product $|\phi \rangle \otimes |\varphi \rangle$ of the form:
\begin{eqnarray}
|\psi \rangle = |\phi \rangle \otimes |\varphi \rangle &=& (\alpha |0 \rangle + \beta |1 \rangle)\otimes (\gamma |0 \rangle + \delta |1 \rangle) \nonumber \\ &\equiv& \alpha\gamma |00 \rangle + \alpha\delta |01 \rangle + \beta\gamma |10 \rangle + \beta\delta |11 \rangle.
\end{eqnarray}
The two-qubit composite state is a vector in 4-dimensional Hilbert space. Naturally, this reasoning generalises to any number of qubits. The most intriguing kind of composite states in quantum mechanics are so-called entangled states. One of the entangled states of two qubits is given by
\begin{equation}
|\psi \rangle_{AB} = \frac{1}{\sqrt{2}} (|0 \rangle_{A}|0 \rangle_{B} + |1 \rangle_{A}|1 \rangle_{B})\, .
\end{equation}
This state together with three other two-qubit entangled states is the so-called set of Bell basis states. What is so special about entangled states? First of all, the entangled states cannot be factored into a tensor product: $|\psi \rangle \neq |\phi \rangle \otimes |\varphi \rangle $ for any basis states. Furthermore, one may notice that if the measurement of the first qubit $A$ yields 0 then the state of the second qubit $B$ is instantaneously collapsed to $|0 \rangle$. The same occurs for the second qubit $B$. For the entangled state $|\psi\rangle_{AB}$, the measurement results are perfectly correlated. Although the qubits may be separated by a large distance, their behaviour is in some sense synchronised, i.e., the measurement of one of them affects instantaneously the state of the other. This non-local character or the so-called ``spooky action at a distance" of the Bell pair is called entanglement. In order to show that qubits share nonclassical correlations, that is, they are entangled, we also need perfect correlations in the $\{|+ \rangle, |- \rangle\}$ basis, where $|- \rangle = \frac{1}{\sqrt{2}} (|0 \rangle - |1 \rangle)$.  The true importance of entanglement is still unclear, however, it is considered essential for quantum computation \cite{wootters,horodecki}. In fact, many ``quantum tricks" such as quantum teleportation or superdense coding rely heavily on the entangled states.

The quantum computer processes information by applying some set of quantum operations on qubits according to a blueprint called a quantum algorithm \cite{nielsen}. These operations consist of linear, unitary evolutions $U$: single and two-qubit operations (the so-called gates), and measurements (a measurement can also ``process" information as can be readily seen in section \ref{mbqc}) \cite{koklecture}. The unitarity of the quantum gates, $U^{\dagger}U = I$, implies that the quantum computation is reversible. The single-qubit operations can be represented graphically in the \emph{Bloch} sphere. A Bloch sphere is a geometrical representation of the state space of a qubit and any unitary single-qubit gate can be described as a rotation in the Bloch sphere. The three most important single-qubit gates are the so-called Pauli operators $X$, $Y$ and $Z$. In matrix notation, Pauli operators have the following representations in the $|0\rangle$, $|1\rangle$ basis
\[ X = \left[ \begin{array}{cc}
0 & \ \ 1 \\
1 & \ \ 0\end{array} \right], \quad Y = \left[ \begin{array}{cc}
0 & -i\\
i & \ \ 0\end{array} \right], \quad Z = \left[ \begin{array}{cc}
1 & \ \ 0 \\
0 & -1\end{array} \right].\]
In the computational basis, the $X$ operator is a bit flip, and the $Z$ operator is a phase flip, that is, a phase rotation in the Bloch sphere. The $Y$ operator can be constructed from $X$ and $Z$ operators \cite{koklecture}. Another extremely useful and essential for quantum computation operation is the Hadamard gate $H$. In matrix notation, the Hadamard operation is given by
\[ H = \frac{1}{\sqrt{2}}\left[ \begin{array}{cc}
1 & \ \ 1 \\
1 & -1\end{array} \right].\]
The Hadamard gate applied to the basis state $|0 \rangle$ and $|1 \rangle$ returns the balanced superposition states $|+\rangle = \frac{1}{\sqrt{2}} (|0 \rangle + |1 \rangle)$ and $|-\rangle = \frac{1}{\sqrt{2}} (|0 \rangle - |1 \rangle)$, respectively. Therefore, the Hadamard gate gives rise to the superposition states of possibly large number of qubits. The last of the crucial single-qubit gates is the general phase shift operation $R(\varphi)$ represented as
\[ R(\varphi) = \left[ \begin{array}{cc}
1 & 0 \\
0 & e^{i \varphi}\end{array} \right].\]
For $\varphi = \pi$ the phase shift gate takes the form of the Pauli $Z$ operator. When $\varphi = \pi/2$ and $\varphi = \pi/4$, the phase shift operator corresponds to the $\pi/2$-phase gate and $\pi/8$ gate, respectively.

Two important two-qubit gates are the con\-trolled-X (CX) and the con\-trolled-Z (CZ), which are applied between the so-called \textit{control} and \textit{target} qubits. The matrix representation of these gates is the following
\[ \mbox{CX} = \left[ \begin{array}{cccc}
1 & \ \ 0 & \ \ 0 & \ \ 0\\
0 & \ \ 1 & \ \ 0 & \ \ 0\\
0 & \ \ 0 & \ \ 0 & \ \ 1\\
0 & \ \ 0 & \ \ 1 & \ \ 0\end{array} \right],\]

\[ \mbox{CZ} = \left[ \begin{array}{cccc}
1 & \ \ 0 & \ \ 0 & \ \ 0\\
0 & \ \ 1 & \ \ 0 & \ \ 0\\
0 & \ \ 0 & \ \ 1 & \ \ 0\\
0 & \ \ 0 & \ \ 0 & -1\end{array} \right].\]
The CX operation flips the state of the target qubit by applying the $X$ operation only when the control qubit is in the basis state $|1\rangle$ (the state of the control qubit is left unchanged). In other words, the CX stores the result of addition modulo 2 of both qubit states in the state of the target qubit. In the case of CZ gate, the $Z$ operation is applied to the target qubit if the control one is present in the basis state $|1\rangle$ (again the state of the control qubit is left unchanged), otherwise states of both qubits are unchanged. The importance of the CX and CZ gates stems from the fact that together with the Hadamard gate $H$, we can create entangled states of any number of qubits initially prepared in one of the basis states. Furthermore, the CX or CZ gates and single-qubit gates serve as a basis building block for any other two-qubit gate \cite{nielsen}.

The linearity of the quantum gates means that qubits in any superposition state of the computational basis can be manipulated by applying these gates. This suggests that a single quantum computer can process information in parallel, a phenomenon known as \textit{quantum parallelism}. Therefore, by means of the superposition principle, linear quantum gates and quantum interference amplitudes of the favoured states that represent the correct answer to the computational problem can be enhanced. In a nutshell, this is why quantum computers are capable of solving some computational problems more efficiently than any classical computer. The phenomena described in this section may possibly constitute the foundation for the power of quantum computation. However, it is still unknown how large is the class of computational problems that can be solved efficiently on a quantum computer with respect to its classical counterpart \cite{nielsen}. Therefore, we are still not confident whether quantum computation is, in principle, more powerful than classical computation.

In the next section, we abandon the abstract way of thinking about quantum computation and introduce an architecture that can be used to physically build a quantum computer, the so-called distributed quantum computer.

\section{Distributed quantum computation}\label{dqc}
There are many physical systems in which a qubit and a quantum computer as a whole can be realised. One can represent a qubit as a spin of an electron, a nucleus or even an atom \cite{loss,burkard,kane,privmana,jaksch98,brennen99}. Other physical representations of qubits are based on Josephson junctions (so-called superconducting qubits) or quantum dots \cite{makhlin,martinis,chiorescu,devoret,loss,imamoglu99,elzerman}. One of the prominent approaches to the physical implementation of a qubit and quantum computer is linear quantum optics \cite{knill,koklinear}. One can use coherent and  squeezed states of light or even a single photon (Fock state or polarisation degree of freedom of a photon) to represent a quantum bit \cite{knill,koklinear}. The drawback of photonic systems for quantum computation is the fact that there is no direct interaction between photons. Nevertheless, photons are perfect carriers of quantum information and can be utilised in the distributed model of quantum computation as quantum communication channels \cite{nielsen,duan}.

At the present time, a number of models of quantum computation exist, such as adiabatic quantum computing, or the most widely used standard circuit model of quantum computation. Regardless of the model of quantum computation, anyone trying to build a quantum computer faces two main challenges:
\begin{enumerate}
\item the problem of decoherence, that is, how well we can suppress the unwanted influence of the environment on our quantum computer,
\item the problem of scalability of basic modules of our quantum computer.
\end{enumerate}
The difficulties associated with the fragility of quantum information (decoherence) and scalability of a quantum computer architecture are one of the most important cornerstones of the distributed version of quantum computation. Decoherence, i.e., the deterioration of the quantum state, affects each qubit and introduces errors to the computation. This has to be suppressed to the lowest level possible, but crucially below the fault tolerance threshold \cite{shor96}. As one would expect, any interesting, i.e., complicated, computational problem usually employs many qubits. The most well known quantum algorithms, Shor's factoring algorithm, Grover's database search algorithm and the Deutsch-Jozsa algorithm have been demonstrated experimentally but only for few qubits \cite{vandersypen,chuang1,chuang2}. These experiments are proof-of-principle experiments of quantum computation power. All of these suggest that a truly useful and powerful quantum computer has to be robust and scalable machine. In the case of many qubits, which may interact with the environment and their neighbours, protection against decoherence becomes quite a challenging task. The scalability and decoherence issue are the main difficulties that are addressed by distributed quantum information processing. It may be much more feasible to build a number of small-scale remotely distributed quantum computers (processors) and connect them together instead of one large machine. In the distributed model of quantum computation, a small number of stationary qubits are placed in the (distant) nodes of a large network. A distributed quantum computer may also be based on a model of quantum computation that is inherently distributed such as a measurement-based model of quantum computation \cite{danos}. Here, the computation is done via single-qubit measurements and feed-forward operations on a large, multi-qubit, entangled graph state \cite{rauss,hein}. The stationary qubits are usually encoded in the ground levels of trapped atoms, ions or quantum dots and therefore can additionally serve as a good quantum memory \cite{lim}. This kind of qubit implementation allows for fast and reliable single qubit operations and rather straightforward measurement techniques. In this setting a possibly large collection of small-scale quantum processors can solve a single computational problem as long as they communicate the outputs of their computations with each other or with a central quantum processor. Robust communication between any two stationary nodes (qubits) is usually provided via flying qubits - single photon qubits \cite{lim_lett}. Computation with a distributed quantum network consists of the preparation of initial states, which may involve exchange of classical and quantum information between nodes. Next, computation at each node is performed and then all the partial results from each node are sent to the central processor \cite{cirac}. The central node gathers results and returns the final answer to the computational problem. Since the quantum computation is probabilistic in nature, one may have to repeat the distributed computation many times until the required result is obtained. The advantages of the distributed model of quantum computation, which result from the spatial separation of stationary qubits, are the following:
\begin{enumerate}
\item each qubit is uniquely addressable. Therefore, control and measurement of an individual quantum processor is completely decoupled from the rest of the computational resources. Naturally, better protection against decoherence originating from the interaction with the environment is more feasible too.
\item enhanced flexibility. By means of the optical elements qubits may interact with each other more easily. Entanglement can in principle be generated between any two stationary qubits. Moreover, the distributed character of the architecture allows for applications not only in quantum computation but in quantum communication too.
\end{enumerate}
Even though each node of a quantum network consists of a small number of qubits, decoherence still will lead to errors and deterioration over time \cite{meter}. In order to avoid this scenario, one may encode logical qubits in many physical qubits and apply error correcting procedures \cite{meter}. The main disadvantage of the distributed model of quantum computation is the lack of local interaction between nodes, therefore the need for entangling procedures. Naturally, the distributed quantum computation has to operate on distributed versions of known standard quantum algorithms. In other words, the centralised quantum algorithm has to be distributed over nodes of a large quantum network too. This adds an additional cost associated with communication to the overall cost of a computation \cite{grover}. Consequently, one has to decide how to partition a single problem between many remotely distributed quantum processors in an optimal way and then how to communicate and collect outputs of these processors, effectively finding the final solution to the computational problem \cite{grover}. This issue was first addressed by Eisert \textit{et al.} where they considered how to distribute the CX and number of other important gates between two quantum processors \cite{eisert}. Eisert \textit{et al.} proved that implementation of distributed version of the CX gate requires one pre-shared EPR pair and communication of two classical bits between two individual quantum nodes. Since the CX gate is a basic building block of any other multi-qubit gate and together with general single-qubit operation it constitutes a universal set of gates for universal quantum commutation, the distributed model of  quantum computation is universal \cite{nielsen,denchev}. Apart from devising the non-local version of gates, Eisert \textit{et al.} addressed the problem of minimal resources, both classical and quantum ones, and optimal procedures that are required to implement these distributed gates \cite{eisert}.

In most models for distributed quantum computing one assumes that all quantum processors work perfectly \cite{denchev}. Moreover, one is able to transfer and store, manipulate, and retrieve quantum states from each of the nodes of an arbitrary quantum network. Concerning communication, there are few possibilities allowed. In some models, communication is done only with qubits or only with classical bits. Commonly, some amount of entanglement is prepared between the qubits when the quantum network is initialised. Often various nodes share EPR pairs and communication is established with only classical bits (quantum teleportation) or classical bits and qubits (super-dense coding) \cite{buhrman}. Obviously, generation of the pre-shared entanglement can be quite challenging especially for large networks. In some cases the cost of entanglement preparation can render the distributed quantum computation with pre-shared entanglement inefficient in comparison with other models of distributed quantum computing based on disentangled states. Nevertheless, use of the pre-shared entangled states under ideal conditions is usually advantageous over uncorrelated ones \cite{cirac}. Furthermore, even for noisy communication channels one can employ purification procedures \cite{cirac}. Naturally, the resources one exploits to solve a computational problem will depend on the problem at hand and available methods for entanglement generation. On the other hand, in some models of distributed quantum computing, nodes communicate with each other without any pre-shared entanglement by means of flying qubits (single photons).

\subsection{Quantum communication}
In quantum communication protocols, photons serve as carriers of quantum information between nodes of communication network. In most of the quantum communication protocols, an important task for photons is to generate perfect entangled states between distant nodes. This is not a trivial task. Each photon that carries quantum information between the nodes of a quantum network is prone to losses. The probability that photon is lost is given by
\begin{equation}
p_{lost} = \exp\left(-\frac{l_{0}}{l_{att}}\right)\, ,
\end{equation}
where $l_{0}$ is the communication distance and $l_{att}$ is the characteristic channel attenuation distance \cite{duan,kok_book}. This implies an exponential attenuation that decreases the fidelity of quantum communication protocols. The solution that addresses these limitations was given in terms of quantum repeater and purification protocols. Some of the well known quantum repeater and purification protocols are probabilistic. This imposes a requirement for a medium that would facilitate an interaction between photons, and store photonic qubits, i.e., a stationary qubit. Hence, the concept of an optical quantum memory realised in atomic vapour (atomic ensemble) was introduced. Consequently, an optical quantum memory is a necessary ingredient in many quantum communication protocols and an essential ingredient in many optical quantum information processing protocols.

In general, a quantum memory has to fulfil the following requirements: efficient mapping of a photon into the memory, long storage times and efficient retrieval of a photon back from the memory. Moreover, one has to be able to control the state of a quantum memory at all times.
The storage time itself has to be much longer than the characteristic time scale of an application in which quantum memory is used. Not all of these requirements have to be met for all quantum applications. In fact, for some applications such as quantum computation, the first and third requirement can be lifted and a quantum memory can serve as a qubit itself, the so-called stationary or matter qubit. Ideally, all operations that concern quantum memories should be highly efficient and deterministic. Unfortunately, this is never the case and all realistic quantum memories are imperfect. Hence, a question arises: how to evaluate the performance of a quantum memory? The most commonly used measure of quantum memory performance is the average fidelity $F$, i.e., state overlap between the input and output quantum states \cite{nielsen,hammerer}. A quantum memory characterised with unit average fidelity perfectly maps the input state, stores it for some time and returns it unchanged. Naturally, a truly quantum memory has to outperform any classical memory for quantum state storage \cite{hammerer}. A classical memory fidelity for quasi-classical bright coherent states is $F_{classical} = \frac{1}{2}$, \cite{braunstein} and for an arbitrary qubit states, the maximal classical fidelity is $F_{classical} = \frac{2}{3}$ \cite{massar}. Therefore, any truly quantum memory has to exceed these classical bounds. Fidelity is not the only measure for quantifying the performance of quantum memories. Similar to the case of the requirements, an appropriate measure for quantum memory performance depends on a particular application \cite{hammerer}.

\subsection{The one-way model of quantum computation}\label{mbqc}
\begin{figure}[t]
\begin{center}
\includegraphics[scale=1.5]{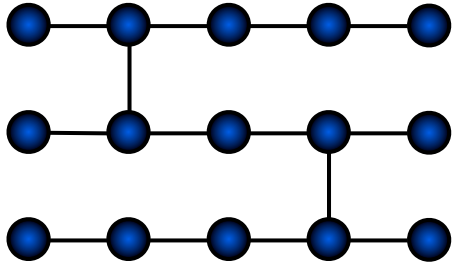}
\end{center}
\caption{A graph state. Nodes represent physical qubits which are connected via the CZ operations. Horizontal strings of
physical qubits constitute logical qubits. The vertical links between logical qubits represent two-qubit CZ gates.\label{graph}}
\end{figure}

A natural candidate for a distributed model of quantum computation is the so-called measure\-ment-based or one-way model of quantum computation realised on graph states \cite{koklecture,danos,nielsen_microcluster}. Although, the very first experimental proposal for a one-way quantum computer was based on optical lattices (where cold atoms are \textit{locally} trapped in a standing-wave potential created by counter-propagating laser fields \cite{briegel09,briegel01,jaksch99}), nevertheless this model of quantum computation is especially well suited for the distributed implementation. What is a graph state? Graph or cluster states are large entangled states that act as a universal resource for a one--way quantum computer \cite{koklecture,rauss,hein}. The cluster states are represented in the form of a lattice or a graph. We associate with every node $j$ of a graph an isolated qubit in the state $| + \rangle_{j} = \frac{1}{\sqrt{2}} (|0\rangle_{j} + |1\rangle_{j})$ subsequently connected, that is, entangled, with adjacent qubits via the CZ$_{jk}$ operations
\begin{equation}
\mbox{CZ}_{jk} = |0\rangle_{j}\langle0| \otimes \mathbb{\hat{I}}_{k} + |1\rangle_{j}\langle1| \otimes Z_{k}\, ,
\end{equation}
where $|0\rangle$, $|1\rangle$ are the computational basis states, $Z$ is the Pauli operator and $\mathbb{\hat{I}}$ denotes the identity matrix. Commonly, graph states are described in terms of the stabilizer operators. A set of commuting operators $S_{j}$ constitutes a stabilizer of the quantum state $|\phi\rangle$ under which the state is invariant. The stabilizer formalism allows us to describe multi-qubit quantum states and their evolution in terms of few stabilizer operators, which usually consist of operators from the Pauli group $G_{n}$ on $n$ qubits. The Pauli group $G_{1}$ on a single qubit is a group under matrix multiplication consisting of the identity matrix and Pauli matrices multiplied by $\pm 1$, $\pm i$ factors. The Pauli group $G_{n}$ on $n$ qubits is an $n$ tensor product of the Pauli group $G_{1}$ \cite{nielsen}. The state $|\phi_{n}^{C}\rangle$ of a cluster $C$ consisting of $n$ qubits is completely specified by the following set of eigenvalue equations:
\begin{equation}
S_{j} |\phi_{n}^{C}\rangle=|\phi_{n}^{C}\rangle,
\end{equation}
with
\begin{equation}
S_{j} = X_{j} \prod_{k \ \in \ \mbox{nghb}(j)} Z_{k}\, ,
\end{equation}
where $\mbox{nghb}(j)$ is the set of all neighbours of qubit $j$ \cite{rauss}. The $S_{j}$  are Hermitian stabilizer operators whose eigenstates, i.e., the graph states, are mutually orthogonal and form a basis in the Hilbert space of the cluster \cite{rauss}. Cluster states and quantum algorithms implemented on them may be related to mathematical graphs \cite{rauss,hein}. A graph $G(V,E)$ is a pair of a finite set $V$ of vertices connected with edges $e$ from the set $E$. A cluster $C$ is identified with the vertices $V_{C}$ of a graph $C=V_{C}$ \cite{hein}. The set $E_{C}$ of edges is given by $E_{C}=\{(a,b)|a,b \in C, b \in \mbox{nghb}(a)\}$ \cite{hein}. Edges $e$ are realised by CZ operations and connect two vertices of a graph (Fig.~\ref{graph}). The well-known graph theory notation is a very useful tool in analysing properties of the cluster states.

Let us now review some details of the one-way model of computation. In the measure\-ment-based model of quantum computing, the entire resource for quantum computation is provided from the beginning as a graph state (Fig.~\ref{graph}). Quantum computation consists of single-qubit measurements on the graph states and every quantum algorithm is encoded in a measurement blueprint. A measurement of a qubit in the $Z$ eigenbasis, i.e., in the computational basis, removes a qubit from a graph and all links to its neighbours are broken. Consequently, a cluster is reduced by one qubit, and possible corrective $Z$ operations are applied to its neighbours depending on the measurement outcome (if the measurement result is 0 then nothing happens, but when the measurement outcome is 1 a phase-flip is applied to all neighbours). By means of a $Z$ measurement, any cluster can be carved out from a generic, fully connected cluster (Fig.~\ref{graph}).

Other single-qubit measurements are performed in the basis
\begin{equation}
B(\alpha) \in \\ \{|\alpha_{+}\rangle,|\alpha_{-}\rangle\}\, , \ \mbox{where} \ |\alpha_{\pm}\rangle = \frac{1}{\sqrt{2}} (|0\rangle \pm e^{i \alpha}|1\rangle)\, .
\end{equation}
For $\alpha = 0$ the measurement is realised in the $X$ eigenbasis. An interesting feature of $X$ measurement is that two neighbouring $X$ measurements in a linear cluster remove measured qubits and connects their neighbours with each other resulting in a shortened cluster. For $\alpha = \pi/2$, the $Y$ measurement is performed. In the case of a $Y$ measurement, the measured qubit is removed from a cluster and its neighbours are connected (up to a corrective phase operation). Measurements in the $X$ and $Y$ eigenbases propagate quantum information through a cluster. In general, any quantum computation proceeds as a series of measurements governed by an appropriate blueprint. The choice of measurement basis for every physical qubit is encoded in this measurement blueprint. Moreover, all measurement bases depend on the outcomes of the preceding measurements. This implements the so-called feed-forward operation. Although the result of any measurement is completely random, information processing is possible because of the feed-forward operations. The feed-forward operations ensure that measurement bases are correlated and a deterministic computation can be realised. In this way quantum information propagates (due to the feed-forwarding which implies time ordering in \textit{one way}) through the cluster until the last column of qubits, which are then ready to be read out. Readouts are performed in the $Z$ eigenbasis up to Pauli corrections and the output of the computation is given as a classical bit string \cite{rauss}.
\begin{figure}[t]
\begin{center}
\includegraphics[scale=1.3]{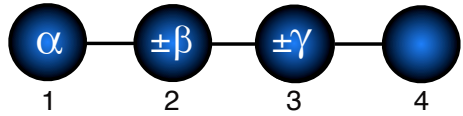}
\end{center}
\caption{A linear 4-qubit cluster. Nodes represent a physical qubits which are connected via the CZ operations. \label{chain}}
\end{figure}

A simple example of a measurement-based computation can presented on a linear 4-qubit cluster given by
\begin{equation}
|\phi_{4}^{C}\rangle = \frac{1}{2} \left(|+ 0 + 0\rangle + |+ 0 - 1\rangle + |- 1 - 0\rangle + |- 1 + 1\rangle\right)
\end{equation}
with $|-\rangle = \frac{1}{\sqrt{2}} (|0\rangle - |1\rangle)$. Although this is a very basic cluster, it allows us to perform
an arbitrary single-qubit rotation in only three (measurement) steps:
\begin{itemize}
\item measure qubit 1 in basis $B_{1}(\alpha)$,
\item measure qubit 2 in basis $B_{2}((-1)^{m_{1}} \beta)$ depending on the outcome $m_{1}$ of the previous measurement,
\item measure qubit 3 in basis $B_{3}((-1)^{m_{2}} \gamma)$ depending on the outcome $m_{2}$ of the previous measurement.
\end{itemize}
Following these measurements an arbitrary single-qubit rotation (up to corrective Hadamard $H$ and Pauli $X$, $Z$ operations) is applied to the fourth qubit in a linear cluster according to the unitary transformation $U_{rot}$ given by \cite{anders08}
\begin{equation}
U_{rot} = R_{z}((-1)^{m_{2}} \gamma) R_{x}((-1)^{m_{1}} \beta) R_{z}(\alpha)
\end{equation}
We again emphasize the importance of the feed-forward operations. The angles of the rotation and by implication the final corrective operations depend on the outcomes of previous measurements \cite{anders08}.

On the basis of cluster states a universal set of quantum gates can be implemented, e.g., single-qubit gates such as the Hadamard, the $\pi$/2-phase gate and $\pi$/8 gate, and a two-qubit CX gate \cite{nielsen,nielsen_microcluster}. Most importantly, the measurement-based model of quantum computation on cluster states is completely equivalent to the standard circuit model, thus the one-way model is capable of efficiently simulating any quantum circuit. Consequently, the measurement-based model of computation is a universal model of quantum computation \cite{nielsen_microcluster}.

Cluster states are a very promising resource for quantum information processing. One possible way of creating large networks of qubits is by trapping small atomic ensembles in optical lattices or placing them in the distributed nodes of a quantum network. Therefore, in Sec.~\ref{atomic}, we introduce the concept of an atomic medium as a quantum memory for light. Since a cluster state consists of a large set of entangled qubits, efficient protocols for generating entanglement between nodes of a network are required. We review some of the well-known entangling procedures in Chapter~\ref{Chapter5} and present a new procedure based on some manipulation techniques for atomic ensembles that are described in detail in Chapter~\ref{Chapter4}.

In the next section, however, we present foundations of another important discipline of quantum information theory, namely quantum metrology.

\section{Quantum metrology}\label{qmet}

Quantum metrology, or quantum parameter estimation theory, is an important and relatively young branch of science that received a lot of attention in recent years. It studies high-precision measurements of physical parameters, such as phase, based on systems and physical evolutions that are governed by the principles of quantum mechanics. The main theoretical objective of this field is to establish the ultimate physical limits on the information we can gain from a measurement. From an experimental perspective, quantum-enhanced metrology promises many advances in science and technology since an optimally designed quantum measurement procedure outperforms any classical procedure. Furthermore, improved measurement techniques frequently lead not only to the technological advancement but also to a fundamentally deeper understanding of Nature. The main figure of merit in the field of quantum metrology for both theorists and experimentalists is the precision with which the value of an unknown parameter can be estimated.

\subsection{The quantum Cram\'{e}r-Rao bound}
In this section, we introduce the two most crucial concepts in quantum metrology, namely the Fisher information and the quantum Cram\'{e}r-Rao bound. The Fisher information is a quantity that measures the amount of information about the parameter we wish to estimate revealed by the measurement procedure. Given the Fisher information, we can bound the minimal value of uncertainty in the parameter with the quantum Cram\'{e}r-Rao bound. Here, we consider the estimation of a single parameter $\varphi$. The most general parameter estimation procedure corresponding to any conceivable experimental setup is shown in Fig.~\ref{PE}. This procedure consists of three elementary steps:
\begin{enumerate}
 \item prepare a probe system in an initial quantum state $\rho(0)$,
 \item evolve it to a state $\rho(\varphi)$ by a unitary evolution $U(\varphi)=\exp(-i\varphi\mathcal{H})$, where the Hermitian operator $\mathcal{H}$ is the generator of translations in the parameter $\varphi$,
 \item subject the probe system to a generalised measurement $M$, described by a Positive Operator Valued Measure ({\sc povm}) that consists of elements $\hat{E}_x$, where $x$ denotes the measurement outcome.
\end{enumerate}
\begin{figure}[t]
\centering
\includegraphics[width=6cm]{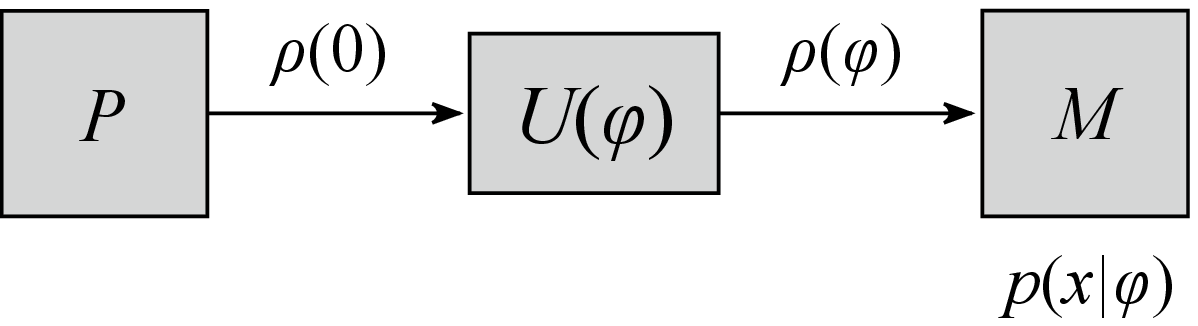}
\caption{The general parameter estimation procedure involving state preparation $P$, evolution $U(\varphi)$, and generalised measurement $M$ with outcomes $x$, which produces a probability distribution $p(x|\varphi)$. \label{PE}}
\end{figure}
The conditional probability $p(x|\varphi)$ of finding measurement outcome $x$ is given by the Born rule
\begin{equation}
p(x|\varphi) = \mbox{Tr}[\hat{E}_x \rho(\varphi)]
\end{equation}
with $\int dx \, \hat{E}_x = \hat{\mathbb{I}}$. Given the probability distribution $p(x|\varphi)$, we can derive the expression for the Fisher information and subsequently the quantum Cram\'{e}r-Rao bound. The following derivation is due to Braunstein and Caves \cite{braunstein94}, and can also be found in Kok and Lovett \cite{kok_book}. We start the derivation by noting that the above measurement procedure returns the measurement outcome $x$ with probability $p(x|\varphi)$ instead of a more desired single value for the parameter with probability $p(\varphi|x)$. Therefore, we need to relate these two values with a help of a special function called an estimator. The estimator $T(x)$ for a parameter $\varphi$ is a function that allows us to find the value of the parameter $\varphi$ given the measurement outcome $x$. For an estimator $T$, we define $\Delta T \equiv T(x) - \langle T \rangle_{\varphi}$ with
\begin{equation}
\langle T \rangle_{\varphi} \equiv \int dx \, p(x|\varphi) T(x) \, .
\end{equation}
When $\langle T \rangle_{\varphi} = \varphi$, the estimator is unbiased. Given $N$ independent measurement outcomes $x_{1}, \ldots, x_{N}$ we can write
\begin{equation}
\int dx_{1} \cdots dx_{N} \, p(x_{1}|\varphi) \cdots p(x_{N}|\varphi) \Delta T = 0 \, , \label{start}
\end{equation}
Following the definition of $\Delta T$, we can easily verify that Eq.~\ref{start} holds for \textit{any} estimator $T$. Next, we take the derivative to $\varphi$ of Eq.~\ref{start} and rewrite it as
\begin{equation}
\int dx_{1} \cdots dx_{N} \, p(x_{1}|\varphi) \cdots p(x_{N}|\varphi) \left(\sum_{i = 1}^{N} \frac{\partial \ln p(x_{i}|\varphi)}{\partial \varphi} \right) \Delta T = \left\langle \frac{d\langle T \rangle_{\varphi}}{d\varphi} \right\rangle \, .
\end{equation}
Now we apply the Cauchy-Schwarz inequality:
\begin{equation}
|\langle f,g \rangle|^{2} \leq \langle f,f \rangle \langle g,g \rangle
\end{equation}
with $f$ and $g$ defined as
\begin{equation}
f = \sum_{i = 1}^{N} \frac{\partial \ln p(x_{i}|\varphi)}{\partial \varphi} \, , \ \ g = \Delta T \, .
\end{equation}
Hence, we obtain
\begin{eqnarray}
&&\int dx_{1} \cdots dx_{N} \, p(x_{1}|\varphi) \cdots p(x_{N}|\varphi) \left(\sum_{i = 1}^{N} \frac{\partial \ln p(x_{i}|\varphi)}{\partial \varphi} \right)^{2} \nonumber \\
&&\times \int dx_{1} \cdots dx_{N} \, p(x_{1}|\varphi) \cdots p(x_{N}|\varphi) (\Delta T)^{2} \geq \left|\left\langle \frac{d\langle T \rangle_{\varphi}}{d\varphi} \right\rangle\right|^{2} \, .
\end{eqnarray}
We identify the first term with the Fisher information $F(\varphi)$ and rewrite this inequality as
\begin{equation}
N F(\varphi) \times \langle(\Delta T)^{2}\rangle_{\varphi} \geq \left| \frac{d\langle T \rangle_{\varphi}}{d\varphi} \right|^{2} \, ,
\end{equation}
where
\begin{equation}
F(\varphi) \equiv  \int dx \, p(x|\varphi) \left(\frac{\partial \ln p(x|\varphi)}{\partial \varphi} \right)^{2} = \int dx \, \frac{1}{p(x|\varphi)} \left(\frac{\partial p(x|\varphi)}{\partial \varphi} \right)^{2} \, . \label{fisher}
\end{equation}
The Fisher information measures the average squared rate of change of the conditional probability distribution (derived from a measurement) with the parameter $\varphi$. Therefore, higher sensitivity of the probe system to the parameter in question implies higher Fisher information. Strictly speaking, the Fisher information quantifies the amount of information about parameter $\varphi$ extracted from the probe system prepared in $\rho(\varphi)$ by a generalised measurement described by the {\sc povm}. The unit of the Fisher information is given by the inverse squared unit of the parameter in question, that is, $[F(\varphi)] = 1/[\varphi]^2$. The above inequality relates the Fisher information $F(\varphi)$ and the average error in the estimator $T$. However, we want to express it in terms of the average error in the actual value of $\varphi$. Therefore, we use the following expression for the error $\Delta \varphi$:
\begin{equation}
\Delta \varphi \equiv \frac{T(x)}{|d\langle T \rangle_{\varphi}/d\varphi|} - \varphi \, . \label{error_phi}
\end{equation}
The derivative accounts for a possible change in the units between the average value of the estimator $\langle T \rangle_{\varphi}$ and parameter $\varphi$. In order to find a relationship between $\langle(\Delta T)^{2}\rangle_{\varphi}$ and $\langle(\Delta \varphi)^{2}\rangle_{\varphi}$, we use $\Delta T \equiv T(x) - \langle T \rangle_{\varphi}$ to calculate
\begin{equation}
\langle(\Delta T)^{2}\rangle_{\varphi} = \langle T^{2}(x) \rangle_{\varphi} - \langle T \rangle^{2}_{\varphi} \, ,
\end{equation}
and we use Eq.~\ref{error_phi} to further find
\begin{eqnarray}
\langle T^{2}(x) \rangle_{\varphi} &=& \left| \frac{d\langle T \rangle_{\varphi}}{d\varphi} \right|^{2} \left( \langle(\Delta \varphi)^{2}\rangle_{\varphi} + 2 \langle \Delta \varphi \rangle_{\varphi} \varphi + \varphi^{2} \right) \, , \\
\langle T \rangle^{2}_{\varphi} &=& \left| \frac{d\langle T \rangle_{\varphi}}{d\varphi} \right|^{2} \left( \langle \Delta \varphi \rangle^{2}_{\varphi} + 2 \langle \Delta \varphi \rangle_{\varphi} \varphi + \varphi^{2} \right) \, .
\end{eqnarray}
Given above equations, we find
\begin{equation}
\langle(\Delta T)^{2}\rangle_{\varphi} = \left| \frac{d\langle T \rangle_{\varphi}}{d\varphi} \right|^{2} \left( \langle(\Delta \varphi)^{2}\rangle_{\varphi} - \langle \Delta \varphi \rangle^{2}_{\varphi} \right) \, .
\end{equation}
This relation together with Eq.~\ref{error_phi} leads to the quantum Cram\'{e}r-Rao bound on the minimum value of the mean squared error in the parameter $\varphi$
\begin{equation}
\langle(\Delta \varphi)^{2}\rangle_{\varphi} \geq \frac{1}{N F(\varphi)} + \langle \Delta \varphi \rangle^{2}_{\varphi} \geq \frac{1}{N F(\varphi)} \, .
\end{equation}
The last inequality holds for unbiased estimators: $\langle \Delta \varphi \rangle_{\varphi} = 0$. The minimal error in $\varphi$ depends on the inverse of $N$ times the measurement procedure is repeated and the Fisher information. The Cram\'{e}r-Rao bound is a theoretical limit and in general it is not tight. In order to attain this bound, we have to use the probe system in an appropriate initial quantum state and then subject it to a suitable measurement. In other words, for a given measurement procedure we need to find an optimal initial quantum state and an optimal measurement observable.

There exist two important regimes of the quantum Cram\'{e}r-Rao bound, the so-called Standard Quantum Limit ({\sc sql}) and the Heisenberg Limit. The {\sc sql} or the shot noise limit is a classical limit for which each measurement reveals a constant amount of information about the parameter. The Heisenberg limit is imposed by the laws of quantum mechanics and for many years it was considered optimal and unbreakable. However, the optimality of the Heisenberg limit has been questioned recently. The Heisenberg limit and its optimality for the most general parameter estimation procedures will be the subject of Chapter~\ref{Chapter2}.

\subsection{The statistical distance}
The Fisher information defined in Eq.~\ref{fisher} is a function of the probability distribution $p(x|\varphi)$. In this section, we introduce the concept of the statistical distance between two probability distributions and relate it to the Fisher information. From a conceptual perspective, this corresponds to a parameter estimation procedure producing two distinct probability distributions $p(x|\varphi)$ and $p(x|\varphi')$ associated with two possible values of the parameter: $\varphi$ and $\varphi'$. The statistical distance measures how different these probability distribution are.

First, we define a space of probability distributions with a distance $s$ between two distributions defined on it \cite{kok_book}. Then, we introduce two infinitesimally close probability distributions $p(x)$ and $p'(x) = p(x) + dp(x)$. The infinitesimal statistical distance for $p(x)$ and $p'(x)$ is given by
\begin{equation}
ds^{2} = \int dx \, \frac{1}{p(x)} \left[ dp(x) \right]^{2} \, .
\end{equation}
We can divide both sides by $d\varphi^{2}$ assuming that $p(x)$ depends on a parameter $\varphi$
\begin{equation}
\left(\frac{ds}{d\varphi}\right)^{2} = \int dx \, \frac{1}{p(x|\varphi)} \left( \frac{\partial p(x|\varphi)}{\partial \varphi} \right)^{2} = F(\varphi) \, .
\end{equation}
This relates the Fisher information to the derivative of the statistical distance over $\varphi$ squared, i.e., the rate of change of the statistical distance with the parameter.

One of the most widely used systems for quantum metrology are optical systems such as interferometers fed with different states of light. A comprehensive description of various states of light can be given in terms of continuous variables. This approach applies not only to the field of quantum metrology but also to the field of optical quantum computation. Given their importance to many distinct subfields of quantum information, we introduce continuous variables in the next section.

\section{Continuous variables}\label{cv}

Continuous variables (CVs) may serve as a useful tool for describing various states of light. More importantly, in the context of quantum computation, CVs present an interesting alternative to discrete quantum systems, such as qubits. In this section, we introduce the notion of continuous variables and some basic operations that can be performed on them.

In general, continuous variables are eigenstates of an operator with a continuous spectrum \cite{kok_book}. There are a number of operators with continuous spectrum such as position, momentum, and quadrature operators of the electromagnetic field whose eigenstates can implement the continuous variables. We are especially interested in the last one, i.e., an optical representation of CVs.

We model a single mode of a free electromagnetic radiation field as a quantum harmonic oscillator. We write down the Hamiltonian of a harmonic oscillator in terms of creation and annihilation operators as
\begin{equation}
\hat{H} = \hbar \omega (\hat{a}^{\dag} \hat{a} + \frac{1}{2})\, ,
\end{equation}
where $\omega$ denotes the frequency of harmonic oscillator. The creation and annihilation operators are field operators that create or annihilate single excitations (quanta) of the radiation field in a well-defined single mode. The annihilation operator is associated with a quantised amplitude of a single excitation \cite{leonhardt}. We can rewrite the Hamiltonian of a harmonic oscillator in terms of the so-called quadrature operators
\begin{equation}
\hat{H} = \frac{1}{2} (\omega \hat{x}^{2} + \hat{p}^{2})\, ,
\end{equation}
with $\hat{x}$ and $\hat{p}$ defined in terms of creation and annihilation operators by
\begin{equation}
\hat{x} = \sqrt{\frac{\hbar}{2 \omega}} (\hat{a} + \hat{a}^{\dag})\, , \ \ \hat{p} = -i \sqrt{\frac{\hbar \omega}{2}} (\hat{a} - \hat{a}^{\dag})\, ,
\end{equation}
For simplicity, we can define dimensionless quadrature operators as
\begin{equation}
\hat{x} = \frac{1}{2} (\hat{a} + \hat{a}^{\dag})\, , \ \ \hat{p} = \frac{1}{2i} (\hat{a} - \hat{a}^{\dag})\, ,
\end{equation}
Given the bosonic commutation relation ($\left[ \hat{a}, \hat{a}^{\dag} \right] = 1$), the dimensionless quadrature operators obey the following commutation relation
\begin{equation}
\left[ \hat{x}, \hat{p} \right] = \frac{i}{2}.
\end{equation}
This commutation relation is reminiscent of the commutation relation for canonically conjugate position and momentum operators with $\hbar = 1/2$ \cite{braunstein05}. Hence, the quadrature operators are traditionally regarded as the position and momentum of the electromagnetic harmonic oscillator. Naturally, the quadratures have nothing to do with the position and the momentum of a single quantum since they are defined in the phase space of a harmonic oscillator \cite{leonhardt}. Since we think about the quadratures as position- and momentumlike quantities, their spectrum is unbounded and more importantly continuous. Therefore, we may use their eigenstates as an implementation of the continuous variables. We introduce eigenstates of single-mode quadrature operators satisfying
\begin{equation}
\hat{x} |x\rangle = x |x\rangle\, , \ \ \hat{p} |p\rangle = p |p\rangle\, .
\end{equation}
The eigenstates are orthogonal: $\langle x|x' \rangle = \delta\left(x - x'\right)$, $\langle p|p' \rangle = \delta\left(p - p'\right)$ and complete
\begin{equation}
\int^{\infty}_{-\infty} |x\rangle \langle x| dx = 1\, , \ \ \int^{\infty}_{-\infty} |p\rangle \langle p| dp = 1\, .
\end{equation}
According to the quantum-mechanical formalism, the eigenstates of canonically conjugate operators are related to each other by the Fourier transform, thus we may write
\begin{equation}
|x\rangle = \frac{1}{\sqrt{\pi}} \int dp \, \exp(-2i x p) |p\rangle\, , \ \ |p\rangle = \frac{1}{\sqrt{\pi}} \int dx \, \exp(2i x p) |x\rangle\,
\end{equation}
with $\hbar = 1/2$. To this end, we have introduced the continuous variables as the eigenstates of quadrature operators (position and momentum) of the electromagnetic field. Now, in order to perform a continuous-variable quantum computation, we need to create an initial CV state, i.e., a register, then apply an appropriate interaction Hamiltonian to induce evolutions on the continuous variables, and perform a measurement that reveals the result of computation \cite{kendon}. The continuous-variable quantum computation was introduced by Lloyd and Braunstein \cite{lloyd}. In principle, there are two distinct types of operations associated with CVs \cite{braunstein05}
\begin{enumerate}
\item Gaussian operations that include linear phase-space displacements, interaction Hamiltonians at most quadratic in $\hat{x}$ and $\hat{p}$ and homodyne detections (measurements of the quadratures of electromagnetic field),
\item non-Gaussian operations that include interaction Hamiltonians at least cubic (non-linear) in $\hat{x}$ and $\hat{p}$ or operations conditioned on non-Gaussian measurements such as photon counting.
\end{enumerate}
First, we focus our attention on the Gaussian operations. We introduce linear (in the quadrature operators) Hamiltonians. The displacement operator that allows us to move between different eigenstates of the position operator can be written as
\begin{equation}
\hat{X}(x) = \exp(-2i \, x \hat{p}).
\end{equation}
A straightforward calculation verifies that $\hat{X}(x)$ truly is a displacement operator. When applied to a position eigenstate $|y\rangle$ it gives: $\hat{X}(x) |y\rangle = |y + x\rangle$. For a momentum eigenstate $|r\rangle$, the effect of $\hat{X}(x)$ is the following
\begin{equation}
\hat{X}(x) |r\rangle = \exp(-2i \, x r) |r\rangle.
\end{equation}
It simply introduces a phase shift in the front of a momentum eigenstate. Since we have two conjugate quadrature operators, the form of another linear operator is given by
\begin{equation}
\hat{Z}(p) = \exp(2i \, p \hat{x}).
\end{equation}
Its action on the eigenstates of quadrature operators is the opposite to the action of $\hat{X}(x)$ and reads
\begin{equation}
\hat{Z}(p) |x\rangle = \exp(2i \, p x) |x\rangle\, , \ \ \hat{Z}(p) |r\rangle = |r + p\rangle\, .
\end{equation}
In summary, the $\hat{X}(x)$ and $\hat{Z}(p)$ linear operators displace the continuous variables to another eigenstate or introduce a state-dependent phase shift \cite{kok_book}. These operators implement phase-space displacements and constitute the continuous-variable generalisation of Pauli bit flip $X$ and phase flip $Z$ operators. Naturally, this set of operations is too limited for a fully functional quantum computer, therefore, we introduce Hamiltonians quadratic in the quadrature operators. One of the most important unitary operators in the field of quantum computation is the Fourier transform given by
\begin{equation}
\hat{F} = \exp\left[-2i \frac{\pi}{4} (\hat{x}^{2} + \hat{p}^{2}) \right]\, .
\end{equation}
When we apply the Fourier transform to a position eigenstate $|x\rangle$, we have
\begin{equation}
\hat{F} |x\rangle = \frac{1}{\sqrt{\pi}} \int dy \, \exp(2i \, x y) |y\rangle = |x\rangle_{p}\, .
\end{equation}
The action of the Fourier transform on a position eigenstate yields a momentum eigenstate with numerical value $x$ (subscript $p$ denotes a momentum domain). Furthermore, with a help of the Fourier transform, a momentum eigenstate can be written as a superposition of all possible position eigenstates. The application of the Fourier transform to a momentum eigenstate has an analogous effect, i.e., it gives a position eigenstate $|p\rangle_{x}$. The Fourier transform is the continuous-variable version of the Hadamard gate for discrete quantum systems. Other useful quadratic Hamiltonians include the phase gate $\hat{\Phi}(\theta)$ (a squeezing operator) applied on a single-mode system:
\begin{equation}
\hat{\Phi}(\theta) = \exp\left(2i \, \theta \hat{x}^2 \right)
\end{equation}
and continuous-variable versions of the CX and CZ gates applied on two CVs:
\begin{equation}
\mbox{CX}_{ij} = \exp\left(-2i \, \hat{x}_{i}\otimes\hat{p}_{j}\right)\, \ \mbox{and} \ \mbox{CZ}_{ij} = \exp\left(2i \, \hat{x}_{i}\otimes\hat{x}_{j}\right)\, .
\end{equation}

A truly powerful quantum computer has to be able to perform a universal quantum computation. Are the above CV operations sufficient to implemented any quantum computation? The generalised Gottesman-Knill theorem states that a CV quantum computer equipped with linear and quadratic Hamiltonians, i.e., the Gaussian operations, and allowing for classical feed-forward can be efficiently simulated on a classical computer. We note that it is interesting that a number of CV protocols which rely heavily on entanglement such as quantum teleportation satisfy the conditions of the Gottesman-Knill theorem and may be simulated efficiently on a classical computer \cite{braunstein98, braunstein05}. However, to move beyond a classical domain and at the same time implement universal quantum computation, we require arbitrary Hamiltonians to induce arbitrary evolutions. Fortunately, we can generate any interaction Hamiltonian corresponding to an arbitrary Hermitian polynomial of $\hat{x}$ and $\hat{p}$ given a small set of elementary interaction Hamiltonians. Before presenting this universal set, let us show why the linear and quadratic operations can never give us the higher-order polynomials. We invoke the Baker-Campbell-Hausdorff relation
\begin{equation}
e^{A} B e^{-A} = B + \frac{1}{1!} [A, B] + \frac{1}{2!} [A, [A, B]] + \ldots
\end{equation}
Here, $A$ and $B$ operators are at most quadratic in $\hat{x}$ and $\hat{p}$. Therefore, the commutator $[A, B]$ and all repeated commutators can produce polynomials of order at most two. In conclusion, to generate an arbitrary polynomial we require interaction Hamiltonians at least cubic in the position and momentum operators $\hat{x}$ and $\hat{p}$. The most well known Hamiltonian of this type is the so-called Kerr Hamiltonian $\hat{H}_{K} = (\hat{x}^{2} + \hat{p}^{2})^{2}$. The higher-order Hamiltonians belong to the class of non-Gaussian operations and, therefore, are much harder to generate. However, to perform universal quantum computation only one of such higher-order Hamiltonians, e.g., $\hat{H}_{K}$, suffices \cite{kok_book}.

The universal set of elementary operations for universal continuous-variable computation consists of
\begin{enumerate}
\item linear operations, e.g., $\hat{X}(x)$, $\hat{Z}(p)$,
\item quadratic operations, e.g., $\hat{F}$, $\hat{\Phi}(\theta)$,
\item a single non-linear (non-Gaussian) operation of higher-order, typically the Kerr Hamiltonian $\hat{H}_{K}$,
\item multi-mode interaction Hamiltonian applied on at least two modes, e.g., CX, CZ operations or the beam splitter interaction,
\item homodyne measurement.
\end{enumerate}
This set of operations can generate any multi-mode Hermitian polynomial in the canonical position and momentum operators. For the continuous variables implemented as the quadratures of the electromagnetic field, the universal set of elementary operations can be generated using linear optical elements, such as a simple $\pi/2$ phase shift (Fourier transform), and the non-linear optical medium such as a Kerr nonlinearity.

The only basic ingredient (omitting error correction \cite{braunstein_errora, braunstein_errorb}) of our continuous-variable quantum computer that is still missing is a physical input state that can be used as a register with which we encode our information. The position and momentum eigenstates represent an idealised implementation of the continuous variables. When one inspects the orthogonality conditions one easily notices that these eigenstates are non-normalisable and, therefore, unphysical, i.e., they cannot be generated in the laboratory. The way to deal with this difficulty is by approximating idealised eigenstates with a normalised Gaussian states. The Gaussian position and momentum  eigenstates centered around the position value $x$ and momentum value $p$ can be written as \cite{kok_book}
\begin{eqnarray}
|G(x)\rangle &=& \int^{\infty}_{-\infty} \frac{dy}{\sqrt[4]{\pi \Delta^2}} \, \exp\left[-\frac{(y - x)^2}{2 \Delta^2}\right] |y\rangle\, , \\
|G(p)\rangle &=& \int^{\infty}_{-\infty} \frac{dr}{\sqrt[4]{\pi/\Delta^2}} \, \exp\left[- 2 \Delta^2 (r - p)^2\right] |r\rangle\, ,
\end{eqnarray}
where $\Delta$ is the width of the Gaussian state with $\hbar = 1/2$. Depending on a value of $\Delta$, the Gaussian state represents various quantum states of light. When $\Delta = 0$, the Gaussian state $|G(x)\rangle = |x\rangle$ corresponds to an infinitely squeezed (in the position domain) state and $|G(p)\rangle$ represents an infinitely anti-squeezed state. For $\Delta = 1/\sqrt{2}$, we associate Gaussian states with coherent states of light. The Gaussian states of light can be generated unconditionally, however, their quality depends on the amount of squeezing applied. Naturally, the coherent states are free from these imperfections. As one expects, all Gaussian operations map Gaussian states onto Gaussian states.

Continuous variables are especially well suited for quantum communication protocols. Therefore, a number of applications have been generalised to CVs. These include quantum teleportation \cite{braunstein98, braunstein, furusawa} and entanglement swapping, quantum super-dense coding, quantum error correction, quantum cryptography \cite{ralph, hillery, gottesman} and entanglement distillation \cite{braunstein05}. On the other hand, continuous-variable quantum computing has received much less attention. In Chapter~\ref{Chapter3}, we present a comprehensive analysis of a parameter estimation protocol and the Deutsch-Jozsa algorithm in the setting of continuous-variable quantum systems. We devise a simple procedure that unifies quantum metrology and the Deutsch-Jozsa algorithm. We are not aware of a counterpart of this protocol existing in the setting of discrete quantum systems.

\section{Atomic ensembles}\label{atomic}

An atomic ensemble or atomic vapour is a gas that consists of several hundred of the same species of atoms, typically alkali atoms such as Cesium or Rubidium, trapped at room temperature or trapped and cooled to $\mu$K temperature. An atomic ensemble may serve as a good quantum memory for light. As the preceding sections may suggest, quantum memories can often be viewed as interfaces for either continuous-variable states or discrete states \cite{hammerer}. The behaviour of continuous-variable memories is described in terms of quadrature operators $\hat{x}$ and $\hat{p}$ subjected to homodyne measurements. The discrete memories are described with a help of $\hat{a}$ and $\hat{a}^{\dagger}$ operators that annihilate or create single quanta of light which are then measured with photon counting detectors \cite{hammerer}. The remainder of this section and Chapter~\ref{Chapter4} are focused on discrete quantum memories, that is, single-photon memories.

Any good and efficient quantum memory has to meet the following requirements. The atoms have to possess a long-lived ground state that is easily populated by optical pumping techniques. Moreover, the macroscopic ensemble should have a large optical depth $d = \rho \sigma L$, where $\rho$ is the atom number density, $\sigma$ is the absorption cross section of an atom and $L$ denotes the length of atomic medium. In other words, the atomic ensemble should easily, i.e., effectively, interact with light pulses. This is in fact one of the main advantages of atomic ensembles for interface purposes. A large number of atoms increases the coupling strength of an interaction between light and matter, and therefore allows us to coherently manipulate the quantum state of the ensemble with light and vice versa. Moreover, a large number of atoms helps to suppress the negative impact of decoherence on information stored in an atomic ensemble \cite{duan,hammerer,fleisch,lukin,barrett}.

\begin{figure}[t]
\begin{center}
\includegraphics[scale=3.5]{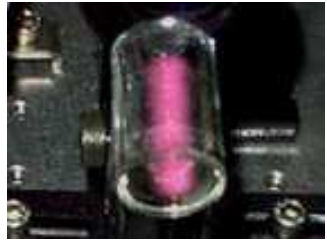}
\end{center}
\caption{A picture of an atomic ensemble consisting of a cloud of atoms trapped in a glass cell (taken from the homepage of the Experimental Quantum Optics Group at the Niels Bohr Institute in Copenhagen). \label{atomic_ens}}
\end{figure}

The simplest way to prepare an atomic ensemble is to trap a cloud of alkali atoms in a glass cell (see Fig.~\ref{atomic_ens}). This is the so-called hot atomic vapour or room temperature atomic vapour. Room temperature atomic ensembles are used extensively because of their simplicity and large optical depth, which is the key figure of merit for quantum memory efficiency. These kinds of interfaces will inherently suffer from thermal motion and therefore from Doppler broadening. Moreover, atoms moving in and out of the interaction region may limit the performance of a quantum memory. One of the widely used methods to overcome this problem is utilisation of a buffer gas \cite{manz,novikova}. A few torr of a noble gas, typically neon or helium, limits the thermal diffusion of atoms inside a vapour \cite{novikova,pugatch}. Another advantage of a buffer gas is the suppression of decoherence from the collisions between alkali atoms and with the walls of a cell. By means of a buffer gas, the atoms can retain coherence for more than $10^{8}$ collisions \cite{manz}. Although a buffer gas seems to be indispensable, too high buffer gas pressure may also introduce some incoherent processes to the operation of a quantum memory \cite{manz}. One of the most recent techniques for suppression of the collisional and motional decoherence involves buffer gas cooled below 7K. In an experiment by Hong \textit{et al.} \cite{thong}, Rubidium atoms are cooled by a buffer gas and the diffusion time is slowed down. Moreover, the optical depth of a medium in this experiment is very large ($d >$ 70). The mentioned setup combines simplicity and large optical depth of a room temperature atomic vapour with slow atomic motion that is characteristic for another technique of trapping alkali atoms, namely so-called magneto-optical trapping (MOT) \cite{thong}.

A MOT technique combines laser cooling and trapping with magnetic fields. Atoms trapped with MOT are cooled down to the $\mu$K temperature, therefore the collisional and motional decoherence becomes negligible in comparison with a typical operational time scale of a quantum memory. The shortcoming of a cold atomic ensemble is rather low optical depth ($d < 10$). The very principle on which the MOT is operating, i.e., the magnetic field, also introduces another difficulty. The magnetic field causes decoherence of the ground states usually realised as magnetic Zeeman sublevels of a ground state. This problem can be overcome by switching off the MOT trap and then performing operations on a quantum memory \cite{hammerer}. However, lack of the magnetic field trapping allows atoms to slowly diffuse and therefore limits the lifetime of a quantum memory. Nevertheless, by means of the MOT trap atomic vapours can be prepared in the form of a ``frozen" gas with lifetime much longer than in the case of a room temperature vapours.

The last widely used method for confining large numbers of atoms to a small sample is called Bose-Einstein condensation. A Bose-Einstein condensate (BEC) has extremely large optical depth. However, the preparation of a BEC is an extremely challenging experiment.

There are a number of effects that influence the overall efficiency of quantum memories based on atomic ensembles. In spite of many efforts
the efficiency of quantum memories reaches at the best 70\% \cite{hammerer}. The main source of low fidelity is a low optical depth $d$. Only an optically thick medium, that is, highly dense and/or large medium, can effectively interact with the light fields. The broadening of the optical transitions, both homogenous and inhomogeneous, is another source of decoherence for quantum memories. The homogenous broadening is mainly due to the spontaneous emission and results in the inefficiency of storage that depends on the optical depth as $1/d_{hombroad}$, where $d_{hombroad}$ is the optical depth without the homogenous broadening \cite{hammerer}. The inefficiency of storage of light pulses based on techniques such as electromagnetically induced transparency or Raman interaction scales as $1/d_{hombroad}$ \cite{hammerer}. For atomic ensembles at room temperature the inhomogeneous broadening is due to the thermal motion and associated with it Doppler broadening of the atomic lines. The Doppler broadening induces shifts in the energy level structure of the atoms in completely incoherent fashion and results in the inefficiency of storage that scales as $1/d_{inhombroad}^{2}$, where $d_{inhombroad}$ is the optical depth in a presence of the homogenous broadening \cite{hammerer}. For a sufficiently dense and/or large medium, inhomogeneous broadening is less dominant than homogenous broadening. Apart from the Doppler broadening, the thermal or atomic motion is responsible for atomic collisions, which are yet another factor that limits fidelity of a quantum memory.  


\def\la{\langle}
\def\ra{\rangle}
\def\a{{\bm{a}}}
\def\b{{\bm{b}}}
\def\x{{\bm{x}}}
\def\Om{{\Omega}}
\def\Ups{{\Upsilon}}
\def\thet{\bm{\vartheta}}
\def\re{\text{Re}}
\def\im{\text{Im}}
\def\tr{\text{Tr}}

\def\be{\begin{equation}}
\def\ee{\end{equation}}
\def\ba{\begin{eqnarray}}
\def\ea{\end{eqnarray}}

\part{Quantum Metrology, the Deutsch-Jozsa Algorithm and Continuous Variables}

\chapter{General Optimality of the Heisenberg Limit for Quantum Metrology}
\label{Chapter2}
\lhead{\textsc{Chapter 2. Optimality of the Heisenberg Limit for Metrology}}

\section{Introduction}
\noindent
Parameter estimation is a fundamental pillar of science and technology, and improved measurement techniques for parameter estimation have often led to scientific breakthroughs and technological advancement. Caves \cite{caves} showed that quantum mechanical systems can in principle produce greater sensitivity over classical methods, and many quantum parameter estimation protocols have been proposed since \cite{kok_book}. The field of quantum metrology started with the work of Helstrom \cite{helstrom67,helstrom76}, who derived the minimum value for the mean square error in a parameter in terms of the density matrix of the quantum system and a measurement procedure. This was a generalisation of a known result in classical parameter estimation, called the Cram\'er-Rao bound. Braunstein and Caves \cite{braunstein94} showed how this bound can be formulated for the most general state preparation and measurement procedures. While it is generally a hard problem to show that the Cram\'er-Rao bound can be attained in a given setup, at least it gives an upper limit to the precision of quantum parameter estimation.

The quantum Cram\'er-Rao bound is typically formulated in terms of the Fisher information, an abstract quantity that measures the maximum information about a parameter $\varphi$ that can be extracted from a given measurement procedure. One of the central questions in quantum metrology is how the Fisher information scales with the physical resources used in the measurement procedure. We usually consider two scaling regimes: First, in the {\em standard quantum limit} ({\sc sql}) \cite{gardiner04} or {\em shot-noise limit} the Fisher information is constant, and the error scales with the inverse square root of the number of times $T$ we make a measurement. Second, in the {\em Heisenberg limit} \cite{holland93} the error is bounded by the inverse of the physical resources. Typically, these are expressed in terms of the size $N$ of the probe system, e.g., (average) photon number. However, it has been clearly demonstrated that this form of the limit is not universally valid. For example,  Beltr\'an and Luis \cite{luis05} showed that the use of classical optical nonlinearities can lead to an error with average photon number scaling $N^{-3/2}$. Boixo {\em et al}.\ \cite{boixo} devised a parameter estimation procedure that sees the error scale with $N^{-k}$ with $k\in\mathbb{N}$, and Roy and Braunstein \cite{Roy08} construct a procedure that achieves an error that scales with $2^{-N}$. The central question is then: What is the real fundamental Heisenberg limit for quantum metrology? We could redefine this limit accordingly to scale as $2^{-N}$, but in practice this bound will never be tight and therefore of limited use.

In this chapter, we give a natural definition of the relevant physical resources for quantum metrology based on the general description of a parameter estimation procedure, and we prove the asymptotical bound on the mean squared error based on this resource count. We will show that the resource count is proportional to the size of the probe system only if the interaction between the object and the probe is non-entangling over the systems constituting the probe. In Sec.~\ref{sec:res}, we study the query complexity of quantum metrology networks, which will lead to a resource count given by the expectation value of the generator of translations in the parameter $\varphi$. In Sec.~\ref{sec:proof}, we prove that the mean error in $\varphi$ is asymptotically bounded by the inverse of this resource count. We argue that this is the fundamental Heisenberg limit for quantum metrology. Furthermore, in Sec.~\ref{sec:con}, we clarify the origin of the term ``Heisenberg limit". Finally, we illustrate how this general principle can resolve paradoxical situations in which the Heisenberg limit seems to be surpassed.

\begin{figure}[t]
\centering
\includegraphics[width=10cm]{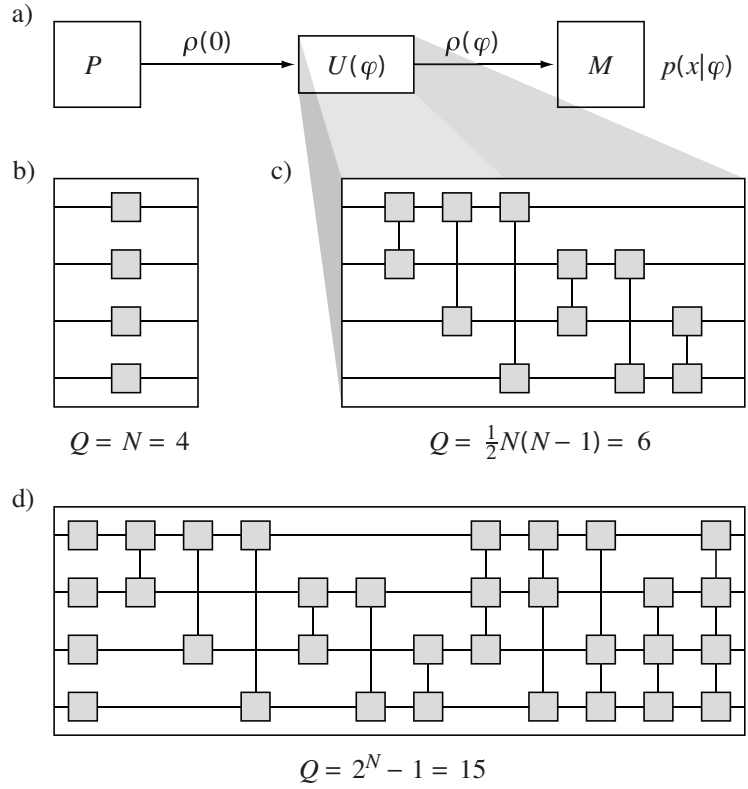}
\caption{\label{fig:setup} a) General parameter estimation procedure involving state preparation $P$, evolution $U(\varphi)$ and generalised measurement $M$ with outcomes $x$, which produces a probability distribution $p(x|\varphi)$. In terms of quantum networks, the evolution can be written as a number of queries of the parameter $\varphi$. b) Example for $N=4$ of the usual situation described by $\mathcal{H}_{\rm GLM}$, where each system performs a single query, and the number of queries equals the number of systems (the grey box represents $O_j(\varphi)$); c) for $\mathcal{H}_{\rm BFCG}$ the number of queries $Q$ does not always equal the number of systems:  any two systems can jointly perform a single query, and the number of queries then scales quadratically with the number of systems; d) for $\mathcal{H}_{\rm RB}$ all possible subsets of systems perform a single query. The number of queries scales exponentially with the number of systems.}
\end{figure}

\section{Parameter estimation and resources} \label{sec:res}
\noindent
The most general parameter estimation procedure is shown in Fig.~\ref{fig:setup}a). Consider a probe system prepared in an initial quantum state $\rho(0)$ that is evolved to a state $\rho(\varphi)$ by $U(\varphi)=\exp(-i\varphi\mathcal{H})$. This is a unitary evolution when we include the relevant environment into our description, and it includes feed-forward procedures. The Hermitian operator $\mathcal{H}$ is the generator of translations in $\varphi$, the parameter we wish to estimate. The system is subjected to a generalised measurement $M$, described by a Positive Operator Valued Measure ({\sc povm}) that consists of elements $\hat{E}_x$, where $x$ denotes the measurement outcome. These can be discrete or continuous (or a mixture of both). The probability distribution that describes the measurement data is given by the Born rule $p(x|\varphi) = \tr[\hat{E}_x\,\rho(\varphi)]$, and the maximum amount of information about $\varphi$ that can be extracted from this measurement is given by the Fisher information
\be
 F(\varphi) = \int dx\, \frac{1}{p(x|\varphi)} \left( \frac{\partial p(x|\varphi)}{\partial \varphi} \right)^2\, .
\ee
This leads to the quantum Cram\'er-Rao bound \cite{helstrom67,braunstein94}
\be\label{eq:cr}
 \delta\varphi \geq \frac{1}{\sqrt{T F(\varphi)}}\, ,
\ee
where $(\delta\varphi)^2$ is the mean square error in the parameter $\varphi$, and $T$ is the number of times the procedure is repeated. The {\sc sql} is obtained when the Fisher information is a constant with respect to $T$, and the Heisenberg limit is obtained in a single-shot experiment ($T=1$) when the Fisher information scales quadratically with the resource count. The {\sc sql} and the Heisenberg limit therefore relate to two fundamentally different quantities, $T$ and $F$, respectively. We need to reconcile the meaning of these two limits if we want to compare them in a meaningful way.

To solve this problem, we can define an unambiguous resource count for parameter estimation by recognising that a quantum parameter estimation protocol can be written as a quantum network acting on a set of quantum systems, with repeated ``black-box'' couplings of the network to the system we wish to probe for the parameter $\varphi$ \cite{Giovannetti06}. The quantum networks arise naturally in the circuit model of quantum computation. A quantum network consists of wires that connect successive quantum gates. The wires represent movement of quantum systems through space or time, and gates perform simple computational tasks on the information carried by these quantum systems \cite{nielsen}. In general, a quantum network involves many quantum systems and many quantum gates. Traditionally, we represent a quantum gate as a function $f$ with fixed number of input parameters and fixed number of output parameters \cite{nielsen}. In the following analysis, we employ a special type of the quantum gate called a black-box or a quantum oracle. A black-box is a unitary operator defined by its action on quantum systems whose internal workings are usually unknown. As any other quantum gate, a black-box is a function that can be univariate or multi-variate. When the function is multi-variate, a {\em query} to the black-box must take the form of multiple input parameters. Likewise, when the operator that describes the fundamental ``atomic'' interaction between the queried system and the probe is a two-body interaction on the probe, then a query can consist only of precisely two input bodies. The scaling of the error in $\varphi$ is then determined by the {\em query complexity} of the network. The number of queries $Q$ is not always identical to the number of physical systems $N$ in the network.

In Fig.~\ref{fig:setup}b-d) we consider three examples. The quantum network with univariate black-boxes in b) was analysed by Giovannetti, Lloyd, and Maccone \cite{Giovannetti06}. Suppose that each grey box in Fig.~\ref{fig:setup} is a unitary gate $O_j(\varphi) = \exp(-i\varphi H_j)$, where $j = 1,\ldots,N$  denotes the system, and $H_j$ is a positive Hermitian operator. It is convenient to define the generator of the joint queries as
\be
\mathcal{H}_{\rm GLM} = \sum_j H_j\, ,
\ee
because all $H_j$ commute with each other. The number of queries $Q$ is then equal to the number of terms in $\mathcal{H}_{\rm GLM}$, or $Q=N$. In Fig.~\ref{fig:setup}c) the black-box is bi-variate. This is a type of Hamiltonian considered by Boixo, Flammia, Caves, and Geremia \cite{boixo}, and takes the form
\be
\mathcal{H}_{BFCG} = \sum_{k=1}^N \sum_{j=1}^{k}  H_j\otimes H_k\, .
\ee
A physical query to a black-box characterised by $O_{jk}(\varphi) = \exp(-i\varphi H_j \otimes H_k)$ must consist of two systems, labeled $j$ and $k$. Since each pair interaction is a single query, the total number of queries is $\binom{N}{2} = \frac12 N(N-1)$. Finally, in Fig.~\ref{fig:setup}d) we depict the network corresponding to the protocol by Roy and Braunstein \cite{Roy08}. It is easy to see that the number of terms in the corresponding generator $\mathcal{H}_{RB}$ is given by $2^N-1$, and the number of queries is therefore $Q=2^{N}-1$.

A similar argument can be made to find the correct number of queries for all types of networks. The key principle is that a physical query to a quantum system consists of probe-systems that \emph{together} undergo an  operation, which can potentially entangle them. The entangling power of the black-box operation over multiple input systems  accounts for the super-linear scaling of $Q$ with $N$. Only when $\mathcal{H}$ does not have any entangling power across the input, we are guaranteed to have $Q=O(N)$. This is in agreement with Refs.~\cite{boixo} and \cite{Roy08} where $\sqrt{F(\varphi)}$ scales super-linearly in $N$, but is always linear in $Q$, as defined here. Since we have a systematic method for increasing $N$ (and $Q$) given the atomic interaction $H_j$, this uniquely defines an asymptotic query complexity of the network. Since both $T$ and $Q$ count the number of queries, this allows us to meaningfully compare the {\sc sql} with the Heisenberg limit.

Given that in Eq.~(\ref{eq:cr}) $\sqrt{F(\varphi)}\lesssim Q$, we have to find a general procedure that bounds $Q$, based on the physical description of the estimation protocol in Fig.~\ref{fig:setup}a). Previously, we showed that $Q$ is the number of black-box terms in $\mathcal{H}$, and a straightforward choice for the resource count is therefore $|\la\mathcal{H}\ra| \leq O(Q)$. An important subtlety occurs when $\mathcal{H}$ corresponds to a proper Hamiltonian. The origin of the energy scale has no physical meaning, and the actual value of $|\la \mathcal{H}\ra|$ can be changed arbitrarily. Hence, we must fix the scale such that the ground state has zero energy (equivalently, we may choose $\la\mathcal{H}-h_{\rm min}I\ra$, where $h_{\rm min}$ is the smallest eigenvalue, and $I$ the identity operator). In most cases, this is an intuitive choice. For example, it is natural to associate zero energy to the vacuum state, and add the corresponding amount of energy for each added photon. Technically, this corresponds to the normal ordering of the Hamiltonian of the radiation field in order to remove the infinite vacuum energy. Slightly less intuitive is that the average energy of $N$ spins in a Greenberger-Horne-Zeilinger state $(\ket{\uparrow}^{\otimes N} + \ket{\downarrow}^{\otimes N})/\sqrt{2}$ is no longer taken to be zero, but rather $N/2$ times the energy splitting between $\ket{\uparrow}$ and $\ket{\downarrow}$.

While the expectation value of $\mathcal{H}$ is easy to calculate, it is not the only way to obtain a bound of $O(Q)$ from $\mathcal{H}$. Other seemingly natural choices are the variance and the semi-norm. For example, if we write $\mathcal{H} \equiv \sum_j^Q A_j$, the variance is
\begin{eqnarray}
(\Delta\mathcal{H})^2 &=& \left\la \left(\sum_j^{Q} A_j\right)^2 \right\ra - \left\la \sum_j^Q A_j\right\ra^2  \cr &=&  \sum_j^{Q^2} \la L_j \ra - \sum_{j,k}^Q \la A_j\ra\la A_k\ra \leq c Q^2
\end{eqnarray}
for some positive number $c$ and positive operator $L_j$. This gives $\Delta\mathcal{H}\leq O(Q)$, where e.g., in Ref.~ \cite{boixo} $Q=O(N^2)$. Similarly, $|\la\mathcal{H}\ra| \leq \sum_j^Q |\la A_j\ra| \leq O(Q)$ since all expectation values are positive and finite. In other words, in terms of the scaling behaviour with $Q$, we can use either the variance or the expectation value. However, there are important classes of quantum systems for which the variance of the energy diverges, such as systems with a Breit-Wigner (or Lorentzian) spectrum \cite{breit,uffink93}. Furthermore, for the \textsc{NOON} states written as $\left(\ket{N, 0} + \ket{0, N}\right)/\sqrt{2}$, the variance of the energy is zero \cite{kok_book}. The variance of a Hermitian operator is upper bounded by the operator semi-norm
\be
(\Delta\mathcal{H})^2 \leq \frac{||\mathcal{H}||^{2}}{4}\, ,
\ee
where the operator semi-norm is defined as $||\mathcal{H}|| = h_{\rm max} - h_{\rm min}$ with $h_{\rm max}$ and $h_{\rm min}$ being the
maximal and minimal eigenvalue of $\mathcal{H}$, respectively. Again, the semi-norm does not exist for a large class of states, such as optical Gaussian states. In these cases the resource count, and by implication the scaling of the error, would be ill-defined.

Also, from a physical perspective the higher-order moments do not describe ``a\-mounts'' in the same way as the first moment does, and refer instead to the shape of the distribution. This is a further argument that $|\la\mathcal{H}\ra|$ is the natural choice for the resource count. Sometimes, it is unclear how the query complexity is defined, for example when the estimation procedure does not involve repeated applications of the gates $O_j(\varphi)$, or when an indeterminate number of identical particles, such as photons, are involved. Nevertheless, the generator $\mathcal{H}$ is always well-defined in any estimation procedure, and we can use its expectation value to define the relevant resource count.

The resource count in terms of $|\la\mathcal{H}\ra|$ is completely general for all possible quantum networks. The most general quantum interaction $U(\varphi)$ acting on the probe system is represented by the unitary transformation
\begin{equation}\label{net}
		\Qcircuit @C=0.66em @R=0.8em @!R  {
	\lstick{U(\varphi) =} & \gate{{V_0}} 	&  \gate{{O(\varphi)}}	& \gate{{V_1}} 	& \gate{{O(\varphi)}} 	& \qw & \ldots & &\gate{{O(\varphi)}} & \gate{{V_Q}} & \qw }
\end{equation}
This general interaction consists of $Q$ applications of $O(\varphi)$, interspersed with $Q+1$ arbitrary unitary gates $V_j$. The arbitrary unitary gates $V_j$ together with ancillary systems may be used to introduce adaptive (feed-forward) strategies to the estimation procedure. For a general interaction $U(\varphi)$, we can use an argument by Giovannetti {\em et al}.\ \cite{Giovannetti06} to show that the expectation value of the generator of $U(\varphi)$ is given by
\be
|\la\mathcal{H}\ra| = \left| \left\langle i \left(\frac{\partial U(\varphi)}{\partial \varphi}\right) U^{\dagger}(\varphi)\right\rangle \right| \leq \sum_{j = 1}^{Q} |\la A'_{j} \ra |\, ,
\ee
where
\be
A'_{j} = V_{Q}\, O(\varphi)\, \ldots\, V_{j+1}\, O(\varphi)\, V_{j}\, \frac{\partial O(\varphi)}{\partial \varphi}\, V^{\dag}_{j}\, O^{\dag}(\varphi)\, V^{\dag}_{j+1}\, \ldots\, O^{\dag}(\varphi)\, V^{\dag}_{Q} \, .
\ee
Since all the $A'_{j}$ have the same spectrum as $A_{j}$ (the spectrum of the generator of $O(\varphi)$ is unchanged by the $V_{j}$'s), then the expectation value $|\la\mathcal{H}\ra|$ is unaffected by the intermediate arbitrary unitary gates, and the scaling is therefore still determined by $Q$.

\section{Optimality proof of the Heisenberg limit} \label{sec:proof}
\noindent
After establishing the appropriate resource count, we are finally in a position to prove the optimality of the Heisenberg limit for quantum parameter estimation in its most general form. The Fisher information can be related to a statistical distance $s$ on the probability simplex spanned by $p(x|\varphi)$. Consider two probability distributions $p(x|\varphi)$ and $p(x|\varphi) + dp(x)$. The infinitesimal statistical distance between these distributions is given by \cite{bhatta43,wootters81}
\be
ds^2 = \int dx\, \frac{1}{p(x|\varphi)}[dp(x|\varphi)]^2 \, .
\ee
Dividing both sides by $(d\varphi)^2$ we obtain
\be\label{eq:sf}
 \left( \frac{ds}{d\varphi} \right)^2 = \int dx\, \frac{1}{p(x|\varphi)} \left( \frac{\partial p(x|\varphi)}{\partial \varphi} \right)^2 = F(\varphi)\, ,
\ee
which relates the Fisher information to the rate of change of the statistical distance (i.e., the speed of dynamical evolution).

When we count the resources that are used in a parameter estimation procedure, we must make sure that we do not leave anything out, and this can be guaranteed by including in our description the environment that the estimation procedure couples to. This reduces the quantum states to pure states.
The distance between the probe state $\rho(0)$ and the evolved state $\rho(\varphi)$ can then be represented by the pure states $\ket{\psi(0)}$ and $\ket{\psi(\varphi)}$, respectively, and the unitary evolution is given by
\be \label{eq:evo}
\ket{ \psi(\varphi)} = \exp\left(-i\varphi\mathcal{H}\right)\ket{ \psi(0)}\, .
\ee
Here, {\em we place no restriction} on $\mathcal{H}$, other than fixing the energy scale if necessary. It was shown by Anandan and Aharonov \cite{anandan} that the derivative of the statistical distance between two pure states is given by the variance of $\mathcal{H}$
\be \label{eq:ml}
\frac{ds}{d\varphi} = 2 \Delta\mathcal{H}\, .
\ee
Combining this equality with Eq.~(\ref{eq:sf}) and Eq.~(\ref{eq:cr}) leads to the Cram\'er-Rao bound
\be
 (\delta\varphi)^2 \geq \frac{1}{T} \left( \frac{ds}{d\varphi} \right)^{-2} \geq \frac{1}{T\, 4 (\Delta\mathcal{H})^2}\, .
\ee
Since both the variance and the expectation value of $\mathcal{H}$ are bounded by a linear function of $Q$, in the asymptotic limit we have:
\be\label{eq:kok}
 (\delta\varphi)^2 \gtrsim \frac{1}{T\, |\la\mathcal{H}\ra|^2}\, .
\ee
When all resources are used in a single-shot ($T=1$) experiment, the error in $\varphi$ is bounded by
\be\label{eq:hl}
 \delta\varphi \gtrsim \frac{1}{|\la\mathcal{H}\ra|}\, .
\ee
Since $|\la\mathcal{H}\ra|$ is the resource count in the parameter estimation procedure, this is the Heisenberg limit. It is always positive and finite, and in the limit where $|\la\mathcal{H}\ra| \to 0$ there are no resources available to estimate $\varphi$, and $\delta\varphi$ cannot be bounded. In general, the bound is not tight. Indeed, only carefully chosen entangled systems can achieve this bound \cite{Giovannetti06}. This completes the proof of the optimality of the Heisenberg limit in the most general case.

\section{Consequences of the new Heisenberg limit} \label{sec:con}
\noindent
In addition to Eqs.~(\ref{eq:sf}) and (\ref{eq:ml}), for mixed states the Fisher information is bounded by the variance of $\mathcal{H}$ according to $F(\varphi) \leq 4 (\Delta\mathcal{H})^2$ \cite{braunstein96}. This leads to a (single-shot) quantum Cram\'er-Rao bound
\be\label{eq:up}
 \delta\varphi \geq \frac{1}{2\Delta\mathcal{H}}\, .
\ee
However, since $\Delta\mathcal{H}$ is not a resource count, such as the average photon number, but rather a variance (or uncertainty) this is not the Heisenberg limit. In fact, it is Heisenberg's uncertainty relation for the parameter $\varphi$ and its conjugate operator $\mathcal{H}$. Any parameter estimation procedure must respect both bounds, and the Heisenberg limit in Eq.~(\ref{eq:hl}) may not be attained for a particular input state because the bound in Eq.~(\ref{eq:up}) prevents it from doing so.

The term ``Heisenberg limit'' was introduced by Holland and Burnett \cite{holland93}, who referred to the number-phase uncertainty relation in Heitler \cite{heitler54}. However, as our optimality proof and the subsequent discussion indicate, the Heisenberg limit is {\em not} an uncertainty relation, since it relates the uncertainty of the parameter to the {\em first} moment of the conjugate observable $\mathcal{H}$, rather than the second.
The (generalised) uncertainty relation can be identified with the Mandelstam-Tamm bound on the time it takes for a quantum system to evolve to an orthogonal state \cite{mandelstam45, kok_book}. To see this, we can formally solve
\be
\frac{ds}{d\varphi} \leq 2 \Delta\mathcal{H}\,
\ee
by separation of variables, yielding
\be
 \int_0^{\varphi} d\varphi' \geq \frac{1}{2\Delta\mathcal{H}} \int_0^{\pi} ds \quad\Rightarrow\quad \varphi \geq \frac{\pi}{2} \frac{1}{\Delta\mathcal{H}}\, .
\ee
We again emphasise that both limits given in terms of the variance and the expectation value of $\mathcal{H}$ are completely general and complement each other.

Finally, we demonstrate that our proof applies to continuous variable systems as well as discrete systems, by considering the procedure of Beltr\'an and Luis \cite{luis05}. The construction is as follows: The evolution $O(\varphi)$ is generated by an optical nonlinearity proportional to the square of the photon number operator $\hat{n}^2$ acting on a single-mode coherent state $\ket{\psi(0)}=\ket{\alpha}$. The evolved state before detection is given by
\be
\ket{\psi({\varphi})} = \exp(-i\varphi\hat{n}^2)\ket{\alpha}\, ,
\ee
and the mean square error in $\varphi$ is calculated as
\be
\delta\varphi \simeq \frac14 \la \hat{n}\ra^{-3/2} = \frac14 |\alpha|^{-3}
\ee
to leading order in the average photon number $\la \hat{n}\ra$. Since here the average energy is directly proportional to the average photon number, this procedure seems to surpass the Heisenberg limit. To resolve this paradox, we note that the generator of translations in $\varphi$ is {\em not} the photon number operator $\hat{n}$, but rather the higher-order nonlinearity $\mathcal{H} = \hat{n}^2$. The appropriate resource count is therefore $|\la\mathcal{H}\ra| = \la \hat{n}^2\ra$, instead of the average photon number $\la\hat{n}\ra$. It is easily verified that to leading order $\delta\varphi$ is theoretically bounded by $1/\la\hat{n}^2\ra = 1/|\alpha|^4$. Hence the parameter estimation procedure does not even attain the Heisenberg limit.

Formally, we can attain the Heisenberg limit for this generator of translations in $\varphi$ with the following modification of the input state and the measurement. Consider the single-mode input state $|\psi_{0}\rangle = \left(|0\rangle + |N\rangle\right)/\sqrt{2}$, where $\ket{0}$ denotes no photons, and $\ket{N}$ denotes $N$ photons. The state of the probe before detection is then given by
\be
|\psi(\varphi)\rangle = \exp(-i\varphi \hat{n}^2) |\psi(0)\rangle = (\ket{0} + e^{-i\varphi N^2}\ket{N})/\sqrt{2}\, .
\ee
We define the measurement observable $X = \ket{0}\bra{N}+ \ket{N}\bra{0}$. Hence, for the final state $\ket{\psi(\varphi)}$ we calculate
\be
\langle X \rangle = \langle \psi_{\varphi}| X |\psi_{\varphi}\rangle = \mbox{cos}(N^2 \varphi) \ \mbox{and} \ \Delta X = \mbox{sin}(N^2 \varphi)\, .
\ee
Using the standard expression for the mean squared error, we find that
\begin{equation}
 \delta \varphi = \frac{\Delta X}{\left| d\langle X \rangle/d\varphi \right|} = \frac{1}{N^2}\, .
\end{equation}
Since $|\la\mathcal{H}\ra| = \la \hat{n}^2\ra = \frac12 N^2$, this attains the Heisenberg limit. This is a formal demonstration that the Heisenberg limit can be attained according to quantum mechanics, even though we currently do not know how to implement it.

\section{Summary}
\noindent
In conclusion, we demonstrated that the Heisenberg limit is optimal for all parameter estimation procedures in quantum metrology, but it requires careful consideration as to which resource is appropriate for expressing the scaling behaviour of the mean square error. The correct identification of the resource count was achieved using computational complexity theory, further strengthening the connection between physics and computer science. The correct resource to take into account is (the expectation value of) the generator of the translations in the parameter. In the case of most optical phase estimation protocols this reduces to the average photon number. Contrary to the origin of the term ``Heisenberg limit'', it is not a generalised uncertainty relation. We can identify a generalised uncertainty relation with the Mandelstam-Tamm bound on the speed of dynamical evolution of quantum systems when $\mathcal{H}$ is the Hamiltonian. Our general approach to quantum metrology resolves paradoxical situations in which the Heisenberg limit seems to be surpassed even when it is unclear how the query complexity is defined. Like other fundamental limits, the new Heisenberg limit increases our understanding of Nature, and will likely lead to new recipes for high-precision measurements.


\chapter{Unifying Parameter Estimation and the Deutsch-Jozsa Algorithm for Continuous Variables}
\label{Chapter3}
\lhead{\textsc{Chapter 3. Parameter Estimation and the DJ Algorithm for CVs}}

\section{Introduction}
\noindent
It is well known that quantum metrology promises many advances in science and technology. Continuous variables (CVs) are natural candidates for optical implementations of quantum metrology protocols \cite{caves,braunstein98,kok_book}. The importance of CVs for quantum metrology stems from the unconditional and efficient character of CV preparation, manipulation, and detection techniques \cite{braunstein05,lloyd}. In this chapter, we devise an optimal parameter estimation procedure for continuous variables. Our procedure employs a single CV and estimates a value of an unknown parameter with Heisenberg-limited precision. Furthermore, for a particular, fixed value of the parameter in question the procedure behaves as the Deutsch-Jozsa algorithm for CVs. In fact, our protocol extends the Deutsch-Jozsa algorithm over continuous variables presented by Pati and Braunstein \cite{pati}. Instead of idealised, non-normalisable (unphysical) states, we employ Gaussian states to represent continuous variables. Moreover, we define Gaussian states on a finite domain, thus removing an unphysical, infinite speed-up over any classical procedure offered by the idealised states. An extensive analysis of the Deutsch-Jozsa algorithm over continuous variables was given by Adcock, H{\o}yer, and Sanders \cite{sanders09}.

The Deutsch-Jozsa algorithm is one of the first quantum algorithms, preceded only by the original Deutsch algorithm \cite{deutsch}. Even though the Deutsch problem is rather artificial, the algorithm drew enormous attention due to the computational speed-up over any classical procedure. The structure of the algorithm is simple enough to determine the source of this speed-up. The quantum superposition principle and consequent quantum parallelism that lie at the heart of quantum mechanics permits the interference of many distinct computational paths and allows the correct answer to the problem to emerge in a single query. In other words, the Deutsch-Jozsa algorithm probes a global property of an unknown function $f(x)$ and returns the result in a single run.

This chapter is organised as follows. In Sec.~\ref{sec:dj_con}, we recall the Deutsch-Jozsa algorithm for discrete quantum systems, that is, the qubits. In Sec.~\ref{sec:dj_con}, we review the Deutsch-Jozsa algorithm over continuous-variable quantum systems and present its simplified version. In Sec.~\ref{sec:qm}, we review basic concepts in quantum metrology. In Sec.~\ref{sec:unify}, we introduce a general procedure that unifies parameter estimation with the Deutsch-Jozsa algorithm, and we analyse it in detail. Finally, we make some concluding remarks in Sec.~\ref{sec:conc}.

\section{Deutsch-Jozsa algorithm}\label{sec:dj}
\noindent
The following algorithm is not the original algorithm proposed by Deutsch and Jozsa \cite{deutsch} (which was probabilistic), but its improved version \cite{cleve}. However, for a historical reason we still refer to the following algorithm as the Deutsch-Jozsa algorithm.
\begin{figure}[t]
\begin{center}
\begin{equation*}
	\Qcircuit @C=1.3em @R=0.8em @!R @!C {
\lstick{\ket{0}^{\otimes n}}			& \gate{H^{\otimes n}}	& \multigate{1}{{U_{f}}} 	& \gate{H^{\otimes n}}	& \meter  \\
\lstick{\ket{1}}	& \gate{H}	& \ghost{{U_f}} 		& \qw 			& \qw
	}
\end{equation*}
\end{center}
\caption{A quantum circuit representing the Deutsch-Jozsa algorithm over qubits. The quantum network ${N_{DJ}}$ consists of the Hadamard gates $H$ and controlled black-box gate $U_{f}$ applied to the $n$ qubit register and single target qubit prepared in $|0\rangle^{\otimes n}$ and $|1\rangle$ states , respectively. The last operation is the Hadamard gate which enables the interference of different computational paths. \label{DJ_q}}
\end{figure}
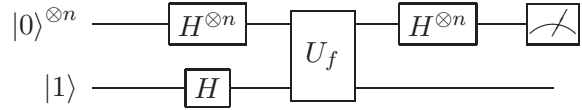
The Deutsch-Jozsa algorithm is an extension of a simple Deutsch algorithm. It addresses the Deutsch problem defined as follows. Imagine two parties, Alice and Bob, playing a game. Alice chooses a number $x$ from 0 to $2^{n} - 1$. Given this number Bob evaluates function $f(x)$ that returns only two values 0 or 1. Furthermore, Bob makes a promise that he will use only two kinds of functions, either constant or balanced. A constant function returns either 0 or 1 value for all input values $x$. A balanced function returns 0 value for exactly half of the values of $x$, and 1 for the remaining half of the values. The objective of this game is for Alice to decide which kind of the function Bob used. How many queries has Alice to submit to learn the property of function $f(x)$ with \textit{certainty}? Classically, the answer is straightforward. Since in the worst case scenario Bob can return $2^{n}/2$ 0s before sending a 1, Alice has to submit $2^{n}/2 + 1$ numbers to decide whether function $f(x)$ is constant or balanced with certainty \cite{nielsen}. However, in the best case scenario only two evaluations of the function suffice. If the second value returned by Bob is different from the first one then Alice concludes that the function is balanced. 
Naturally, when Alice is allowed to make her guess with some probability of error then she needs only few random queries.

In the quantum domain, the answer to the Deutsch problem can be given following only one query. Here, Alice stores her queries in the $n$ qubit register. Additionally, Bob has a single target qubit which serves as a repository for every value that can be returned by function $f(x)$. In order to evaluate the value of the function Bob uses a controlled black-box unitary operator $U_{f}$ whose action on the state of the register and target qubits is given by
\begin{equation}
U_{f} |x\rangle |y\rangle = |x\rangle |y \oplus f(x)\rangle\, ,
\end{equation}
where $y \oplus f(x)$ is modulo 2 addition. Therefore, following the action of $U_{f}$ the state of the target qubit is flipped or remains unchanged depending on the state of the register. The $U_{f}$ operator represents the CX gate.  The Deutsch-Jozsa algorithm is shown in Fig.~\ref{DJ_q} and is implemented by the following quantum network:
\begin{equation}
N_{DJ} = H^{\otimes (n+1)} U_{f} H^{\otimes n}\, .
\end{equation}
The elementary steps of the Deutsch-Jozsa algorithm are listed below:
\begin{enumerate}
\item prepare the $n$ qubit register and single target qubit in $|0\rangle^{\otimes n}$ and $|1\rangle$ states, respectively. Therefore, the input state is $|\psi\rangle = |0\rangle^{\otimes n} |1\rangle$;
\item apply the Hadamard gate to all $n+1$ qubits, thus creating a complete superposition state of the $n$ qubit register and balanced superposition state of the target qubit according to
\begin{equation}
|\psi\rangle = \frac{1}{\sqrt{2^{n}}} \sum_{x = 0}^{2^{n}-1} |x\rangle \left(\frac{|0\rangle - |1\rangle}{\sqrt{2}}\right)\, ;
\end{equation}
\item next, Bob applies a controlled black-box operator $U_{f}$ to the register and target qubits
\begin{equation}
|\psi\rangle = \frac{1}{\sqrt{2^{n}}} \sum_{x = 0}^{2^{n}-1} (-1)^{f(x)}|x\rangle \left(\frac{|0\rangle - |1\rangle}{\sqrt{2}}\right)\, ;
\end{equation}
\item subsequently, the Hadamard gate is applied to the $n$ qubit register, giving
\begin{equation}
|\psi\rangle = \frac{1}{2^{n}} \sum_{y = 0}^{2^{n}-1} \sum_{x = 0}^{2^{n}-1} (-1)^{x \cdot y + f(x)}|y\rangle \left(\frac{|0\rangle - |1\rangle}{\sqrt{2}}\right)\, ,
\end{equation}
with $x \cdot y$ being the inner product of $x$ and $y$, taken modulo 2.
\item finally, Alice measures her $n$ qubit register by projecting on the $|0\rangle^{\otimes n}$ state. The probability of finding all her qubits in the $|0\rangle$ state is given by \cite{blumel}
\begin{equation}\label{prob}
P_{0} = \left| \frac{1}{2^{n}} \sum_{x = 0}^{2^{n}-1} (-1)^{f(x)} \right|^{2}\, .
\end{equation}
We note that the probability $P_{0}$ depends on the character of function $f(x)$. If the function is \textit{constant} (taking either 0 or 1 value) then Eq.~(\ref{prob}) reduces to
\begin{equation}
P_{0} = \left| \frac{\pm 1}{2^{n}} \sum_{x = 0}^{2^{n}-1} 1 \right|^{2} = 1\, .
\end{equation}
All other probabilities are exactly zero. Thus, when the function $f(x)$ is constant, the register is found in the $|0\rangle^{\otimes n}$ state with \textit{certainty}. If, however, the function $f(x)$ is \textit{balanced} then Eq.~(\ref{prob}) reduces to
\begin{equation}\label{prob}
P_{0} = \left| \frac{1}{2^{n}} \sum_{x = 0}^{2^{n}-1} (-1)^{f(x)} \right|^{2} = \left| \frac{1}{2^{n}} 0 \right|^{2} = 0\, .
\end{equation}
For the balanced function, the positive part of the probability cancels the negative one resulting in a zero value for probability $P_{0}$. Therefore, Alice never observes the register in the $|0\rangle^{\otimes n}$ state, that is, at least one of the register qubits must be in the $|1\rangle$ state.
\end{enumerate}
In summary, if Alice finds all register qubits present in the $|0\rangle$ state then function $f(x)$ is constant, otherwise function $f(x)$ is balanced. The key ingredient of the Deutsch-Jozsa algorithm (and many other well know quantum algorithms) is embodied by the Hadamard operation that enables parallel processing and interference of different computational paths. In other words, the quantum superposition principle that gives rise to a quantum parallelism allows for the answer to the Deutsch problem to emerge in a single execution of the algorithm. In the next section, we show how the Deutsch-Jozsa algorithm can be extended to continuous-variable quantum systems.

\section{Deutsch-Jozsa algorithm over continuous variables}\label{sec:dj_con}
\noindent
The generalisation of the Deutsch-Jozsa algorithm to continuous variables was devised by Pati and Braunstein \cite{pati}. This generalisation was implemented with idealised continuous variables defined on an infinite domain. However, any practical CV implementation of the Deutsch problem can be realised only in a finite domain. Nevertheless, for simplicity and clarity, we first recall the Deutsch-Jozsa algorithm over continuous variables as originally stated in Ref.~\cite{pati}.

As already stated, the objective of the Deutsch problem is to determine whether some function $f(x)$ is constant or balanced. Similarly as in the discrete case, this is achieved by Alice and Bob playing the following game. Alice submits a value of $x$ from $-\infty$ to $+\infty$ to Bob. Then Bob evaluates $f(x)$, which can take only two values: 0 or 1. Bob also promises Alice to use either balanced or constant functions. A constant function is either 0 or 1 for all values of $x \in (-\infty,+\infty)$. A balanced function is 0 for half of the values of $x$, and 1 for the remaining values of $x$. This is defined in terms of the Lebesgue measure $\mu$ on $\mathbb{R}$: $\mu ({x \in \mathbb{R}|f(x) = 0})=\mu ({x \in \mathbb{R}|f(x) = 1})$ \cite{pati}. The goal of this game is the same as the objective of the traditional Deutsch problem, that is, to establish if the function used by Bob is constant or balanced. Classically, Alice would have to submit infinitely many values of $x$ to learn the global property of $f(x)$ with certainty. However, if Bob can use a unitary black-box operation to calculate function $f(x)$, then only a single function evaluation is sufficient to reveal the global property of $f(x)$. In the setting of idealised CVs, this would imply an infinite speed-up over any classical procedure.

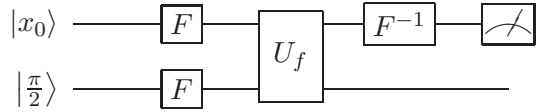
\begin{figure}[t]
\begin{center}
\begin{equation*}
	\Qcircuit @C=1.3em @R=0.8em @!R @!C {
\lstick{\ket{x_0}}			& \gate{F}	& \multigate{1}{{U_{f}}} 	& \gate{F^{-1}}	& \meter  \\
\lstick{\ket{\frac{\pi}{2}}}	& \gate{F}	& \ghost{{U_f}} 		& \qw 			& \qw
	}
\end{equation*}
\end{center}
\caption{A quantum circuit representing the Deutsch-Jozsa algorithm over continuous variables. The quantum network ${N_{DJ}}$ consists of the Fourier transforms $F$ and controlled black-box gate $U_{f}$ applied to the register and target CVs prepared in the idealised position eigenstates $|x_{0}\rangle$ and $|\pi/2\rangle$, respectively. The last operation is an inverse Fourier transform that enables the interference of different computational paths. \label{DJ_2CV}}
\end{figure}

The ideal Deutsch-Jozsa algorithm over continuous variables is shown in Fig.~\ref{DJ_2CV}. This implementation of the Deutsch-Jozsa algorithm employs two CVs, the so-called register and target CVs. Alice stores her query in the register CV, and the target CV is used by Bob during function evaluation. The register is prepared in the position eigenstate $|x_{0}\rangle$ and the target in the position eigenstate $|\pi/2\rangle$. The quantum network ${N_{DJ}}$ implementing the Deutsch-Jozsa algorithm is given by the following unitary transformation:
\begin{equation}
N_{DJ} = F^{-1}_{r} U_{f} F_{r} F_{t}\, ,
\end{equation}
where $F$ is the Fourier transform and $r$ and $t$ indicates the register and target CV, respectively. The Fourier transform applied to a CV in some position eigenstate $|x\rangle$ creates a superposition of all position eigenstates according to
\begin{equation}
F|x\rangle = \frac{1}{\sqrt{\pi}} \int^{\infty}_{-\infty} dy \, e^{2 i x y} |y\rangle\, ,
\end{equation}
where we used photon number units in which $\hbar=\frac12$. The unitary black-box operator $U_{f}$ evaluates a value of function $f(x)$ and stores it in the state of the target CV: $|x \rangle |y \rangle \xrightarrow{} |x \rangle |y + f(x) \rangle$. Let us analyse the Deutsch-Jozsa algorithm step by step:
\begin{enumerate}
\item prepare the register and target CVs in an ideal position eigenstate $|x_{0}\rangle$ and $|\pi/2\rangle$, respectively;
\item apply the Fourier transform $F$ to the register and target CVs,
\begin{equation}
|s\rangle = F_{r} F_{t} |x_{0} \rangle |\pi/2 \rangle = \frac{1}{\pi} \int^{\infty}_{-\infty} \int^{\infty}_{-\infty} dx dy \, e^{2i x x_{0}+ i \pi y} |x\rangle |y \rangle\, ; \nonumber
\end{equation}
\item following the action of a unitary black-box operator $U_{f}$, the state of the CVs is given by
\begin{equation}
U_{f}|s\rangle = \frac{1}{\sqrt{\pi}} \int^{\infty}_{-\infty} dx \, e^{2i x x_{0}} e^{-i \pi f(x)}  |x\rangle F_{t}|\pi/2\rangle\, ; \nonumber
\end{equation}
\item the quantum network ${N_{DJ}}$ is finalised with an inverse Fourier transform $F^{-1}$ applied to the register CV. Therefore, the state of the CVs can be written as
\begin{eqnarray}
F^{-1}_{r} U_{f} |s\rangle = \frac{1}{\pi}\int^{\infty}_{-\infty} \int^{\infty}_{-\infty} dx dx' \, e^{2i x (x_{0}-x')} e^{-i \pi f(x)}  |x' \rangle F_{t}|\pi/2 \rangle\, ; \nonumber
\end{eqnarray}
\item following the quantum network ${N_{DJ}}$, the property of the function $f(x)$ is determined by projecting the state of the register CV onto the original position eigenstate $|x_{0}\rangle$.
\end{enumerate}
The continuous-variable projection operator for idealised states can be written as
\begin{equation}
P_{x_{0}} = \int^{x_{0}+\varepsilon}_{x_{0}-\varepsilon} dy \, |y\rangle \langle y|\, ,
\end{equation}
where $\varepsilon$ is the spread around $x_{0}$ value, that is, the CV measurement cannot be performed with infinite precision. The orthogonal complement of $P_{x_{0}}$ is given by
\begin{equation}
P_{\bar{x}_{0}} = I - P_{x_{0}} = I - \int^{x_{0}+\varepsilon}_{x_{0}-\varepsilon} dy |y\rangle \langle y|.
\end{equation}
By construction, a complete set of orthogonal projectors $P_m$ satisfies the completeness relations $\sum_{m} P_{m} = I$ and $P_{m}P_{m'} = \delta_{m m'}P_{m}$. If $f(x)$ is constant, then the measurement statistics based on the preceding set of orthogonal projection operators (and taking $\varepsilon \rightarrow 0$) is given by
\begin{eqnarray}
p(x_{0}) &=& \mbox{Tr}[\hat{P}_{x_{0}} \rho_{DJ}] = 1, \\
p(\bar{x}_{0}) &=& \mbox{Tr}[\hat{P}_{\bar{x}_{0}} \rho_{DJ}] = 0\, ,
\end{eqnarray}
where $p(x_{0})$ is the probability of measurement outcome to be $x_{0}$, $p(\bar{x}_{0})$ is the probability of a measurement outcome different than $x_{0}$, and $\rho_{DJ} = N_{DJ} |r\rangle |t\rangle \langle t| \langle r| N^{-1}_{DJ}$. Conversely, if $f(x)$ is balanced, then the measurement statistics assuming $\varepsilon \rightarrow 0$ is given by
\begin{eqnarray}
p(x_{0}) &=& \mbox{Tr}[\hat{P}_{x_{0}} \rho_{DJ}] = 0, \\
p(\bar{x}_{0}) &=& \mbox{Tr}[\hat{P}_{\bar{x}_{0}} \rho_{DJ}] = 1.
\end{eqnarray}
Therefore, if the state of the register CV remains unchanged, then the function $f(x)$ is definitely constant, and if the state of the register CV is not $|x_{0}\rangle$, then the function $f(x)$ is balanced. A single function evaluation solves the Deutsch problem.

The core of the preceding implementation of the Deutsch-Jozsa algorithm is represented by a unitary, controlled black-box operator $U_{f}$ applied between the Fourier-transformed register and target CVs. Here, the Fourier-transformed target CV, together with a black-box operator, induces a phase shift, which depends on the global property of the function $f(x)$:
\begin{equation}
U_{f}(|x\rangle F_{t} |\pi/2\rangle) = e^{-2i f(\hat{x}) \hat{p}_{t}} |x\rangle F_{t}|\pi/2\rangle = e^{-i \pi f(x)} |x\rangle F_{t} |\pi/2\rangle\, . \nonumber
\end{equation}
Notice that the state of the target CV is not changed following the action of $U_{f}$. In fact, $F_{t} |\pi/2\rangle$ is an eigenstate of $U_{f}$ with an eigenvalue $e^{-i \pi f(x)}$ ``kicked back" in front of the register CV \cite{cleve}. Conventionally, the Deutsch-Jozsa algorithm employs multiple quantum systems; however, as the preceding simple analysis of the action of $U_{f}$ indicates, the target CV can be omitted. It is easy to show that a single register CV together with a redefined black-box operator $U_{f} \equiv e^{-2 i \, \pi/2 \, f(\hat{x})}$ is enough to implement the Deutsch-Jozsa algorithm over continuous variables. In Ref.~\cite{sanders09}, the authors arrived at the same conclusion; however, they used a slightly different approach. We emphasise that a direct consequence of employing a single system is that this protocol does not use any entanglement to determine the global property of the function in a single run. Moreover, the preceding implementation of the Deutsch-Jozsa algorithm is expressed in terms of the idealised position eigenstates. However, a more realistic and physically meaningful representation of a continuous variable is given by, for example, Gaussian states.

Similar to the setting of discrete quantum systems (e.g., qubits), some features of the Deutsch-Jozsa algorithm can serve as a starting point for developing other quantum algorithms. A slightly modified black-box operator $U_{f} \equiv e^{-2 i \, \pi/2 \, f(\hat{x})}$ for a simplified Deutsch-Jozsa algorithm can be used as the core of a protocol capable of estimating an unknown parameter that under appropriate conditions still retains the capabilities of the Deutsch-Jozsa algorithm. Before introducing this protocol, we recall some basic concepts in quantum parameter estimation theory.

\section{Parameter estimation}\label{sec:qm}
\noindent
The most general parameter estimation procedure is shown in Fig.~\ref{PE}, and consists of three elementary steps:
\begin{enumerate}
\item prepare a probe system in an initial quantum state $\rho(0)$;
\item evolve it to a state $\rho(\varphi)$ by a unitary evolution $U(\varphi)=\exp(-i\varphi\mathcal{H})$;
\item subject the probe system to a generalised measurement $M$, described by a Positive Operator Valued Measure ({\sc povm}) that consists of elements $\hat{E}_x$, where $x$ denotes the measurement outcome.
\end{enumerate}
Here, the Hermitian operator $\mathcal{H}$ is the generator of translations in $\varphi$, the parameter we wish to estimate. The amount of information about $\varphi$ that can be extracted by a measurement of the probe system is given by the Fisher information,
\begin{equation}
F(\varphi) = \sum_{x} \, \frac{1}{p(x|\varphi)} \left( \frac{\partial p(x|\varphi)}{\partial \varphi} \right)^2\, ,
\end{equation}
where $p(x|\varphi) = \mbox{Tr}[\hat{E}_{x} \rho(\varphi)]$ is the probability distribution given by the Born rule that describes the measurement data, and $x$ is a discrete measurement outcome.
\begin{figure}[t]
\centering
\includegraphics[width=6cm]{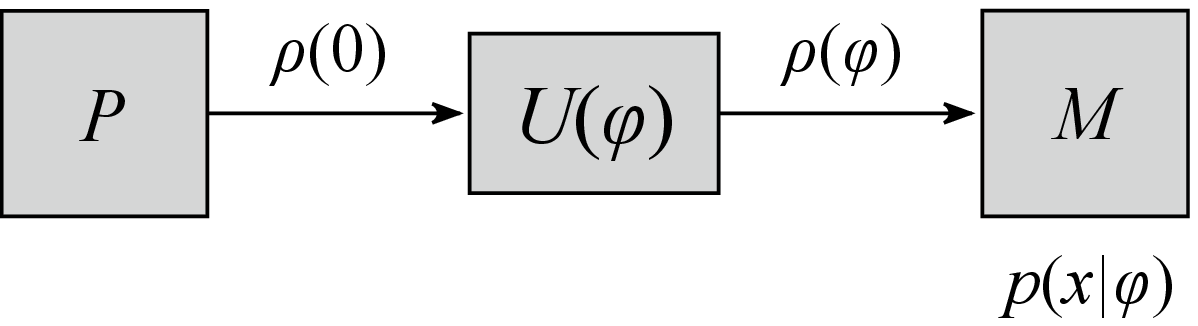}
\caption{The general parameter estimation procedure involving state preparation $P$, evolution $U(\varphi)$, and generalised measurement $M$ with outcomes $x$, which produces a probability distribution $p(x|\varphi)$. \label{PE}}
\end{figure}
Based on the Fisher information, one can bound a minimal value of the uncertainty in $\varphi$ with the quantum Cram\'{e}r-Rao bound \cite{helstrom67,helstrom69,braunstein94},
\begin{equation}
(\delta\varphi)^2 \geq \frac{1}{T F(\varphi)}\, ,
\end{equation}
where $(\delta\varphi)^2$ is the mean squared error in the parameter $\varphi$, and $T$ is the number of times the procedure is repeated. The ultimate limit of the quantum Cram\'{e}r-Rao bound depends on how the Fisher information is bounded from above. The Fisher information can be bounded in two ways: by the variance of $\mathcal{H}$ \cite{braunstein96} or by the expectation value of $\mathcal{H}$ \cite{zwierz10},
\begin{equation}
F(\varphi) \leq 16(\Delta \mathcal{H})^2  \quad\mbox{and}\quad  F(\varphi) \lesssim 4 \left| \langle\mathcal{H}\rangle \right|^2\, , \label{bounds}
\end{equation}
where we again used $\hbar = \frac12$. Since both bounds are completely general and complement each other, any parameter estimation procedure must respect them. Typically, the Fisher information may be related to various resource counts such as the average photon number, the average energy of the probe system, or the number of fundamental ``atomic'' unitary evolution gates that are used in the estimation procedure. As we have established in Chapter~\ref{Chapter2}, all these different resource counts are encompassed by the expectation value of $\mathcal{H}$ which plays the role of a proper resource count \cite{zwierz10}. We usually consider two scaling regimes of the quantum Cram\'{e}r-Rao bound. The first regime, the so-called {\em standard quantum limit} ({\sc sql}) \cite{gardiner04} or {\em shot-noise limit}, is obtained when the Fisher information is a constant with respect to $T$ and the resource count. The {\sc sql} is typically given by
\begin{equation}
\delta\varphi \gtrsim \frac{1}{\sqrt{T}}\, .
\end{equation}
The second regime, the so-called {\em Heisenberg limit} \cite{holland93}, is obtained in a single-shot experiment ($T=1$) when the Fisher information scales quadratically with the resource count. The Heisenberg limit is then given by
\begin{equation}
\delta\varphi \geq \frac{1}{\sqrt{F(\varphi)}}\, .\label{HL}
\end{equation}
Therefore, the uncertainty in the parameter $\varphi$ scales linearly inversely with the resource count. Both scaling regimes of the quantum Cram\'{e}r-Rao bound can be compared directly in terms of an appropriate resource count \cite{zwierz10}.

\section{General procedure with Gaussian states} \label{sec:unify}
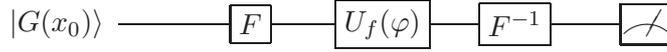
\begin{figure}[t]
\begin{center}
\begin{equation*}
	\Qcircuit @C=1.3em @R=0.8em @!R @!C {
\lstick{\ket{G(x_0)}}			& \gate{F}	& \gate{U_{f}(\varphi)} 	& \gate{F^{-1}}	& \meter
	}
\end{equation*}
\end{center}
\caption{A quantum circuit representing the general protocol over continuous variables. The quantum network consists of the Fourier transform $F$ and black-box gate $U_{f}(\varphi)$ applied to a single register CV prepared in the Gaussian state $|G(x_{0})\rangle$. The last operation is an inverse Fourier transformation that enables the interference of different computational paths. \label{estimation}}
\end{figure}

\noindent
In this section, we present a general procedure capable of determining the value of a single parameter $\varphi \in [0,2\pi)$ or implementing the Deutsch-Jozsa algorithm (see Fig.~\ref{estimation}).
Here, the black-box operator is defined in the following way:
\begin{equation}
U_{f}(\varphi) \equiv \exp\left[-2i \varphi f(\hat{x})\right]\, ,
\end{equation}
where $f(\hat{x}) |x\rangle = f(x) |x\rangle$. The function $f(x)$ again takes only two values: 0 and 1. Without loss of generality, ideal, non-normalisable continuous variable states are regularised to Gaussian input states. Similar to the case of the Deutsch-Jozsa algorithm, any physical continuous-variable parameter estimation protocol can be implemented only on a finite domain. Therefore, we introduce the semi-Gaussian input state defined on a finite domain given by
\begin{equation}
|G(x_{0})\rangle = \int^{T}_{-T} \frac{dx}{N_{x}} \, \exp\left[-\frac{(x - x_{0})^2}{2 \Delta^2}\right] |x\rangle\, ,
\end{equation}
where $\Delta$ is the variance of the state and $N_{x}$ is the normalisation constant given by
\begin{equation}
N_{x}^{2} = \frac{\sqrt{\pi \Delta^2}}{2} \, \left[\mbox{erf}\left(\frac{T + x_{0}}{\Delta}\right) + \mbox{erf}\left(\frac{T - x_{0}}{\Delta}\right)\right]\, . \nonumber
\end{equation}
We note that for $\Delta \ll T$ we recover the normalisation constant in the form of $N_{x}^2 = \sqrt{\pi \Delta^2}$ which is characteristic for a Gaussian state defined on an infinite domain, that is, from $-\infty$ to $+\infty$. The Fourier-transformed semi-Gaussian state defined on a finite domain can be written as
\begin{equation}
|G(p_{0})\rangle = \int^{P}_{-P} \frac{dp}{N_{p}} \, \exp\left[- 2 \Delta^2 (p - p_{0})^2\right] |p\rangle\, ,
\end{equation}
where $1/(2 \Delta)$ is the variance of the Fourier-transformed semi-Gaussian state and $N_{p}$ is given by
\begin{equation}
N_{p}^{2} = \frac{\sqrt{\pi/ 4 \Delta^2}}{2} \, \left[\mbox{erf}(2 (P + p_{0}) \Delta) + \mbox{erf}(2 (P - p_{0}) \Delta)\right]\, .\nonumber
\end{equation}
For $P \gg 1/(2 \Delta)$ the normalisation constant takes the form of $N_{p}^2 = \sqrt{\pi/ 4 \Delta^2}$, characteristic for a Fourier-transformed Gaussian state define on an infinite domain. The relationship between domains of the semi-Gaussian input state and its Fourier-transformed counterpart is given by $P = 1/(2T)$.

The general procedure consists of the following instructions:
\begin{enumerate}
\item prepare the register CV in the normalised semi-Gaussian state $|r\rangle = |G(x_{0})\rangle$, and apply the Fourier transform $F$ defined by
\begin{equation}
F|x\rangle = |x\rangle_{p} =  \frac{1}{\sqrt{2T}} \int^{T}_{-T} dy \, e^{2 i x y} |y\rangle \, ,
\end{equation}
where $|x\rangle_{p}$ is the Fourier-transformed position eigenstate, that is, the momentum eigenstate;
\item subsequently, a black-box operator $U_{f}(\varphi)$ is applied. Then the state of the system is
\begin{eqnarray}
U_{f}(\varphi) F|r\rangle &=& \int^{T}_{-T} \frac{dx}{N_{x}} \, \exp\left[{-\frac{\left(x - x_{0}\right)^2}{2 \Delta^2}}\right] e^{-2i \varphi f(\hat{x})} |x\rangle_{p} \nonumber \\
&=& \frac{1}{\sqrt{2 T}} \int^{T}_{-T} \frac{dx dy}{N_{x}} \, \exp\left[{-\frac{\left(x - x_{0}\right)^2}{2 \Delta^2}}\right] \nonumber \\ && \times e^{2i y x} e^{-2i \varphi f(y)} |y\rangle \nonumber \, ;
\end{eqnarray}
\item finally, an inverse Fourier transform $F^{-1}$ is applied followed by a measurement. The state of the register CV is measured by projecting onto the original semi-Gaussian state centered around $x_{0}$.
\end{enumerate}
The measurement is described by a {\sc povm} $\{P_{x_{0}}, P_{\bar{x}_{0}}\}$, where
\begin{equation}
P_{x_{0}} = \int^{T}_{-T} dx dy \, g_{xy} |x\rangle \langle y|, \; \mbox{and} \; P_{\bar{x}_{0}} = \mathbb{I} - P_{x_{0}} \label{POVM}
\end{equation}
with
\begin{equation}
g_{xy} = \frac{1}{N_{\varepsilon}^{2}} \, \exp\left[-\frac{\left(x - x_{0}\right)^2}{2 \varepsilon^2}\right] \exp\left[-\frac{\left(y - x_{0}\right)^2}{2 \varepsilon^2}\right],
\end{equation}
and $\varepsilon$ is the intrinsic precision of the measurement apparatus; that is, any CV measurement must have finite precision if it is to be physical, and $N_{\varepsilon}$ is the normalisation constant given by
\begin{equation}
N_{\varepsilon}^{2} = \frac{\sqrt{\pi \varepsilon^2}}{2} \, \left[\mbox{erf}\left(\frac{T + x_{0}}{\varepsilon}\right) + \mbox{erf}\left(\frac{T - x_{0}}{\varepsilon}\right)\right]\, . \nonumber
\end{equation}
The optimal measurement which corresponds to the initial semi-Gaussian register state has $\varepsilon = \Delta$; thus $N_{\varepsilon} = N_{x}$.

Now let us calculate the measurement statistics. Analytical expressions for the measurement statistics are hard to find due to the presence of error functions. However, for the semi-Gaussian states with $\Delta \ll T$ the calculations simplify considerably. Under this regime, the limits of integration for the integrals containing terms that depend on $\Delta$ range from $-\infty$ to $+\infty$. Necessarily, the normalisation constants have to be changed and are expressed as $\sqrt{2T} N_{x} = \sqrt{\pi} \sqrt[4]{\pi \Delta^2}$. In other words, a semi-Gaussian input state defined on a finite domain is approximated with a Gaussian state defined on an infinite domain. Therefore, the measurement statistics based on the preceding {\sc povm} are given by the following expression:
\begin{eqnarray}
p(x_{0}|\varphi) &=& \frac{4 \Delta^2}{\pi} \int^{P}_{-P} dz dy \, e^{-4 \Delta^2 (z^2 + y^2)} e^{2i \varphi (f(z)-f(y))}, \nonumber \\
p(\bar{x}_{0}|\varphi) &=& 1 - p(x_{0}|\varphi). \label{pro}
\end{eqnarray}
Here, the interval $(-P, P)$ is a finite domain of the Fourier-transformed semi-Gaussian state $|G(x_{0})\rangle$ and denotes the interval where for this particular procedure function $f(x)$ is defined.

At this point, we have to give an explicit definition of the function. Functions $f(x)$ defined on a finite domain returning only two values $\left(\{0,1\}\right)$ fall into three distinct categories: constant, balanced, and neither constant nor balanced. We recall that the objective of the Deutsch-Jozsa algorithm is to probe whether an unknown function $f(x)$ is constant or balanced. We parametrize the three possibilities for defining $f(x)$ by introducing a parameter $r$. The preceding integrals can then be evaluated for any function $f(x)$ behaving as a step function, with the parameter $r$ marking the point where $f(x)$ changes its value. Hence, for $r = 0$ and $r = \pm P$ the function $f(x)$ is balanced and constant, respectively. For $0 < r < P$ (or $-P < r < 0$), the function $f(x)$ is neither constant nor balanced. We consider only positive values of $r$ due to the symmetry of the setup. This leads to
\begin{eqnarray}
p(x_{0}|\varphi) &=& \frac{1}{2} \left[\mbox{erf}^{2}(2 P \Delta) + \mbox{erf}^{2}(2 r \Delta)\right] + \nonumber \\
&& \frac{1}{2} \left[\mbox{erf}^{2}(2 P \Delta) - \mbox{erf}^{2}(2 r \Delta)\right]\cos(2 \varphi), \nonumber \\
p(\bar{x}_{0}|\varphi) &=& 1 - p(x_{0}|\varphi)\, , \nonumber
\end{eqnarray}
where $p(x_{0}|\varphi)$ is the probability of measurement outcome to be in the interval $x_{0} \pm \varepsilon$ and $p(\bar{x}_{0}|\varphi)$ is the probability of measurement outcome not to be in the interval $x_{0} \pm \varepsilon$.

\subsection{Representations of $f(x)$}
\noindent
\begin{figure}[t]
\begin{center}
\includegraphics[scale=0.90]{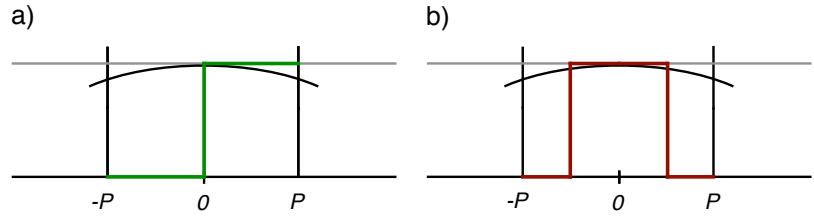}
\caption{Representations of function $f(x)$ in the momentum domain $z \in (-P, P)$. a) a simple step-function representation of a balanced function with $r = 0$, b) the ``hat" representation of a balanced function that changes its value twice at points $r_{1} = -P/2$ and $r_{2} = P/2$. The curved line represents the Fourier transformed Gaussian state and the straight, grey line corresponds to the Fourier transformed idealised state. For more details see text. \label{function_thesis}}
\end{center}
\end{figure}
Our choice to represent $f(x)$ as a step function simplified our calculations. However, we can imagine more elaborate behaviour patterns for $f(x)$. In principle, since in the case of the Fourier-transformed idealised CVs all terms have amplitudes of equal magnitude, all finite subintervals where the function takes value 0, can be added up to a single interval. The same applies to all subintervals where function takes value 1. Therefore, one ends up with two intervals and a relationship between them given by the parameter $r$. However, in the setting of semi-Gaussian states defined on a finite domain, the preceding reasoning is not quite as straightforward. The amplitudes of the Fourier-transformed Gaussian states have a slightly different magnitude. One may notice this feature by inspecting Eq.~(\ref{pro}). Since in our calculations we favour a step-function representation over any other, let us estimate the maximum error we make with this assumption. Due to a trivial nature of a constant function, in the following analysis we consider a balanced function. We consider the step-function representation of a balanced function with $r = 0$ (see Fig.~\ref{function_thesis}a)). The biggest deviation from this representation is offered by a balanced function that changes its value twice at points $r_{1} = -P/2$ and $r_{2} = P/2$ (see Fig.~\ref{function_thesis}b)). Both representations produce two distinct probability distributions, $p_{step}(x_{0}|\varphi)$ and $p_{hat}(x_{0}|\varphi)$, respectively, that differ by the error $\varepsilon_{P\Delta}$ given by
\begin{equation}
\varepsilon_{P\Delta} = \left|1 - \cos(2 \varphi)\right| \times \left|- \frac{8}{\pi} (P \Delta)^6 + \frac{24}{\pi} (P \Delta)^8 + O\left((P \Delta)^{10}\right)\right| \, . \nonumber
\end{equation}
The error tends to zero with $P \Delta \rightarrow 0$. This is natural since when $\Delta \rightarrow 0$ all amplitudes of the Fourier-transformed idealised position eigenstate have the same magnitude; that is, the spectrum is flat.

\subsection{Analysis}
\begin{figure}[t]
\centering
\includegraphics[width=8.5cm]{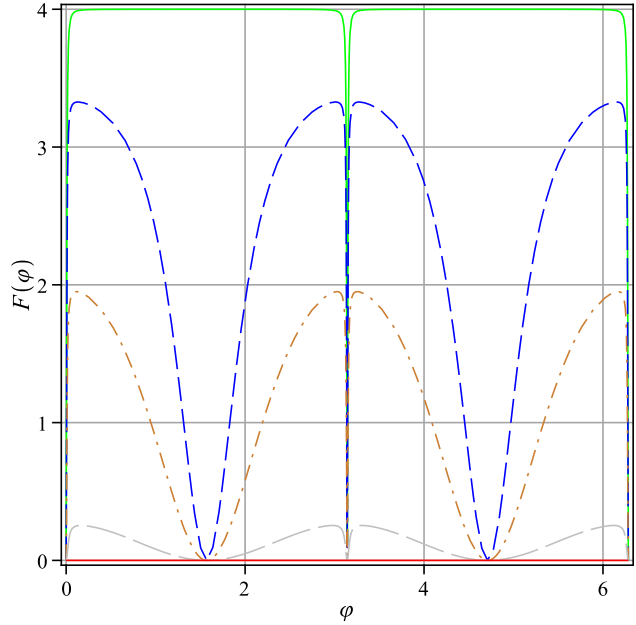}
\caption{General dependence of the Fisher information $F(\varphi)$ for five values of the parameter $r$: $r = 0$ corresponds to the uppermost solid line (green), $r = P/8$ corresponds to the dashed line (blue), $r = P/4$ corresponds to the dashed-dotted line (brown), $r = P/2$ corresponds to the long-dashed line (grey), and $r = P$ corresponds to the lowermost solid line (red). Here $P = 3/(2 \Delta)$ with $\Delta = 1/\sqrt{2}$. \label{dep_r}}
\end{figure}

\noindent
Our procedure can be analysed in two ways. As expected, from one perspective it behaves as a parameter estimation protocol. From the other, it  behaves as the Deutsch-Jozsa algorithm. First, we analyse the behaviour of the parameter estimation part of the procedure. Based on the preceding measurement statistics, we calculate the Fisher information $F(\varphi)$. The minimal value of $F(\varphi) = 0$ occurs when function $f(x)$ is constant ($r = P$) with the corresponding measurement statistics:
\begin{eqnarray}
p(x_{0}|\varphi) &=& \mbox{erf}^2(2 P \Delta), \nonumber \\
p(\bar{x}_{0}|\varphi) &=& 1 - \mbox{erf}^2(2 P \Delta). \nonumber
\end{eqnarray}
Conversely, the maximal value of the Fisher information,
\begin{equation}
F(\varphi) = \frac{4 \, \mbox{erf}^2(2 P \Delta) \left[\cos(2 \varphi)-1\right]}{\mbox{erf}^2(2 P \Delta) \left[\cos(2 \varphi) + 1\right] - 2}\, ,
\end{equation}
occurs when function $f(x)$ is balanced ($r = 0$) with the corresponding measurement statistics:
\begin{eqnarray}
p(x_{0}|\varphi) &=& \frac{1}{2} \, \mbox{erf}^2(2 P \Delta) \left[1 + \cos(2 \varphi)\right], \nonumber \\
p(\bar{x}_{0}|\varphi) &=& 1 - \frac{1}{2} \, \mbox{erf}^2(2 P \Delta) \left[1 + \cos(2 \varphi)\right]. \nonumber
\end{eqnarray}
Here the optimal value of the Fisher information $F(\varphi) = 4$ is given for $\mbox{erf}^2(2 P \Delta) = 1$. This condition imposes a lower bound on $P$:
\begin{equation}
\mbox{erf}^2 x = 1 \ \ \mbox{for} \ \ x \geq 3 \ \Rightarrow \ P \geq 3/(2 \Delta)\, ,
\end{equation}
which, in general, implies $P \gtrsim 1/(2 \Delta)$ and is consistent with the approximation applied earlier. The general dependence of the Fisher information $F(\varphi)$ on parameter $r$ with $P = 3/(2 \Delta)$ and $\Delta = 1/\sqrt{2}$ (the variance of the coherent state) is shown in Fig.~\ref{dep_r}. The dips that are especially visible for the balanced function appear because the Fisher information $F(\varphi)$ retains some dependence on the parameter $\varphi$ since for $P = 3/(2 \Delta)$, $\mbox{erf}^2(2 P \Delta) \approx 1$.
Based on the general dependence of $F(\varphi)$ on $r$, we conclude that the maximal value of the Fisher information is indeed obtained for a balanced function.

In order to address the optimality of our parameter estimation protocol, we analyse the behaviour of the generator of translations in the parameter $\varphi$: $\mathcal{H} \equiv f(\hat{x})$. The expectation value of the generator $\mathcal{H}$ in the state of the register CV preceding application of the black-box operator, that is, $|\psi_{in}\rangle = F|r\rangle$ with $\Delta \ll T$, is given by
\begin{equation}
\left| \langle \mathcal{H} \rangle \right|^2 = \left| \langle f(\hat{x}) \rangle \right|^2 = \frac{1}{4} \left[\mbox{erf}(2 P \Delta) - \mbox{erf}(2 r \Delta)\right]^2. \nonumber
\end{equation}
The last equality holds for positive values of $r$. Since $f^2(x) = f(x)$ the variance of the generator $\mathcal{H}$ in $|\psi_{in}\rangle$ can be written as
\begin{eqnarray}
(\Delta \mathcal{H})^2 = (\Delta f(\hat{x}))^2 &=& \frac{1}{2} \left[\mbox{erf}(2 P \Delta) - \mbox{erf}(2 r \Delta)\right] \nonumber \\
&& \times \left[ 1 - \frac{1}{2} \left[\mbox{erf}(2 P \Delta) - \mbox{erf}(2 r \Delta)\right] \right]. \nonumber
\end{eqnarray}
The maximal expectation value of the generator $\mathcal{H}$ occurs for a balanced function ($r = 0$) with $P \geq 3/(2 \Delta)$ and is given by $\left| \langle \mathcal{H} \rangle \right|^2 = 1/4$. On the other hand, the maximal variance of the generator $\mathcal{H}$ is $(\Delta \mathcal{H})^2 = 1/4$. Hence, the Fisher information is bounded by $F(\varphi) \leq 16(\Delta \mathcal{H})^2 = 4$. Therefore, we note that according to Eqs.~(\ref{bounds}) and (\ref{HL}) our procedure attains the scaling regime of the Heisenberg limit. We also note that even though, for this setup, the Fisher information is bounded by the variance of $\mathcal{H}$, asymptotically both bounds given in Eq.~(\ref{bounds}) coincide. In order to establish the optimality of the procedure, we must calculate whether $\delta \varphi = 1/\sqrt{F(\varphi)}$. We use the standard expression for the mean squared error given by
\begin{equation}
\delta \varphi = \frac{\Delta X}{\left| d\langle X \rangle/d\varphi \right|}\, ,
\end{equation}
where $X$ is the measurement observable defined as $X = P_{x_{0}}$ [see Eq.~(\ref{POVM})]. Hence, for the final state $\ket{\psi_{\varphi}} = F^{-1} U_{f}(\varphi) F |r\rangle$ with $\varepsilon = \Delta$, we calculate
\begin{equation}
\langle X \rangle = \langle \psi_{\varphi}| P_{x_{0}} |\psi_{\varphi}\rangle = \frac{1}{2} \, \mbox{erf}^2(2 P \Delta) \left[1 + \cos(2 \varphi)\right]\, . \nonumber
\end{equation}
Based on the property $P^{2}_{x_{0}} = P_{x_{0}}$, we find that $\langle X^{2} \rangle = \langle X \rangle$. For $P \geq 3/(2 \Delta)$ the mean squared error is $\delta \varphi = 1/2$. Hence, we conclude that for a balanced function our parameter estimation procedure over continuous variables attains the ultimate limit of the quantum Cram\'{e}r-Rao bound, and therefore is optimal. This result constitutes an analogy to the phase estimation with a qubit realised as a single photon placed in the arms of the Mach-Zender interferometer. Here the balanced property of function $f(x)$ plays a role of two distinct paths in a balanced Mach-Zender interferometer.

Next, let us analyse the Deutsch-Jozsa side of the procedure. Under appropriate conditions the developed procedure can determine the character of function $f(x)$. If a value of the parameter $\varphi$ is fixed, $\varphi = \pi/2$, then the measurement statistics are given by
\begin{eqnarray}
p(x_{0}) &=& \mbox{erf}^{2}(2 r \Delta), \nonumber \\
p(\bar{x}_{0}) &=& 1 - \mbox{erf}^{2}(2 r \Delta), \nonumber
\end{eqnarray}
It is clear that for a constant and balanced function $f(x)$ the corresponding measurement statistics of the Deutsch-Jozsa algorithm are recovered. Indeed, when function $f(x)$ is constant ($r = P$), then
\begin{eqnarray}
p(x_{0}) &=& \mbox{erf}^{2}(2 P \Delta), \nonumber \\
p(\bar{x}_{0}) &=& 1 - \mbox{erf}^{2}(2 P \Delta), \nonumber
\end{eqnarray}
and when function $f(x)$ is balanced ($r = 0$), then $p(x_{0}) = 0$ and $p(\bar{x}_{0}) = 1$. The Deutsch-Jozsa algorithm over the semi-Gaussian states defined on a finite domain becomes a probabilistic procedure. This is consistent with the conclusions found in Ref.~\cite{sanders09}. However, when the size of the domain is sufficiently large with $P \geq 3/(2 \Delta)$, then a definite distinction between constant and balanced functions can be made. Nevertheless, even for large-enough domains this implementation of the Deutsch-Jozsa protocol does not offer an unphysical, infinite speed-up over the classical procedures. We note that for ideal, non-normalisable position eigenstates ($\Delta \rightarrow 0$), the constant function measurement statistics is retained for $P \rightarrow \infty$, rendering $P$ and $r$ unphysical, thus making a meaningful distinction between the balanced and constant functions impossible.

\begin{figure}[t]
\centering
\includegraphics[width=8.5cm]{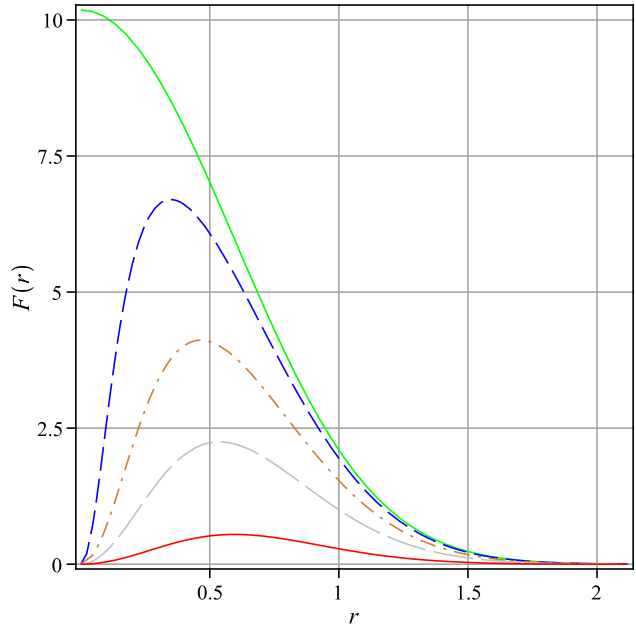}
\caption{General dependence of the Fisher information $F(r)$ (with $P = 3/(2 \Delta)$ and $\Delta = 1/\sqrt{2}$) for four values of the parameter $\varphi$: $\varphi = \pi/2$ corresponds to the uppermost solid line (green), $\varphi = 5 \pi/12$ corresponds to the dashed line (blue), $\varphi = \pi/3$ corresponds to the dashed-dotted line (brown), $\varphi = \pi/4$ corresponds to the long-dashed line (grey), and $\varphi = \pi/8$ corresponds to the lowermost solid line (red). The optimal value of $r$ shifts from balanced to constant. \label{dep_phi}}
\end{figure}

We also calculated the Fisher information $F(r)$ and plotted it against $r \in (0, P)$ for five different values of the parameter $\varphi = \{\pi/2, 5 \pi/12, \pi/3, \pi/4, \pi/8\}$ with $P = 3/(2 \Delta)$ and $\Delta = 1/\sqrt{2}$ (see Fig.~\ref{dep_phi}).
The maximal value of the Fisher information $F(r)$ is obtained for $\varphi = \pi/2$ corresponding to a simplified Deutsch-Jozsa algorithm. We note that the optimality changes from balanced to more constant when $\varphi \neq \pi/2$. Any further analysis of this side of the procedure is problematic due to a lack of the generator of translations in $r$.

One possible application of the Deutsch-Jozsa part of our procedure is to test the quality of the implementation of function $f(x)$ employed in the parameter estimation protocol. Whenever the function is balanced or constant the quality of its implementation can be established by probing the parameter $r$. We also stress that since we are employing a single continuous variable, no entanglement is present at the preparation stage and none is created during the computation. The quantum superposition principle itself is responsible for speed-up over any classical procedure. Even though, in principle, a single continuous variable is quite sufficient, a practical implementation of the Deutsch-Jozsa algorithm may require more continuous variables. Traditionally, operators of type $U_{f}$ and $U_{f}(\varphi)$, which introduce a phase factor in front of the register CV, are applied between two quantum systems, that is, the register and target CVs.

\section{Summary}\label{sec:conc}
\noindent
In conclusion, we developed a general procedure capable of performing two distinct tasks. For one mode of operation the protocol estimates a value of an unknown parameter with Heisenberg-limited precision. On the other hand, for a fixed value of the parameter in question the procedure addresses the Deutsch problem in a single run. Our procedure employs Fourier transforms and black-box unitary operator applied to a single continuous variable represented as the semi-Gaussian state defined on a finite domain. Consequently, for this setup, the parameter estimation side of the procedure is optimal and the Deutsch-Jozsa algorithm offers finite, that is, physically feasible, speed-up over any classical procedure. Furthermore, no entanglement is present at any stage of the procedure. A similar conclusion concerning the quantum metrology part of our procedure can be found in Refs.~\cite{tilma10,pinel10}. We emphasise a special role played by balanced functions $f(x)$. The procedure equipped with the black-box operator that introduces the parameter $\varphi$ via the balanced function attains the ultimate limit of the quantum Cram\'{e}r-Rao bound. This behaviour can be linked to the phase estimation with a qubit realised as a single photon placed in the arms of the Mach-Zender interferometer.  


\part{Atomic Ensembles}

\chapter{Techniques for Atomic Ensembles}
\label{Chapter4}
\lhead{\textsc{Chapter 4. Techniques for Atomic Ensembles}}

\section{Introduction}
Many well known techniques from quantum optics and atomic physics can be used for a coherent manipulation of the quantum states of light and matter.
These include electromagnetically induced transparency (EIT) and stimulated Raman interaction. Both techniques are associated with quantum interfaces, that is, interactions between atomic vapours and light, and can be employed to affect the behaviour of an atomic medium and also the propagation of optical pulses inside a medium. Under certain conditions, EIT and resulting propagation phenomena, such as the reduced group velocity of the optical pulse, may enable the storage of single-photon pulses inside an atomic medium, and therefore, implement an atomic-based quantum memory. Another technique that offers a remarkable control over a collective behaviour of an atomic vapour is the so-called dipole blockade mechanism.

This chapter is organised as follows. In Sec.~\ref{sec:EIT}, we review in some detail a well known techniques for coherent manipulation of atomic ensembles, namely electromagnetically induced transparency and stimulated Raman interaction. In the same section, we review the concept of an atomic medium as single-photon quantum memory. In Sec.~\ref{sec:rydberg}, we introduce a notion of the Rydberg state and the dipole blockade mechanism, which also may be used to induce coherent behaviour of a macroscopic atomic medium.

\section{Electromagnetically induced transparency} \label{sec:EIT}
One of the most important and interesting phenomena in quantum optics is electromagnetically induced transparency, a term coined by Harris \textit{et al.} in 1990 \cite{harris,harris_phys}. Its importance stems from the range of new potential applications it promises for non-linear optics and quantum information applications. EIT is a phenomenon resulting from the modification of the optical properties of an atomic medium, i.e., an atomic ensemble, driven by an optical laser pulse \cite{fleisch,lukin,munro}. The laser field induces coherent behaviour of an atomic medium and leads  to the vanishing absorption and rapidly varying refraction of a resonant signal field (Fig.~\ref{abs_ref}) \cite{arimondo,boller}. The prominent optical properties of the atomic medium are determined by the first-order linear susceptibility $\chi^{(1)}$. The imaginary part of the susceptibility Im[$\chi^{(1)}$] represents the absorption of the optical field by a medium and the real part Re[$\chi^{(1)}$] represents the refractive index $n$.

The optical properties of any atomic medium are mostly determined by its level structure \cite{fleisch}. The behaviour of a two-level atomic medium in  the presence of a resonant optical pulse seems rather straightforward. The laser pulse induces Rabi oscillations, i.e., atomic population is transferred between two levels in a coherent fashion. The addition of a third level to the level structure of atoms dramatically changes this picture. This slight change leads to a number of new and non-intuitive phenomena, such as appearance of dark state polaritons and EIT itself. The phenomenon of EIT is based on quantum interference in the amplitudes of excitation pathways, which results in destructive interference of the imaginary part of the linear susceptibility. In other words, at resonance linear response of a medium is canceled and the atomic medium is completely transparent to the signal field. The idea of interference between different excitation channels was first introduced by Fano \cite{fleisch}. Apart from the transparency window, a number of new possibilities emerge such as opportunity to ``stop" a light pulse inside a medium.

\begin{figure}[t]
\begin{center}
\includegraphics[scale=0.85]{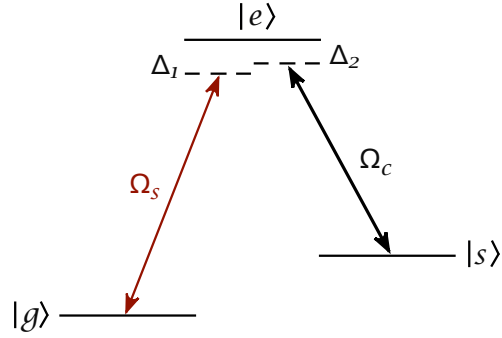}
\end{center}
\caption{The three-level $\Lambda$-type scheme for EIT. $|g \rangle$ and $|s \rangle$ are a lower, long-lived energy states and $|e \rangle$ is an excited state. $\Omega_{s}$ is the Rabi frequency of the signal field detuned from the atomic transition by $\Delta_{1}=\omega_{eg}-\omega_{s}$. $\Omega_{c}$ is the Rabi frequency of the control field detuned from the atomic transition by $\Delta_{2}=\omega_{es}-\omega_{c}$. \label{EIT}}
\end{figure}

To understand the essence of EIT, let us consider an ensemble of atoms with a $\Lambda$-type three-level structure driven by two optical fields. Each atom in an ensemble has  a pair of lower, long-lived energy states $|g \rangle$ and $|s \rangle$. These states can be realised by the electronic ground state of alkali atoms and the transition between them is always dipole-forbidden. A state $|g \rangle$ is coupled to an excited state $|e \rangle$ through the signal optical field. A second strong control field is applied to the transition between states $|s \rangle$ and $|e \rangle$ (Fig.~\ref{EIT}). In this setting, the only way to absorption is by means of the $|e \rangle$ level.
The EIT understood as a lack of the absorption emerges by means of destructive quantum interference between different absorption pathways - the direct one $|g \rangle$-$|e \rangle$ and the indirect pathways such as  $|g \rangle$-$|e \rangle$-$|s \rangle$-$|e \rangle$ \cite{fleisch}. If the control field is much stronger than a signal field and both are detuned by the same amount, amplitudes of these different pathways have the same magnitude but the opposite sign and cancel each other \cite{fleisch}. In this picture atoms are said to be in a so-called dark superposition of the states $|g \rangle$ and $|s \rangle$, which leads to vanishing light absorption. Let us examine the Hamiltonian of the atomic $\Lambda$-type three-level system driven by a pair of near-resonant optical fields. In terms of the Hamiltonian $\hat{H}$, the system can be described as a sum of the free evolution atom Hamiltonian $\hat{H}_{0}$ and the interaction Hamiltonian $\hat{H}_{int}$ \cite{fleisch}. Within the dipole approximation and in the rotating wave approximation, the interaction Hamiltonian $\hat{H}_{int}$ is given by
\begin{equation}
\hat{H}_{int} = -\frac{\hbar}{2} \left[ \begin{array}{ccc}
0 & 0 & \Omega_{s} \\
0 & -2(\Delta_{1}-\Delta_{2}) & \Omega_{c} \\
\Omega_{s} & \Omega_{c} & -2 \Delta_{1} \end{array} \right]\, ,
\end{equation}
where $\Omega_{s}$ is the Rabi frequency of the signal field with frequency $\omega_{s}$ detuned from the corresponding atomic transition by $\Delta_{1}=\omega_{eg}-\omega_{s}$ and $\Omega_{c}$ is the Rabi frequency of the control field with frequency $\omega_{c}$ detuned from the corresponding atomic transition by $\Delta_{2}=\omega_{es}-\omega_{c}$ \cite{fleisch}. The dynamics of the system as a whole are captured by the Hamiltonian $\hat{H}$. For two-photon resonance ($\Delta_{1}=\Delta_{2}=\Delta$), the Hamiltonian $\hat{H}$ has a set of three eigenstates. In terms of the bare atom states $|g \rangle$, $|s \rangle$ and $|e \rangle$, one of the eigenstates has the form $|\psi (\theta) \rangle = \mbox{cos}\theta |g \rangle - \mbox{sin}\theta |s \rangle$, where $\theta$ is the so-called mixing angle given by $\tan \theta = \Omega_{s}/\Omega_{c}$ \cite{fleisch,fleisch_dark}. Under a two-photon resonance, the $| \psi \rangle$ is a stationary state. The state $| \psi \rangle$ is called a dark state because it has no contribution from $|e \rangle$, hence there is no possibility of absorption. Consequently, an opaque, optically thick atomic medium is completely transparent to the signal field in the presence of a strong control field. In general, appearance of the transparency is independent of the detuning $\Delta$ of the signal optical field \cite{fleisch}. Naturally, the ideal transparency occurs at the exact resonance. However, the increased control field strength can circumvent the limitations that are imposed by the resonance condition and even away from the resonance transparency can be observed. The reader should note an interesting feature of the dark state: $| \psi \rangle$ depends on the mixing angle $\theta$. This opens a route to extraordinary possibilities and applications.

\subsection{Stimulated Raman adiabatic passage}
The state of a system described above can be easily manipulated with an appropriate change of Rabi frequencies of the signal and control optical fields \cite{fleisch}. An adiabatic evolution known as stimulated Raman adiabatic passage (STIRAP) can be applied to the system to prepare it in a dark state $| \psi \rangle$. The STIRAP technique is governed by the interaction Hamiltonian $\hat{H}_{int}$ given above with $\Delta_{1}=\Delta_{2}=\Delta$. The adiabatic passage starts when $\Omega_{s} \ll \Omega_{c}$ and the system is in the ground state $| \psi \rangle = |g \rangle$. Then $\Omega_{s}$ is adiabatically increased and  $\Omega_{c}$ is adiabatically decreased up to the point when $\sin \theta = 1 \ (\cos \theta = 0)$ and the dark state $| \psi \rangle = -|s \rangle$. Consequently, by choosing appropriate Rabi frequency for both optical fields, it is possible to transfer a whole atomic population to a maximally coherent dark state $|\psi \rangle = \frac{1}{\sqrt{2}} (|g \rangle - |s \rangle)$ \cite{fleisch}. More importantly, the STIRAP technique is immune to spontaneous emission losses since an excited state $| e \rangle$ is never populated, and therefore the number of photons in the optical field is conserved. The STIRAP technique allows us to prepare the system in one of the bare states ($|g \rangle $ or $|s \rangle$) and in any intermediate superposition. Hence, the STIRAP procedure is a widely used technique for quantum state preparation in atomic ensembles.

\subsection{The propagation phenomena}
\begin{figure}[t]
\begin{center}
\includegraphics[scale=0.67]{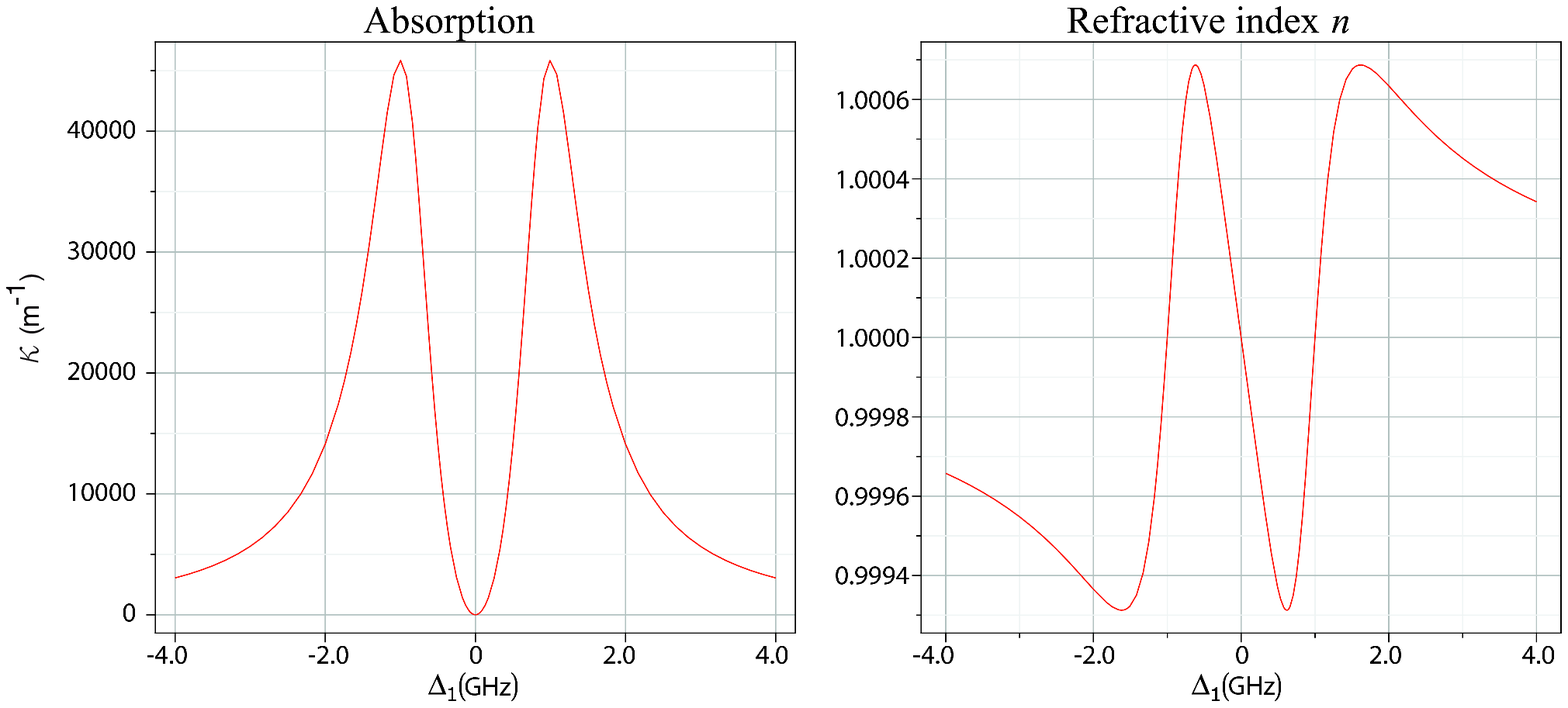}
\end{center}
\caption{Absorption coefficient $\kappa$ (Im[$\chi^{(1)}$]) and refractive index $n$ (Re[$\chi^{(1)}$]) of the optical signal field interacting with an atomic medium in the presence of strong control field. $\Delta_{1}$ is the detuning between the signal field and the atomic transition. The strong control field is on resonance with the appropriate atomic transition, i.e., $\Delta_{2} = 0$. The above figures were prepared with a help of chapter on atomic ensembles in quantum information processing in Ref.~\cite{kok_book}. \label{abs_ref}}
\end{figure}

The EIT technique modifies not only the optical properties of an atomic medium, but the propagation of optical pulses inside a medium is affected as well. These special propagation effects are the source for a variety of applications. First of all, the group velocity of a signal field, i.e., the velocity of the envelope of a wave packet, is changed \cite{harris_PRA}. Under EIT conditions, the group velocity, is reduced since the refractive index $n$ is varying rapidly in the neighbourhood of the two-photon resonance as shown in Fig.~\ref{abs_ref} (the derivative of the refractive index with respect to the frequency is positive and large) and
\begin{equation}
v_{gr} = \frac{c}{n + \frac{dn}{d\omega}\omega} = \frac{c}{1 + n_{gr}},
\end{equation}
with $n_{gr} \sim \rho \sigma c/\Omega_{c}^2$ is the group index and $\sigma = 3 \lambda^2/2 \pi$ is the absorption cross section of an atom and $\rho$ is the atom number density \cite{fleisch}.
Moreover, at resonance the refractive index is equal to unity therefore the phase velocity, i.e., the velocity of a phase front, is equal to the speed of light in vacuum $c$. For high atomic densities and low Rabi frequency of the control field, the group velocity can be lowered to very small values. Different groups performed experiments in which slow group velocities were obtained. In some of the experiments an ultra-cold and dense vapour Na atoms were used \cite{liu}, in others a light pulse was stopped in a hot Rb vapour \cite{phillips} or even in solids \cite{turukhin}. The most remarkable result was obtained in an experiment by Hau \emph{et al.} where the optical pulse was slowed to 17 m/s in a Bose-Einstein condensate of Na atoms \cite{hau}. Naturally, all these experiments suffer from low transfer and storage efficiency due to decoherence effects that are intrinsic to the atomic system. All challenges concerning the atomic vapours that are encountered by experimentalists were described in Chapter~\ref{Chapter1}. One may ask if it is possible to fully stop the optical pulse in the medium, i.e., ``freeze it". Unfortunately, the decreasing group velocity leads to the decreasing transparency window which at some point vanishes and absorption of the signal pulse occurs again. To overcome this limitation one may use a non-stationary, time-dependent control field which dynamically narrows the frequency spectrum of a signal pulse proportional to the group velocity \cite{fleisch}. In other words, the group velocity has to be reduced adiabatically and this allows for the frequency spectrum of the signal pulse to reside within the transparency window. As one would expect, the trapping of the signal pulse, i.e., gradual reduction of the control field intensity, should commence when the entire pulse is within the medium which requires $T_{signal}<L/v_{gr}$ to avoid leakage of the front edge of the signal pulse \cite{lukin}. This again requires an optically dense atomic medium. The fact that one can slow down and confine an entire optical pulse in atomic ensemble for some time may seem unheard-of. Although this effect is much more comprehensible when viewed from the point of view of an atomic medium. During the slowdown of an optical pulse, many additional and interesting effects happen. The reduced velocity introduces a time delay of the light pulse in an atomic medium $\tau_{d} = n_{gr}L/c \sim \rho \sigma L$ and a spatial compression of the signal pulse in the propagation direction. The longer propagation time may be very advantageous in the case of non-linear medium enhancing the non-linear effects. The time delay is proportional to the optical depth of a medium $d = \rho \sigma L$, and therefore a substantial time delay requires an optically thick medium. The spatial compression is associated with different propagation velocities inside and outside the medium. The front edge of a pulse propagates in the medium with a different velocity than its back edge, which propagates outside of the medium with the velocity $c$. This gives rise to the spatial compression by a ratio of the group velocity to the speed of light outside the atomic ensemble \cite{fleisch,lukin}. The spatial compression means that the part of photons from the signal pulse is temporarily stored in the medium in the form of excitations. It is important to point out that no energy carried by photons is stored in the medium only the quantum state of light and the excess energy is transferred to the control field \cite{lukin}. This process resembles stimulated Raman passage. When the optical pulse enters the medium the total number of photons is reduced and the state of atomic system is adiabatically changed to a superposition between the bare states $|g \rangle $ or $|s \rangle$. When the pulse starts to leave the medium this process is reversed. The atomic excitations are turned back to the signal photons with the help of the control field and the state of the system comes back to the bare state $|g \rangle $. Naturally, this adiabatic process depends on the strength of the control field.

All these effects associated with slow light propagation can also be analysed from the point of view of the atomic medium. Under these conditions, a system consisting of an atomic medium driven by optical fields can be described by introducing a new quantum field
$\hat{\Psi}(z,t)$ that is a coherent mixture of electromagnetic and atomic excitations ($|g \rangle $-$|s \rangle$ excitation) \cite{fleisch,fleisch_dark}. The field $\hat{\Psi}(z,t)$ has the form
\begin{eqnarray}
\hat{\Psi}(z,t) = \mbox{cos}\theta \hat{E}(z,t) - \mbox{sin}\theta \sqrt{N} \hat{S}(z,t)\, ,  \\
\mbox{cos}\theta = \frac{\Omega_{c}}{\sqrt{\Omega_{c}^2+\Omega_{s}^2}}\, , \ \ \ \mbox{sin}\theta = \frac{\Omega_{s}}{\sqrt{\Omega_{c}^2+\Omega_{s}^2}}\, , \nonumber
\end{eqnarray}
where $\hat{E}(z,t) = \sum_{k} \hat{a}_{k}(t) e^{ikz}$ is the electric field operator of the signal field consisted of the free-space modes with bosonic operators $\hat{a}_{k}$ and wave vectors $k$, $\hat{S}(z,t) = 1/\sqrt{N} \sum_{i=1}^{N} |g_{i}\rangle \langle s_{i}| e^{-i\omega_{gs}t}$ corresponds to the atomic wave, where $N$ is the number of atoms in the medium and $\omega_{gs}$ is the frequency between long-lived levels $|g \rangle$ and $|s \rangle$. The excitation of the field is called a polariton. The field $\hat{\Psi}(z,t)$ obeys the following equation of motion:
\begin{equation}
\left[\frac{\partial}{\partial t} + c \cos^{2} \theta \frac{\partial}{\partial z}\right] \hat{\Psi}(z,t)=0\, ,
\end{equation}
and propagates with group velocity $v_{gr} = c \ \mbox{cos}^2\theta$. By gradually changing the intensity of the control field one can modify the properties of the polariton from electromagnetic $\hat{E}(z,t)$, with propagation velocity close to the speed of light $c$,
to purely atomic $\hat{S}(z,t)$ with propagation velocity close to zero. It is important to stress that for low group velocities not all but almost all photons from the optical pulse are transferred to the atomic medium. The character of the polariton depends on the intensity of the control field and the density of the atomic medium. In other words, when the control field strength is adiabatically lowered, the signal field is transferred to the atomic medium and propagates as an atomic wave, therefore in some sense the signal pulse is ``stopped". After some time this ``write" process can be reversed. When an intensity of the control field is increased, the signal pulse is retrieved from the atomic medium. Consequently, under EIT conditions the atomic medium acts as a quantum memory capable of slowing down, storing  and releasing optical laser pulses or even single photon wave packets with high fidelity. The EIT enables to reverse the storing procedure and retrieve written information. Most importantly, since the transfer and retrieval of the light field is an adiabatic and coherent process, all properties of the light pulse are conserved at all times.
The applications of electromagnetically induced transparency, i.e., the stopping of light pulses by means of stimulated Raman adiabatic passage, for quantum information processing seem natural. This technique gives the capability for coherent transfer of quantum states between optical light fields and matter system such as atomic vapours. Therefore, EIT can be used for preparation of specific non-classical and entangled states of the atomic medium \cite{lukin_PRL}.

\subsection{Atomic medium as single-photon quantum memory}
EIT and all associated phenomena can also be observed for single-photon wave packets. Hence, one can imagine a single-photon coherently mapped onto an atomic medium \cite{fleisch_darkPRA,mewes}. The subsequent state of a medium is described by the symmetric and collective atomic state given by
\begin{equation}
|s\rangle = \frac{1}{\sqrt{N}} \sum_{j=1}^{N} |g_{1},g_{2}, \ldots ,s_{j}, \ldots ,g_{N}\rangle,
\end{equation}
with high fidelity \cite{fleisch_darkPRA}. The state $|s\rangle$ is a coherent W state. As the following discussion indicates, the optically dense EIT medium may serve as a good quantum memory that can be employed as a node in a quantum network or quantum repeater \cite{kuzmich,wal}. Typically, we make the assumption that all atoms in a medium have the same probability of absorbing a single photon. This approximation implies a very attractive feature of the collective state: it is impossible to learn which atom really absorbed a photon \cite{lukin}. Therefore, within small error the loss of one atom has no effect on the fidelity of resulting state. This remarkable property of collective states make them very robust with respect to decoherence and losses \cite{lukin}. In general, an EIT based quantum memory is capable of storing not only single photon states but any superposition of photonic states, e.g., entangled states.
Although EIT based applications for quantum information science are very promising, one has to remember many potential limitations associated with atomic quantum memories. In many experimental trials, it has been proved that for high transfer-storage-retrieval efficiency, one has to use an atomic ensemble with a very large optical depth, i.e., high density of atoms or large sample size \cite{novikova}. Consequently, higher density will introduce stronger collisional and dephasing effects, which are one of the most severe decoherence causes \cite{laurat,felinto}. Collisions during write and read processes may substantially limit the fidelity of the quantum memory \cite{manz}. One way of dealing with the decoherence processes such as collisions and diffusion is exploitation of a cold atomic vapours in strong optical traps. Other ways are specific, entangled states of light as input fields and optimal input pulse shapes \cite{novikova,pugatch,laurat}.

In spite of these difficulties, recent advances in quantum memories have been remarkable. In recent experiments, truly quantum optical memories that preserve quantum features of light such as entanglement have been demonstrated \cite{chaneliere,eisaman,choi}. In these experiments, a quantum memory was fed by a non-classical field of light originating from an atomic ensemble. An atomic ensemble serving as a source is prepared by a weak laser pulse so that only one of the atoms is in the excited state. This is the so-called weak excitation regime. The excited atom relaxes to one of the lower energy levels, emitting a single photon, the so-called Stokes photon, that carries less energy than the absorbed photon. Next, the strong retrieve laser pulse brings the atomic medium back to its ground state and atomic excitation is converted into an anti-Stokes photon \cite{hammerer,eisaman,choi}. The described technique proved to be extremely useful not only as a way of generating single-photon pulses but in many different applications, such as the quantum repeater protocol, i.e., the DLCZ protocol described in detail in Chapter~\ref{Chapter5} \cite{duan}. Subsequently, the non-classical character of the source was verified with a help of the correlation function. Conditioned on the detection of one Stokes photon, after the retrieve pulse one observes either no anti-Stokes photons or exactly one anti-Stokes photon as the output of the source. The single-photon pulses are then stored and released by means of EIT techniques, i.e., the control field is turned off and after a delay time reapplied again. In an experiment by Choi \textit{et al.} the single-photon pulses are stored for 1 $\mu$s in cold Rb atoms trapped in a magneto-optical trap (MOT) with overall transfer-storage-retrieval efficiency of 17\%. In other experiments by Chaneli\`{e}re \textit{et al.} and Eisaman \textit{et al.} the overall efficiency was close to 6\%. The experiment by Choi \textit{et al.} stands out because it exploits the entangled state of a photonic qubit. A single-photon from the atomic source is split on a beam splitter so that the two components of the input state of the form $| \psi_{in}\rangle = \frac{1}{\sqrt{2}}(|0 \rangle_{A}|1\rangle_{B} + e^{i\varphi}|1\rangle_{A}|0\rangle_{B})$ are directed into two atomic vapours \cite{hammerer,choi}. The EIT procedure is repeated now for two atomic ensembles. Subsequent tomography of the retrieved state verifies that the quantum memory conserved entanglement. The overall efficiency of transfer-storage-retrieval of entanglement is 20\%. The described experiments are proof-of-principle experiments rather than reliable implementations of quantum memories. Still they demonstrate significant progress. Naturally, for applications in a distributed quantum network the overall efficiency has to be much higher. The difficulty of relatively low efficiency of the transfer-storage-retrieval process can be circumvented by the exploitation of an atomic medium with increased optical depth $d$ and optimising the shape of the control field with respect to the signal field. The efficiency of the optical quantum memory depends mostly on the optical depth $d$. However, the retrieval efficiency can be sharply increased if one uses the control field that stores the given signal field in an optimal way \cite{hammerer}. In a recent experiment, Novikova \textit{et al.} used an iterative optimisation procedure that maximised the storage and retrieval efficiency \cite{novikova}. First an initial optical pulse was stored and retrieved. Then a time-reversed profile of the retrieved pulse was used as the next input for the atomic memory. The whole procedure was repeated and converged very quickly to the optimal input pulse profile. The overall efficiency of the transfer-storage-retrieval process for optimal input field was close to 45\%. Moreover this experiment was performed for warm $^{87}$Rb vapour with relatively low optical depth $d \simeq 9$. The exploitation of cold atomic vapours with increased optical depths should boost the overall light-storage efficiency. The experiment of Novikova \textit{et al.} confirmed again that the optical depth is the key figure of merit for the efficiency of quantum memories.

Finally, we would like to mention a quite interesting application of EIT, namely the possibility of building atomic-vapour-based high efficiency photon detectors with an estimated detection efficiency of $\eta_{D} \approx 99.8 \% $ \cite{james}. The single photons stored inside a medium can be counted by means of resonant fluorescence. Moreover, if the detection of light stored in an atomic ensemble does not alter the state of a medium, this kind of detector could then realise a quantum non-demolition measurement of the photon-number operator since one can retrieve the photons stored in the atomic medium \cite{imam}.

Apart from EIT and Raman interactions, one may induce coherent behaviour of a macroscopic atomic medium with the help of Rydberg atoms. In the following section we introduce a notion of the Rydberg state and the dipole blockade mechanism.

\section{Rydberg state and dipole blockade mechanism} \label{sec:rydberg}
Although the concept of a Rydberg atom has been known for more than 100 years, physicists are able to study them in laboratory only since the nineteen seventies. Despite this relatively short period of experimental studies, we know already that Rydberg atoms allow for a number of interesting applications. The Rydberg state is a state of an alkali atom characterised by a high principal quantum number $n$ \cite{gallagher}. Rydberg atoms possess a number of remarkable properties. To begin with, Rydberg atoms are very large compared to normal atoms. The radius of a Rydberg atom scales as $n^2 a_{0}$, where $a_{0}$ is the Bohr radius, and the binding energy of a Rydberg state is given by
\begin{equation}
E = -\frac{R}{(n-\delta)^2}=-\frac{R}{n^{*^{2}}}\, ,
\end{equation}
where $R$ is the Rydberg constant, $n^{*}$ is the effective quantum number, and $\delta$ is the quantum defect which corrects for the deviation
from the hydrogen atom \cite{li}. This implies that the valence electron is very weakly bound to the nucleus. Moreover, the Rydberg states have an incredibly long lifetime, which scales as $\tau_{0} n^5$ where $\tau_{0}$ is the typical lower level lifetime of around $\sim$10 ns. Hence, Rydberg states possess lifetimes of the order of ms and even longer.

Because of the very weak binding energy, Rydberg atoms are extremely sensitive to external electric fields. The Rydberg energy levels are easily perturbed by modest electric fields. Higher electric fields can even ionise Rydberg atoms. In fact, the ionisation is commonly used as one of the detection methods. This sensitivity to electric fields is the source of a phenomenon called the \emph{dipole blockade mechanism}. Atoms in Rydberg states have large dimensions and large dipole moments, resulting in a strong dipole--dipole interaction \cite{gerry}. Under certain circumstances the effect of strong dipole--dipole interaction can be observed in the laboratory. The dipole blockade mechanism was observed experimentally in small clouds of alkali atoms, such as Rubidium in a vapour cell \cite{johnson,van}. This mechanism prevents populating states of an atomic ensembles with two or more atoms excited to the Rydberg level \cite{lukin}. A single atom in a micron-sized atomic ensemble excited to a Rydberg state with a narrowband laser can inhibit excitation of the other atoms in the sample if the long range Rydberg-Rydberg interactions are much larger than a linewidth of the Rydberg state.

\begin{figure}[t]
\begin{center}
\includegraphics[scale=0.85]{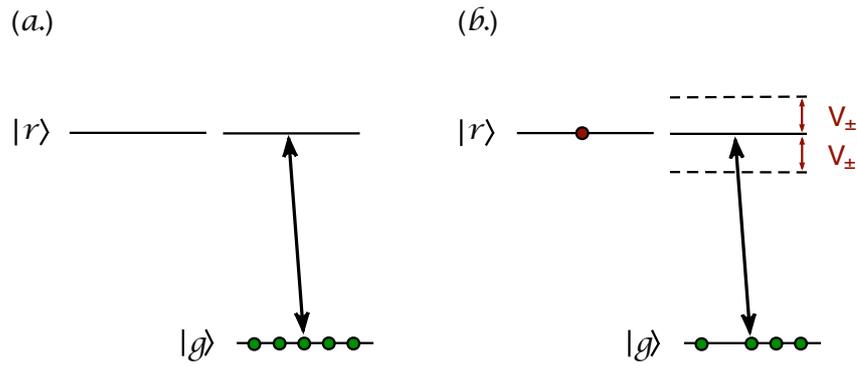}
\end{center}
\caption{Diagram representing the dipole blockade mechanism. The ground state $|g \rangle$ and Rydberg state $|r \rangle$ are coupled by means of a narrowband laser. (a.) After an appropriate interaction time one of the atoms in atomic medium is excited to the Rydberg state $|r \rangle$. (b.) Presence of a single atom in the Rydberg state $|r \rangle$ shifts energy levels of all other atoms located within the long range Rydberg-Rydberg interactions distance and blocks any further excitations. \label{dipole}}
\end{figure}

The physics of the dipole blockade mechanism is presented in Fig.~\ref{dipole}. An optical pulse resonant with a transition to the Rydberg state  $|r \rangle$ will create a Rydberg atom with a very large dipole moment (Fig.~\ref{dipole} (a.)). For sufficiently short separations, the long range Rydberg-Rydberg interactions (dipole interactions) between the Rydberg atom and the other atoms will cause a shift in the Rydberg transition energy of the other atoms. Therefore, the optical pulse becomes off-resonant with the other atoms, and the ensemble is transparent to the pulse. Under dipole blockade conditions, the mesoscopic vapour behaves as one superatom with a two-level structure. A single excitation is coherently shared by all atoms in a sample and one is able to observe Rabi oscillations. Naturally, effectiveness of the blockade depends on an average strength of the interaction between atoms in the ensemble.

The long range Rydberg-Rydberg interactions have different types depending on the separation between atoms. The usual van der Waals interaction of type $C_{6}/R^{6}$ can be enhanced by a static electric field or F\"{o}rster processes to the $C_{3}/R^{3}$ long range interaction. Here, the $C$'s coefficients depend on atomic energy level structure \cite{walkerbloc}.\\
In the absence of an external electric field, the Rydberg-Rydberg interactions are of the van der Waals type $C_{6}/R^{6}$ \cite{walkerbloc,walker}. In a static electric field, a Rydberg atom possesses a large permanent dipole moment $p$ oriented in space along the applied electric field, which scales as $\sim q a_{0} n^2$ with $q$ the electron charge, which leads to a much stronger and longer $C_{3}/R^{3}$ interaction. A pair of Rydberg atoms $i$ and $j$ interact with each other via dipole--dipole potential $V_{dd}$,
\begin{equation}
V_{dd} = \frac{\textbf{p}_{i}\textbf{p}_{j} - 3(\textbf{p}_{i}\cdot\textbf{e}_{ij})(\textbf{p}_{j}\cdot\textbf{e}_{ij})}{4 \pi \epsilon_{0} |\textbf{r}_{i} - \textbf{r}_{j}|^3} = \frac{p^2}{4 \pi \epsilon_{0} R^3}(1 - 3 \cos^2\theta)\, ,
\end{equation}
where $\textbf{e}_{ij}$ is a unit vector along the interatomic direction, $\theta$ is the angle between the interatomic separation $R = |\textbf{R}| = |\textbf{r}_{i} - \textbf{r}_{j}|$ and the electric field
$\textbf{z}$ direction. In general, the interaction between Rydberg atoms can be quite strong. However, for some angles $V_{dd}$ vanishes, which is undesirable for dipole blockade purpose \cite{walker}.
Fortunately, there is another method to induce a strong, isotropic interaction between Rydberg atoms, comparable to $V_{dd}$ through the F\"{o}rster process (in practice, however, the shape and dimensionality of the atomic ensemble may introduce the angular dependence and therefore the F\"{o}rster interaction may no longer be isotropic \cite{saffman10}). The resonant collisional process (F\"{o}rster process) transfers energy between two atoms through the dipole--dipole interaction with strength
$\sim \rho_{1} \rho_{2}/R^3$, where $\rho_{1}$ and $\rho_{2}$ are the dipole matrix elements between initial and final energy states of the
interacting atoms \cite{stoneman}. Therefore, the usual van der Waals interaction can be resonantly enhanced by F\"{o}rster processes such as $nl + nl \rightarrow n'l' + n''l''$ when the $nl + nl$ states are degenerated in energy with the $n'l' + n''l''$ states. The F\"{o}rster process induces an interaction potential of the form
\begin{equation}
V_{\pm}(R) = \frac{\delta}{2} \pm \sqrt{\frac{4 U_{3}(R)^2}{3} + \frac{\delta^2}{4}}\, ,
\end{equation}
where
\begin{equation}
U_{3}(R) = q^2 \langle nl || r || n'l' \rangle \langle nl || r || n''l'' \rangle / R^3\, ,
\end{equation}
with $\delta = E(n'l') + E(n''l'') - 2E(nl)$ as the F\"{o}rster energy defect. There is no angular dependence for the potential
$V_{\pm}(R)$ so an interaction is isotropic. For a perfect F\"{o}rster degeneracy ($\delta = 0$), $V_{+}(R)$ would be of similar strength
and range to $V_{dd}$ \cite{walker}. Although at large separations, a non-zero F\"{o}rster energy defect reduces long-range interaction
between the atoms to be van der Waals $C_{6}/R^{6}$ type. However, if the F\"{o}rster energy defects are smaller compared to the fine-structure splitting then a strong $C_{3}/R^{3}$ interaction can even occur at longer range.

Although F\"{o}rster processes are very promising as a method to induce very long-range $C_{3}/R^{3}$ interactions, there are some selection
rules that need to be fulfilled for obtaining high fidelity dipole blockade. Only for $l' = l'' = l+1$ there are no so-called F\"{o}rster zero states with $C_{3} = 0$ \cite{walker}. Therefore, the fidelity of the dipole blockade mechanism is highly dependent on the weakest interactions between degenerate Rydberg states and may be reduced under unfortunate circumstances. In the case of the F\"{o}rster zero states, strength of the interaction between Rydberg atoms is not enhanced and reduces to the usual van der Waals long-range type. Therefore, a strong dipole blockade requires tuning of the resonances by means of an electric field \cite{walker}. The other possibility for attaining strong dipole blockade is to rely on the van der Waals interaction which at smaller distances, less than 5 $\mu$m, is large enough to mix the fine-structure levels together, so the interaction is of the $V_{dd}$ type \cite{walkerbloc}.

As one would expect, the dipole blockade mechanism fuelled a number of interesting proposals such as a method to entangle large numbers of atoms \cite{lukin}. Fortunately, the exact strength of the dipole blockade in these proposals is not important as long as it is greater than the linewidth of a Rydberg state. Therefore, the atoms can be located at random distances $R$ from each other \cite{walker}. Moreover, with the dipole blockade mechanism at hand, one can avoid a problem of mechanical interactions between atoms, since states with two or more atoms in the Rydberg state are never populated. Therefore, the atoms avoid heating and the internal states of the atoms are decoupled from the atomic motion \cite{lukin}.

The range and quality of the dipole interaction has been studied extensively. In papers by Walker and Saffman the primary errors with respect to the idealized blockade process were analysed \cite{van,walkerbloc}. Naturally, the two most common errors are the occurrence of doubly-excited Rydberg states and singly-excited states outside the desired two-level system. In the case of Rubidium atoms with principal quantum number $n=70$, the blockade energy shift is approximately 1 MHz. Hence, a strong and reliable blockade is possible for two atoms with separation up to $\sim$10 $\mu$m \cite{walkerbloc}. Moreover, the decoherence associated with spontaneous emission from long-lived Rydberg states can be quite low ($\sim$1 ms). The dipole blockade mechanism can be used to build fast quantum gates, i.e., a two qubit phase gate \cite{jaksch,lukingate,brion}. The long-range dipole--dipole interaction between atoms can be employed to realise a universal phase gate between pairs of single-photon pulses \cite{friedler,mohapatra,petrosyan}. Most importantly, the ideas based on the dipole blockade mechanism are experimentally feasible.

The single quantum sensitivity suggests that the dipole blockade mechanism can be used to create cluster (graph) states:
The blockade mechanism can be used in a heralding type of entangling operations and render them nearly deterministic \cite{zwierz}. In Chapter~\ref{Chapter5}, we introduce nearly deterministic entangling protocol based on the dipole blockade mechanism, first however, we review several schemes for probabilistic entanglement generation between atomic vapours followed by a scheme capable of implementing any single-qubit operation on the qubit defined as collective states of mesoscopic ensemble.

\section{Summary}
In conclusion, we have reviewed in some detail electromagnetically induced transparency, Raman interaction and associated propagation phenomena in atomic va\-pours. The EIT technique allows us to induce coherent behaviour of a macroscopic atomic medium under certain conditions. The reduced group velocity and dark state-polaritons are a remarkable propagation phenomenon associated with propagation of an optical pulse in an atomic medium under EIT conditions that has lead to the concept of quantum memory. In principle, an atomic medium is capable of storing single-photon pulses. Apart from EIT, one can employ Rydberg atoms and the dipole blockade mechanism to induce collective behaviour of atomic vapour. On the basis of the above techniques, probabilistic entanglement generation between atomic vapours is feasible.  


\chapter{Atomic Ensembles in Distributed Quantum Computing}
\label{Chapter5}
\lhead{\textsc{Chapter 5. Atomic Ensembles in DQC}}

\section{Introduction}
Initially, atomic vapours were proposed as fast quantum memories. However, it is also possible to define a qubit (stationary qubit or quantum processor) in an atomic ensemble, and the question remains how to implement the entangling operations between the qubits
that enable universal distributed quantum computation. One may choose to create a large network of spatially separated quantum processors and connect them with quantum communication channels. However, it suffices to create a inherently distributed, large
entangled multi-qubit resource ---the graph state--- after which the entire computation proceeds via single-qubit measurements \cite{rauss,hein}. Graph states are large arrays of isolated qubits connected (entangled) via CZ operations. They are a scalable resource and can be built up with probabilistic entangling operations with $p_{success} > 0$ \cite{kok}. When the success probability of entangling operation is low, a very large overhead in optical elements is required. Moreover, finite coherence times of the qubits limit practical use of the graph states. Hence, it is extremely important to build them up in an efficient way.

This chapter is organised as follows. In Sec.~\ref{sec:entinae}, we review several schemes for probabilistic entanglement generation between atomic vapours such as DLCZ protocol and double-heralding protocol. In Sec.~\ref{sec:aeassq}, we review the concept of an atomic ensemble as single qubit system and analyse in detail a scheme for single-qubit operations in atomic ensembles. In Sec.~\ref{sec:newprotocol}, we give a description of a new entangling operation and consider its usefulness for generation of the GHZ and cluster states. In Sec.~\ref{sec:errors}, we consider all major errors and decoherence mechanisms that enter the entangling procedure and propose several experimental implementations.

\section{Entanglement in atomic ensembles} \label{sec:entinae}
In this section, we are going to focus on the probabilistic entanglement generation between two distant qubits implemented as atomic ensembles. We are especially interested in the heralded entanglement generation, i.e., detection of an object such as a single photon heralds the creation of entanglement between two distant macroscopic objects such as atomic ensembles \cite{vanEnk}. The heralded protocols work with some success probability $p_{success}$, which in principle depend on structure of the protocol, the efficiency of detection method and the physical implementation. Therefore, for efficient entanglement generation one has two choices: either prepare many copies of physical systems or repeat entanglement procedure sufficient number of times. If $p_{success}$ is small, it takes on average $1/p_{success}$ copies or repetitions to create entanglement between two distant ensembles. The probabilistic nature of the heralded entanglement procedures imposes some limitations on its practical use in quantum computation but not in quantum communication.

\subsection{The DLCZ protocol}
\begin{figure}[t]
\begin{center}
\includegraphics[scale=0.75]{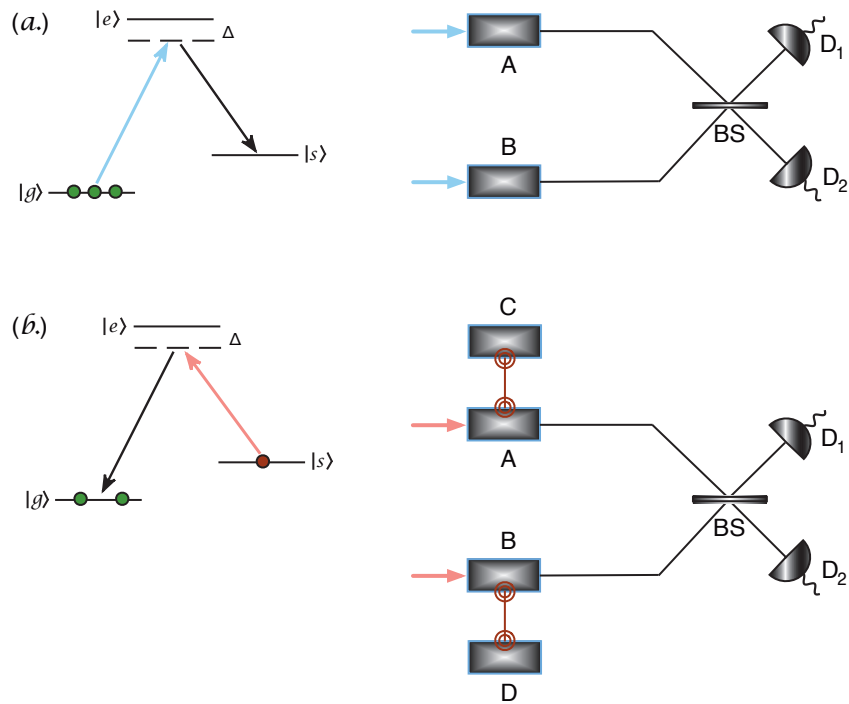}
\end{center}
\caption{(a.) The relevant three-level $\Lambda$-type structure and diagram of the DLCZ protocol. $|g \rangle$ and $|s \rangle$ are a lower, long-lived energy states and $|e \rangle$ is an excited state. The blue line represents a weak, write laser pulse. Conditionally on the detector click entanglement is created between A-B ensembles. (b.) Diagram of the entanglement swapping procedure. The red line represents a read-out laser pulse. Conditionally on the detector click entanglement is extended to C-D ensembles. \label{DLCZ}}
\end{figure}

One of the well known entangling protocols is the DLCZ protocol. It was devised by Duan \textit{et al.} as a quantum repeater protocol. Quantum repeaters are essential for long-distance quantum communication. The DLCZ protocol is realised on two macroscopic atomic ensembles, a balanced beam splitter and two single-photon photo detectors. The relevant atomic level structure is shown in Fig.~\ref{DLCZ}. The $N$ atoms in an ensemble have two lower, long-lived energy states $|g\rangle$, $|s \rangle$ (Zeeman sublevels of the ground state), and an excited state $|e \rangle$. The protocol begins with all atoms prepared in the ground state $|g\rangle$. Then a weak laser pulse that addresses off-resonantly the $|g \rangle$-$|e \rangle$ transition transfers, preferably a single atom to the state $|s \rangle$ and simultaneously produces a single, forward-scattered Stokes photon. This process resembles stimulated Raman passage (STIRAP) and the whole state of the ensemble-light system is given by
\begin{equation}
|\phi\rangle_{EL} = |0\rangle_{E} |0\rangle_{L} + \sqrt{p_{e}} |S\rangle_{E} |1\rangle_{L} + \mathcal{O}(p_{e})\, ,
\end{equation}
where $|0\rangle_{E}$ is the ensemble collective ground state given by $|0\rangle_{E} = |g_{1},g_{2}, \ldots ,g_{N}\rangle$, $|0\rangle_{L}$ is the vacuum state of light, $|S\rangle_{E}$ is the collective state of the ensemble given by $|S\rangle_{E} = \frac{1}{\sqrt{N}} \sum_{j=1}^{N} |g_{1},g_{2}, \ldots ,s_{j}, \ldots ,g_{N}\rangle$, $|1\rangle_{L}$ represents the single forward-scattered Stokes photon state and $p_{e}$ is the excitation probability that, because of the weak excitation laser pulse, is small. The above state represents the very heart of the DLCZ protocol. The STIRAP procedure can be applied simultaneously to two ensembles. In the result, a single forward-scatted Stokes photon is produced in one of the ensembles. It is not possible to learn which ensemble is the source of a Stokes photon. The light modes from both ensembles are then combined on the balanced beam splitter (BS) to erase which-path information (see Fig.~\ref{DLCZ} (a.)). Following the detector click on one of the photo detectors (D$_{1}$,D$_{2}$) the maximally entangled state of two ensembles $\Psi_{\pm} = \frac{1}{\sqrt{2}} (|S\rangle_{A} |0\rangle_{B} \pm |0\rangle_{A} |S\rangle_{B}) $ is created. This scheme works with the success probability given by $p_{success} = p_{e}$. Hence, the entangled state will be generated on average after $1/p_{success}$ procedure repetitions.
As mentioned above, the DLCZ protocol is in fact a quantum repeater protocol. The DLCZ protocol enables the entanglement of two atomic ensembles and then through entanglement swapping, the connection can be established between distant sites \cite{duan}. In other words, the DLCZ protocol enables the distribution of entanglement between distributed quantum network nodes. If one prepares two pairs of atomic ensembles (A-C and B-D) in the maximally entangled state, then by means of a read-out laser pulse, applied to the $|s \rangle$-$|e \rangle$ transition, stored atomic excitation of a single ensemble in each pair can be converted into light modes (see Fig.~\ref{DLCZ} (b.)). These light modes are again combined on the balanced beam splitter to erase which-path information and, conditionally on the detector click, entanglement is extended to the more distant C-D ensembles. This procedure is called entanglement swapping, and can in principle be applied many times creating a communication channel between distant nodes.

The interesting feature of the DLCZ protocol is the fact that it has built-in entanglement purification. The fidelity imperfection of the protocol is proportional to $p_{e}$ and can be lowered close to zero for small excitation probabilities \cite{duan}. The DLCZ protocol is scalable and highly efficient in comparison with direct communication methods. Apart from the communication applications, the DLCZ protocol can be used for quantum teleportation, cryptography and demonstrating the violation of a Bell inequality.

The DLCZ protocol drew a lot of attention from experimental groups all around the world. The first experimental attempts to realise a quantum repeater were limited to the generation of non-classical photon pairs originating from a single atomic ensemble \cite{kuzmich,wal,chou}. In these experiments by means of a write pulse (Raman adiabatic passage) a collective atomic state is created together with a single Stokes photon. After some programmable delay time, the read pulse is applied to the atomic ensemble resulting in a generation of a second (anti-Stokes) photon. The quantum (non-classical) character of  correlations between both photons is confirmed by the violation of a Cauchy-Schwarz inequality \cite{kuzmich,wal,chou}. Although none of the mentioned experiments implemented the DLCZ protocol, techniques used in these experiments are considered a first and a crucial step in the realisation of the protocol \cite{kuzmich}. Shortly after these initial experiments, the full quantum repeater implementations were realised \cite{laurat,choi,chou_nature,chou_science,yuan}. The DLCZ protocol was realised on the atomic ensembles consisting of $\sim10^{5}$ atoms separated by a few meters \cite{chou_nature}. The experiments involve preparation of the collective atomic states and the read out of quantum memories after some delay time. The quality of the entanglement between quantum nodes is given in terms of concurrence $C$ \cite{laurat,choi,chou_nature} or validated by the violation of the Bell inequality \cite{chou_science,yuan}. There are several factors that limit the performance of DLCZ protocol. The main one is low retrieval efficiency varying in range from 30\% to 60\% and decoherence of the collective atomic states \cite{laurat}.

An interesting extension of the quantum repeater protocol to include quantum teleportation was devised by Chen \textit{et al.} They demonstrated teleportation between photonic and atomic qubits \cite{chen}. The quantum state of a single photon was teleported onto an atomic ensemble, stored for up to 8$\mu$s and then converted back to a photonic state. The main advantage of this scheme over other teleportation protocols is the prospect of storing the teleported state and reusing it for further quantum applications. Although this technique makes large-scale communication and distributed quantum computation more feasible, it is not yet useful for practical applications due to many experimental limitations such as short quantum memory lifetime.

\subsection{The double-heralding protocol}
Another protocol for probabilistic entanglement generation between spatially separated quantum nodes is the double-heralding protocol. This protocol is due to Barrett and Kok \cite{kok}. Here, entanglement is established after two consecutive single photon detections, hence the name of the protocol. The double-heralding protocol is based on matter qubits and linear optics. Let us consider two spatially separated matter qubits, e.g., single atoms or atomic ensembles, each having two lower energy levels $|g \rangle$ and $| s \rangle$, and an excited level $| e \rangle$, which is coupled only to the $| g \rangle$ level by means of an optical pulse \cite{kok,koklecture}. If a matter qubit is realised by an atomic ensemble, the above energy levels are represented by collective atomic states. The energy levels $|g \rangle$ and $| s \rangle$ constitute the qubit states. The protocol begins with both matter qubits prepared in the separable state $|\phi\rangle_{LR}=\frac{1}{2}(|s\rangle + |g\rangle)_{L}(|s\rangle + |g\rangle)_{R}$. We apply an optical $\pi$ pulse to each qubit which results in a single photon being emitted when a matter system spontaneously goes back to the $| g \rangle$ level. Following the above manipulations, the total state of the matter qubits and output modes of light is given by
\begin{equation}
|\Phi\rangle=\frac{1}{2} (|ss\rangle |00\rangle + |sg\rangle |01\rangle + |gs\rangle |10\rangle + |gg\rangle |11\rangle)\, ,
\end{equation}
where $|0\rangle$ and $|1\rangle$ denote the vacuum and a single photon state respectively. The modes of light are then combined on the balanced beam splitter (BS) to erase which-path information, which results in the state:
\begin{equation}
|\Phi\rangle=\frac{1}{2} \{|ss\rangle |00\rangle + \frac{1}{\sqrt{2}}[(|sg\rangle + |gs\rangle)|01\rangle + (|sg\rangle -|gs\rangle) |10\rangle + |gg\rangle (|20\rangle + |02\rangle)]\}.
\end{equation}
Following the beam splitter, the light modes are coupled to the regular photo detectors that must have a low dark count rate. Conditional on a single detector click $D_{\pm}$ the state of the matter qubits is given by the following density operator:
\begin{equation}
\rho^{(\pm)} = \frac{1}{2-\eta} |\Psi^{\pm}\rangle\langle \Psi^{\pm}| + \frac{1-\eta}{2-\eta} |gg\rangle\langle gg|\, ,
\end{equation}
where $|\Psi^{\pm}\rangle =  \frac{1}{\sqrt{2}}(|sg\rangle \pm |gs\rangle)$ and $\eta$ is the combined photon collection and detection efficiency \cite{kok}. The above state is a mixed state. To remove the second, separable part of the mixed state $\rho^{(\pm)}$, a bit flip must be applied to both matter qubits. In the case of a matter qubit implemented as atomic ensemble, a bit flip is not a trivial operation. In fact, for a reliable bit flip operation one has to make use of the dipole-blockade mechanism. In the next section, we review the concept of an atomic ensemble as single qubit system and analyse in detail a scheme for single-qubit operations in atomic ensembles. After the bit flip, we repeat the whole procedure. Therefore, after a second measurement event (single detector click in $D_{\pm}$) the total state of two qubits is projected onto the pure maximally entangled state
\begin{equation}
|\Psi^{\pm}\rangle =  \frac{1}{\sqrt{2}}(|sg\rangle \pm |gs\rangle),
\end{equation}
with success probability $p = \eta^2/2$ and unit fidelity. The double-heralding protocol can be used to efficiently create multi-qubit graph states with only moderate overhead in physical resources, which together with the one-way model of computation can be used to implement universal quantum computation \cite{kok,benjamin}. The procedure is a fully scalable scheme for universal quantum computation assuming that the physical implementation allows high-fidelity single qubit operations and measurements.

The double-heralding protocol possesses many attractive features. The scheme is based on a simple level structure and a simple optical network, which imply  rather straightforward  phase stabilisation. Moreover, the protocol works for distributed qubits that facilitate control of decoherence and permit applications in quantum communication such as quantum repeaters. The main disadvantage of the double-heralding protocol is the success probability ($p = \eta^2/2$) depending on the collection of photons, which makes it sensitive to photon loss. This problem was analyzed by Barrett and Kok \cite{kok}. The authors concluded that ``[...] the photon loss does not reduce the fidelity of the entangled states, but merely adds to the overhead cost". The problem of losses was also addressed by the broker-client model devised by Benjamin \textit{et al.} \cite{benjamin_broker}. In the broker-client model two qubits are placed in each node. One of them is used for entanglement generation between nodes and the other one serves for storing of the entanglement when double-heralding procedure succeeds. In this way, influence of an extreme photon loss is suppressed at the cost of a more complicated level scheme and effective graph state generation with small overhead in physical resources is feasible.

From the experimental point of view, one of the main challenges in implementing the double-heralding protocol is the generation of the indistinguishable photons. However, the following experiments prove that this is possible. There is a number of physical systems that may be used to represent a matter qubit such as trapped ions or atoms, NV centers in diamond and Pauli blockade quantum dots \cite{koklecture}. In the experiment by D. L. Moehring \textit{et al.} \cite{moehring} two trapped $^{171}$Yb$^{+}$ ions are separated by one meter. Each of the ions emits a single photon which polarisation is entangled with each ion. The single photons are then combined on the balanced beam splitter and detected by photon-counting photomultiplier tubes (PTMs). In this experiment entanglement is generated for the system with more complex level structure. However, in its essence the mentioned experiment and the double-heralding protocol are analogous. The entanglement between ions was confirmed by a violation of a Bell inequality \cite{matsukevich}. Unfortunately, when all experimental limitations are taken into account, heralded entanglement between ions is established every 8.5 min. This result is consistent with a general observation that for $\eta \ll 1$ the success probability of the generation of a maximally entangled state of even two qubits can be quite low. Therefore, generation of multi-qubit graph states for quantum computation, or quantum communication has to be based on protocols with higher success probabilities.

A deterministic protocol for implementing a universal two-qubit gate between two atoms placed in optical cavities was proposed by Lim \textit{et al.} \cite{lim_lett}. The two-qubit interactions are induced using single-photons, that originate from atom-cavity matter systems, linear optics and photo detectors. The qubits are encoded in two atomic ground states and prepared in an arbitrary state $|\Psi \rangle = \alpha |0 \rangle + \beta |1 \rangle$. Subsequently, an encoding operation is applied to each matter system that transforms the state of each qubit to $|\Psi \rangle = \alpha |0,E \rangle + \beta |1,L \rangle$. In other words, each atom in an optical cavity emits a single photon at an early (\textsc{E}) or a late (\textsc{L}) time. The atom-cavity system acts as ``on-demand" single photon source. Therefore, the encoded state contains both the initial state of an atom-cavity matter system and the state of single photon. This is the so-called time-bin encoding. Consequently, one can prepare two atom-cavity systems and two single photons in the arbitrary encoded state. At this point, the measurement in the appropriate basis (mutually unbiased basis formed by single photon basis states $|E \rangle$ and $|L \rangle$) can realise any universal two-qubit gate \cite{lim}. However, some certain experimental implementations and measurement bases may involve rather complicated linear optical setups. The main strength of this protocol is its repeatability. If the proposed gate fails then the original input state can be recovered by local operations and the whole procedure can be repeated until successful operation is achieved \cite{lim}. On average two repetitions are required to realise the gate operation. This repeat-until-success \textit{modus operandi} leads to the deterministic two-qubit operations. Unfortunately, in the presence of unavoidable photon loss, the above procedure becomes probabilistic \cite{lim}. The successful implementation of the two-qubit gates requires detection of a photon pair in appropriate outputs of optical network and the failure associated with photon emission, collection or detection leaves the matter qubits in an unknown state. Although, the failure of the scheme is heralded and scalable quantum computing is still possible, the overall overhead cost associated with the procedure may be increased significantly \cite{lim}.

Here, we present another entangling protocol that in principle is also deterministic. This protocol employs dipole blockade mechanism between Rydberg atoms. We show how to efficiently create graph states using single photons interacting with atomic ensembles via the dipole blockade mechanism. The protocol requires identical single-photon sources, one atomic medium per physical qubit placed in the arms of a Mach--Zehnder interferometer, and regular photo detectors. We present a general entangling procedure, as well as a procedure that generates $Q$-qubit GHZ states with success probability $p_{success}\sim\eta^{Q/2}$, where $\eta$ is the combined detection and source efficiency. This is significantly more efficient than any known robust probabilistic entangling operation \cite{lim,kok}. The GHZ states are locally equivalent to the graph state and form the basic building block for universal graph states. Our protocol significantly reduces an overhead in optical elements and leads to better quantum computing prospects. However, before giving a detailed description of a new entangling operation, let us review a scheme for implementing single-qubit operations on the qubit defined in atomic ensemble and analyse it in detail.

\section{Atomic ensemble as single qubit system} \label{sec:aeassq}
\begin{figure}[t]
\begin{center}
\includegraphics[scale=0.75]{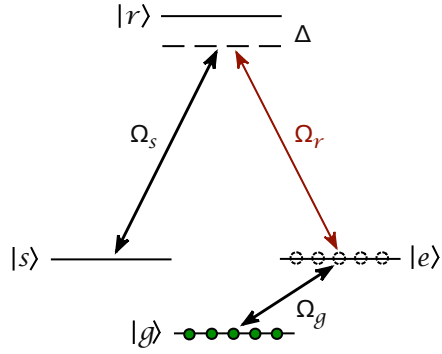}
\end{center}
\caption{Diagram of atomic level structure with allowed atomic transitions. States $|g \rangle$, $|e \rangle$, and $|s \rangle$ can be
realised by a lower, long-lived energy states of alkali atoms. $\Omega_{g}$ denotes Rabi frequency of a laser pulse coupling $|g \rangle$ and $|e \rangle$ states. A second laser pulse $\Omega_{s}$ is applied to the transition between the highly excited Rydberg level $|r\rangle$ and the state $|s \rangle$. This transition may possibly be a two-photon process. $\Omega_{r}$ denotes Rabi frequency of a weak laser pulse coupling $|e \rangle$ and $|r \rangle$ states. Both laser pulses $\Omega_{r}$ and $\Omega_{s}$ may be detuned from the corresponding atomic transition by $\Delta$. \label{levels}}
\end{figure}

\begin{figure}[t]
\begin{center}
\includegraphics[scale=0.70]{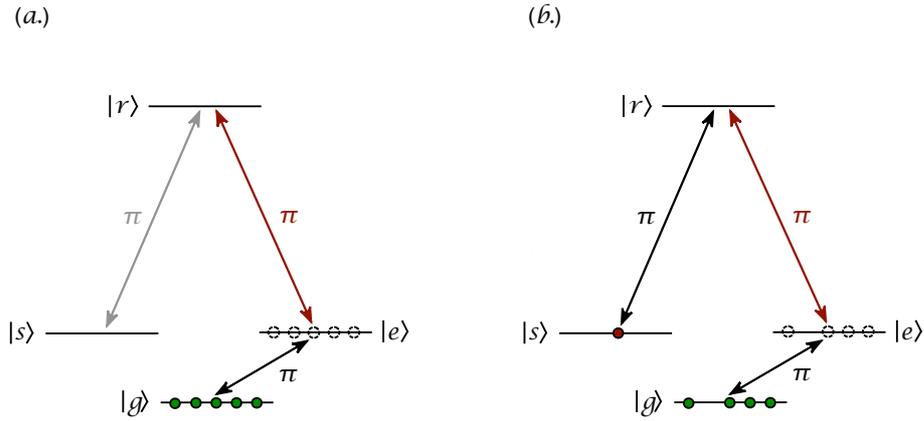}
\end{center}
\caption{Diagram representing the bit flip operation ($X$). (a.) Rotation from $| 0 \rangle_{L}$ to $| 1 \rangle_{L}$ and (b.) rotation from $| 1 \rangle_{L}$ to $| 0 \rangle_{L}$. See text for explanation. \label{X&H}}
\end{figure}

Until now we have avoided the issue of single qubit operations in atomic media. An atomic ensemble can serve as a qubit as long as one is able to apply single qubit rotations. A qubit may be represented by a micron-sized atomic ensemble, cooled to $\mu$K temperatures by the far off-resonant optical trap (FORT) or magneto-optical trap (MOT). The $N$ atoms at positions $r_{j}$ in an ensemble have three lower, long-lived energy states $|g
\rangle$, $|e \rangle$, and $|s \rangle$ (see Fig.~\ref{levels}). The qubit states in a mesoscopic ensemble are collective states defined as
\begin{eqnarray}
|0\rangle_{L} &\equiv& |g\rangle = |g_{1},g_{2}, \ldots ,g_{N}\rangle\, , \\
|1\rangle_{L} &\equiv& |s\rangle = \frac{1}{\sqrt{N}} \sum_{j=1}^{N}
e^{i \textbf{k} \cdot \textbf{r}_{j}} |g_{1},g_{2}, \ldots ,s_{j}, \ldots ,g_{N}\rangle\, .
\end{eqnarray}
Energy levels $|g \rangle$ and $|s \rangle$ play the role of storage states and transition between these states is always dipole-forbidden. These qubit states have a very desirable property of long coherence times. However, in the case of the qubit states defined as collective states of mesoscopic ensemble, the single-qubit manipulations are more complex than in the case of a qubit realised on a single atom. Moreover, one cannot use the weak excitation regime to implement reliable single-qubit operations. In fact, the simplest approach to this problem is to realise single-qubit rotations by means of classical optical pulses and the dipole blockade mechanism. In a paper by Brion, M{\o}lmer, and Saffman \cite{brion}, the single-qubit rotations are performed in only three elementary steps (see Fig.~\ref{X&H}). The laser pulses illuminate the entire ensemble and excite all atoms with equal probability \cite{lukin}. The states $|e \rangle$ and $|r\rangle$ participate in the interaction part of the scheme and states $|g \rangle$ and $|s \rangle$ serve as the storage levels. The single-qubit operations can be implemented with the following laser sequence:
\begin{enumerate}
\item first, two simultaneous $\pi$ pulses are applied to the transitions between levels $|s \rangle$ and $|r \rangle$, and $|g \rangle$ and $|e \rangle$. The first $\pi$ pulse may transfer a single atom from $|s \rangle$ to $|r \rangle$, and the remaining atomic population is transferred from $|g \rangle$ to $|e \rangle$ by the second pulse;
\item then a coherent coupling of states with zero and one Rydberg excited atom is applied for an appropriate amount of time;
\item finally, two simultaneous $\pi$ pulses may transfer a single atom back from $|r\rangle$ to $|s \rangle$, and the remaining atomic population from $|e \rangle$ to $|g \rangle$ .
\end{enumerate}
Therefore, in the case of a bit flip operation ($X$) the coherent coupling is just a $\pi$ pulse with a real Rabi frequency, and the Hadamard gate ($H$) can be performed by a $\pi/2$-pulse on the same transition. An arbitrary phase gate $\Phi(\phi) = \exp(-i\phi Z/2)$ is realised by a detuned optical pulse applied to the transition between $|s\rangle$ and an auxiliary level $|a\rangle$ (not shown in Fig.~\ref{X&H}). The gates $\Phi(\phi)$, $X$, and $H$ generate all single-qubit operations. The readout of a qubit is based on the resonance fluorescence and again requires an auxiliary level $|a\rangle$. An optical laser drives a transition between $|s\rangle$ and $|a\rangle$ producing a large number of fluorescence photons. If the measurement gives no fluorescence photons, the qubit is in  $|0\rangle_{L}$. Otherwise, a state of the qubit is projected into $| 1 \rangle_{L}$.

In summary, all single-qubit manipulations can be implemented in the dipole blockade regime with a laser pulse of a well-defined length and phase resonant with transition between a lower energy level and Rydberg state. The single-qubit manipulations can be executed rather fast through collective enhancement. The collective enhancement emerge from the fact that although only one atom is excited to a Rydberg state, all $N$ atoms in atomic medium interact with the laser field. In general, the above technique for implementing single-qubit manipulation is capable of generating any superpositions of collective qubit states of mesoscopic ensembles.

The scheme for implementing single-qubit operations relies heavily on the dipole blockade mechanism. We analyse the above scheme for the case of a bit flip operation $X$. Therefore, we need to carefully consider the evolution of the system under a $\pi$ pulse applied to the transition between $|e\rangle$ and $|r\rangle$. In the following discussion, we interchangeably use both levels $|g\rangle$ and $|e\rangle$ to denote a low-lying (ground) level.
\begin{figure}[t]
\begin{center}
\includegraphics[scale=0.75]{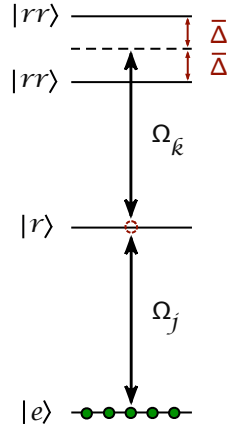}
\end{center}
\caption{The dipole blockade mechanism. The level structure consists of collective states of a mesoscopic atomic ensemble. The state $|e \rangle$ is the collective low-lying state, $|r \rangle$ is the singly excited Rydberg state and $|rr \rangle$ is the doubly excited state. $\Omega$ is the Rabi frequency of a weak laser pulse that is out of resonance with the transition between single and double excited states. $\bar{\Delta}$ is the mean dipole shift induced by a presence of a single atom in Rydberg state. \label{collective}}
\end{figure}
In general, the interaction of atoms with an optical laser pulse, within the dipole approximation and in the rotating frame approximation, is
governed by the interaction Hamiltonian $\hat{H}_{int}$
\begin{eqnarray}
\hat{H}_{int} &=& -i \hbar \sum_{j=1}^{N} \Omega_{j} \ \sigma^{j}_{re} \ \mbox{exp}[i(\omega_{re}-\omega)t] \nonumber \\
&& -i \hbar \sum_{j,k>j}^{N} \Omega_{k} \ \sigma^{jk}_{rr} \ \mbox{exp}[i(\omega_{re}-\omega)t] + \mbox{H.c.}\, ,
\end{eqnarray}
where $\Omega_{j} = \Omega e^{i \textbf{k} \cdot \textbf{r}_{j}}$ is the Rabi frequency, $\omega = k c$ is the frequency of an optical laser pulse,
$\sigma^{j}_{re} = |r_{j} \rangle \langle e|$
and $\sigma^{jk}_{rr} = |r_{j} r_{k} \rangle \langle r_{j}|$ are the atomic transition operators (see Fig.~\ref{collective}) \cite{saffman}.
The first transition operator $\sigma^{j}_{re}$ corresponds to the transition between the collective state $|e \rangle$
and the singly excited state
\begin{equation}
|r \rangle = \frac{1}{\sqrt{N}} \sum_{j=1}^{N} e^{i \textbf{k} \cdot \textbf{r}_{j}} | r_{j} \rangle\, ,
\end{equation}
where $| r_{j} \rangle = |e_{1}, e_{2}, \ldots ,r_{j}, \ldots , e_{N}\rangle$. The second one corresponds to the transition between the singly excited state $|r \rangle$ and the doubly excited state
\begin{equation}
|rr \rangle = \sqrt{\frac{2}{N(N-1)}} \sum_{j,k>j}^{N} e^{i (\textbf{k} \cdot \textbf{r}_{j} + \textbf{k} \cdot \textbf{r}_{k})} | r_{j} r_{k} \rangle\, ,
\end{equation}
where $| r_{j} r_{k} \rangle = |e_{1}, e_{2}, \ldots ,r_{j}, \ldots ,r_{k}, \ldots , e_{N}\rangle$. We assume that the optical laser pulse is resonant with a transition between $|e\rangle$ and $|r\rangle$ ($\omega_{re}-\omega = 0$). Then, the dipole interaction between two Rydberg atoms is given by
\begin{equation}
\hat{V}_{dd} = \hbar \sum_{j,k>j}^{N} \Delta_{jk} |r_{j} r_{k} \rangle \langle r_{j} r_{k}|\, ,
\end{equation}
where $\Delta_{jk} = \frac{C_{6}}{|\textbf{r}_{j} - \textbf{r}_{k}|^6} $ is the dipole shift of the weakest van der Waals type.
Hence, the coupling of levels $|e \rangle$ and $|r \rangle$ is described by the Hamiltonian $\hat{H} = \hat{H}_{int} + \hat{V}_{dd}$.
The state vector of an atomic ensemble is given by
\begin{equation}
|\psi(t) \rangle = c_{g} |g \rangle + \sum_{j=1}^{N} c_{j} e^{i \textbf{k} \cdot \textbf{r}_{j}} |r_{j} \rangle + \sum_{j,k>j}^{N} c_{jk} e^{i (\textbf{k} \cdot \textbf{r}_{j} + \textbf{k} \cdot \textbf{r}_{k})} |r_{j} r_{k} \rangle.
\end{equation}
In the limit where the dipole shift is much larger than the Rabi frequency of an optical laser pulse $\Delta_{jk} \gg \Omega_{j}$,
the Schr\"{o}dinger equation for amplitudes of the state vector gives
\begin{eqnarray}
\dot{c}_{g} &=& \sqrt{N} \Omega c_{r}, \\
\dot{c}_{r} &=& -\sqrt{N} \Omega c_{g} + \frac{\Omega}{\sqrt{N}} \sum_{j,k>j}^{N} c_{jk}, \\
\sum_{j,k>j}^{N} \dot{c}_{jk} &=& - \sum_{j,k>j}^{N} \Omega c_{j} - i \sum_{j,k>j}^{N} c_{jk} \Delta_{jk}, \label{double}
\end{eqnarray}
with $c_{r} = \sqrt{N} c_{j}$ (we assume that all $c_{j}$ coefficients are equal) \cite{saffman}. Elimination of the doubly excited Rydberg state described by Eq. (\ref{double}) by means of an adiabatic approximation $\dot{c}_{jk} \approx 0$ which leads to 
\begin{equation}
c_{jk} = \frac{i \Omega}{\sqrt{N} \Delta_{jk}} c_{r}
\end{equation}
yields
\begin{eqnarray}
\dot{c}_{g} &=& \sqrt{N} \Omega c_{r}, \\
\dot{c}_{r} &=& -\sqrt{N} \Omega c_{g} + \frac{i \bar{\Delta} \Omega^2}{N} c_{r}\, , \label{solution}
\end{eqnarray}
where $\bar{\Delta} = \sum_{j,k>j}^{N} \Delta_{jk}^{-1}$ is the mean dipole shift. The solution of Eq. (\ref{solution}) for $c_{g}(0) = 1$
(initially all atoms are in a low-lying state $|g\rangle$, or equivalently $|e\rangle$) reads as
\begin{equation}
|c_{r}(t)|^2 = \mbox{sin}^2(\sqrt{N l} \Omega t)/l,
\end{equation}
with $l = 1 + \frac{\bar{\Delta}^2 \Omega^2}{4 N^3}$. The evolution from the collective state $|e \rangle$ to the singly excited state $| r \rangle$ in time $t = \frac{\pi}{2 \sqrt{N l} \Omega}$ occurs with probability $P_{1} = 1/\l$. In the limit of finite dipole blockade, the probability of unwanted double excitations after the $\pi$ pulse is given by
\begin{equation}
P_{2} = \sum_{j,k>j}^{N} |c_{jk}|^2 = \frac{\bar{\Delta}_{P_{2}} \Omega^2}{N}, \label{P2}
\end{equation}
with $\bar{\Delta}_{P_{2}} = \sum_{j,k>j}^{N} \frac{1}{\Delta_{jk}^{2}}$. A finite blockade also implies a frequency shift of the effective two-level system ($|e\rangle$ and $|r\rangle$). The resonance frequency is shifted by $\delta \omega = \Omega^2 \bar{\Delta}/N$. The above results can be applied to the case of any single-qubit operation.

The numerical values based on the above model for single-qubit rotation are obtained for the following situation. We assume that a qubit is realised by a quasi one-dimensional (cigar shaped) atomic vapour consisting of $\sim$500 $^{87}$Rb atoms.
The spatial distribution (probability density) of an atomic cloud is given by
\begin{equation}
P(z) = (2 \pi \sigma_{z}^2)^{-1/2} \mbox{exp}(-z^2/2 \sigma_{z}^2) \label{distribution}\, ,
\end{equation}
where $z$ is the dimension along the ensemble, $\sigma_{z} = 3.0$ $\mu$m is the variance in the $z$ direction and $\sigma_{xy} = 0.5$ $\mu$m is the variance in the transverse directions. The level $|r\rangle$ may correspond to $43D_{5/2}$ or $58D_{3/2}$ state. The probability of double excitation given by Eq. (\ref{P2}) can be rewritten in terms of the mean blockade shift $B$ with $1/B^{2} = 2 \bar{\Delta}_{P_{2}}/N (N-1)$ \cite{walkerbloc}. Hence, the probability of a double excitation is given by
\begin{equation}
P_{2} =  \frac{\Omega^2_{N} (N-1)}{2 N B^2}\, ,
\end{equation}
where $\Omega_{N} = \sqrt{N} \Omega$. For the $43D_{5/2}$ and $58D_{3/2}$ states, the mean blockade shift is $B = 2\pi \times 0.25 \ \mbox{MHz}$ and $B = 2\pi \times 2.9 \ \mbox{MHz}$ in a trap with $\sigma = 3.0$  $\mu\mbox{m}$, respectively \cite{walkerbloc}. For $\Omega = 2\pi \times 1$ kHz, the probability of a double excitation for the $43D_{5/2}$ level is $P_{2} \cong 4.0 \times 10^{-3}$ and for the $58D_{3/2}$ level is $P_{2} \cong  3.0 \times 10^{-5}$. The probability of doubly-excited states and singly-excited states outside the desired two-level system resulting from a shifted resonance frequency are similar.

The time of a $\pi$ pulse applied to the transition between $|e \rangle$ and $| r \rangle$ is $t \cong 11.2$ $\mu$s. We estimate that the rest of the $\pi$ pulses which are necessary to realise any single-qubit rotation (see Fig.~\ref{X&H}) can be applied in time significantly shorter than the time $t$. In summary, the above single-qubit rotations can be carried out on a microsecond timescale.

The spontaneous emission from the Rydberg state and the black-body transfer (to other Rydberg states) occur with low rates of order $10^{3}$ Hz (or even $10^2$ Hz for higher Rydberg states) and may introduce small error $P_{decay} \cong 1 - \exp(- 10^{3} t) = 0.01$ \cite{day}. Other sources of errors such as atomic collisions and Doppler broadening are negligible because of the low temperature of the atomic vapour.

The fidelity of the single-qubit rotations can be as high as $F_{single} = \exp[-(2 P_{2} + P_{decay})] \cong 0.99$, where $P_{2} = 3.0 \times 10^{-5}$. This fidelity is given for the worst case scenario when the separation of atoms is maximal and the dipole-dipole interaction is of the weakest (van der Waals) type.

Single-qubit rotations are one of the basic operations that are necessary in any model of quantum computation. The above fast and reliable implementations of the single-qubit operations open the possibility for a realisation of the measurement-based model of quantum computation. However, we are still lacking a scheme for efficient generation of the cluster states, a resource for the one-way quantum computer.

\section{New entangling protocol based on the dipole blockade} \label{sec:newprotocol}
\begin{figure}[t]
\begin{center}
\includegraphics[scale=0.70]{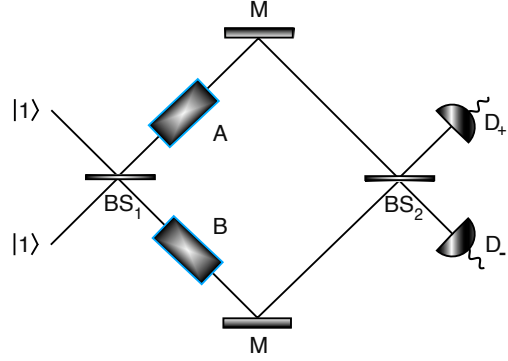}
\end{center}
\caption{Diagram of the entangling protocol. A pair of entangled photons in the state $|\phi\rangle_{light}=\frac{i}{\sqrt{2}} (|02\rangle + |20\rangle)$ interact with atomic vapours placed in the arms of a Mach--Zehnder interferometer. One and only one alkali atom in the ensemble is excited by one of the photons to the Rydberg state $|r\rangle$. Absorption of the second photon is prohibited by the dipole blockade mechanism. Detection of a single photon will leave the atomic ensembles in an entangled state $|\psi^{\pm}\rangle = \frac{1}{\sqrt{2}}(|re\rangle \pm i|er\rangle)$. \label{scheme}}
\end{figure}

We propose a scheme for efficient and reliable cluster state generation, based on the dipole blockade mechanism \cite{zwierz}. The entangling operation between two mesoscopic atomic ensembles takes place in the arms of a Mach--Zehnder interferometer shown in Fig.~\ref{scheme}.
The protocol begins with both ensembles $A$ and $B$ prepared in the collective state $|\phi\rangle_{A,B} = |e\rangle \equiv |e_{1},e_{2}, \ldots
,e_{N}\rangle$ (see Fig.~\ref{levels}). Next, two indistinguishable photons enter each input mode of the interferometer. After the first beam splitter (BS$_{1}$), due to the Hong-Ou-Mandel (HOM) effect two photons propagate in the maximally entangled state:
\begin{equation}
|\phi\rangle_{light}=|11\rangle \xrightarrow{\mbox{BS}_{1}} \frac{i}{\sqrt{2}}(|02\rangle + |20\rangle)\, ,
\end{equation}
where $|0\rangle$ and $|2\rangle$ denote the vacuum and a two-photon state respectively \cite{hong,ghosh}. The photons can be maximally entangled only if prior to a beam splitter interaction both photons were exactly the same in all possible senses. Subsequently, two photons interact with the atomic ensembles: One and only one atom in the ensemble is excited by one of the photons to the Rydberg state $|r\rangle$, and the absorption of the second photon is prohibited by the dipole blockade mechanism \cite{zwierz}. Following the dipole blockade interaction, the total state of the atomic ensembles and light fields is given by
\begin{equation}
|\phi\rangle_{int}=\frac{i}{\sqrt{2}} (|er\rangle |01\rangle +
|re\rangle |10\rangle).
\end{equation}
We omit in the following discussion the overall phase factor introduced to the total state by reflections from mirrors $M$. After the second beam splitter (BS$_{2}$), the total state reads
\begin{equation}
|\phi\rangle_{out}=\frac{i}{\sqrt{2}} (|\psi^{+}\rangle |01\rangle +
|\psi^{-}\rangle |10\rangle)\, ,
\end{equation}
where $|\psi^{\pm}\rangle = \frac{1}{\sqrt{2}}(|re\rangle \pm i |er\rangle)$. Conditional on a single click at photo detectors $D_{+}$ or $D_{-}$, the atomic ensembles are projected onto a maximally entangled state. After establishing entanglement, the qubits are transferred to their
computational basis states $|0\rangle_{L} \equiv |g\rangle$ and $|1\rangle_{L} \equiv |s\rangle$ by classical optical pulses $\Omega_{g}$ and $\Omega_{s}$. Ideally every run of the protocol gives an entangled state of two atomic ensembles with success probability $p_{success} = \eta$, where $\eta = \eta_{D} \eta^{2}_{S}$ is the combined detection and source efficiency. This is a significant improvement compared to the success probability $p_{success} = \eta^2/2$ of the double-heralding protocol in Ref.~\cite{kok}.

\begin{figure}[t]
\begin{center}
\includegraphics[scale=0.60]{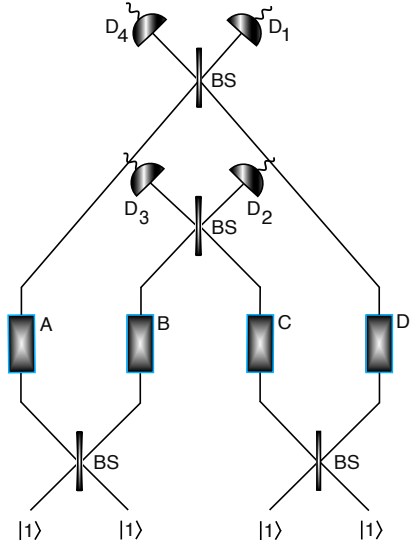}
\end{center}
\caption{The scheme for creating the 4-qubit GHZ state. Four ensembles $A$, $B$, $C$, and $D$ prepared in the state $|\phi\rangle_{ABCD}=|eeee\rangle$ interact with two pairs of entangled, indistinguishable photons. Conditional on photo detector clicks at the photo detector pair $(D_{1},D_{2})$,
$(D_{1},D_{3})$, $(D_{4},D_{2})$ or $(D_{4}, D_{3})$, the state of four qubits is projected onto the 4-qubit GHZ state (up to phase correcting operations) with success probability $p_{success} = \eta^2/2$. \label{GHZ}}
\end{figure}

\subsection{Generation of the GHZ and cluster states}
The entangling operation can be used to efficiently create arbitrary cluster states of any degree of connectivity, including 2D universal resource states for a one-way quantum computer. However, a modification of the entangling procedure yields an even more dramatic improvement in the efficiency of cluster state generation. By arranging the atomic ensembles in a four-mode interferometer as shown in Fig.~\ref{GHZ}, the detection of two photons will create the four-qubit GHZ state \emph{in a single step} with the success probability $p_{success} = \eta^2/2$. Moreover, since only two photons are detected, the protocol is relatively insensitive to detector losses. Higher GHZ states can be created by a straightforward extension. Subsequent GHZ states are generated with success probability
\begin{equation}
p_{success} = \eta^{Q/2} (Q-2)/2^{Q-2}\, ,
\end{equation}
where $Q = 4, 6, \ldots$ is the number of the qubits.

As already mentioned, the GHZ states are locally equivalent to cluster states. The efficiently generated large GHZ states may serve as building blocks for universal graph states. By entangling small clusters with the above entangling procedure, large cluster states can be constructed. A single photon applied to a pair of qubits (each from two different 4-qubit cluster states) followed by a single photo detector click creates an 8-qubit cluster state with success probability $p_{success} = \eta'/8$. This procedure can be repeated in an efficient manner \cite{kieling}. In the case of failure, the two qubits that participated in linking are measured in the computational basis, and the rest of the cluster state is recycled \cite{nielsen_microcluster}.

\section{Errors, decoherence mechanisms and fidelity} \label{sec:errors}
The dominant errors and decoherence mechanisms that enter the entangling operation are the following:
\begin{enumerate}
\item the imperfect mode matching that results in the unwanted coincidence events in the HOM effect,
\item the spontaneous emission rate of the Rydberg state,
\item the black-body transfer rate (to other Rydberg states),
\item the atomic collision rate,
\item the doubly-excited Rydberg states and singly-excited states outside the desired two-level system,
\item no absorption event,
\item the dark count rate of the photo detectors.
\end{enumerate}
We analyse in more detail the above dominant error and decoherence mechanisms on the following experimental implementation. First, let us consider the coincidence events in the HOM effect. The single indistinguishable photons that recombine at the first beam splitter (BS$_{1}$) can be generated by means of spontaneous parametric down-conversion (SPDC) process or sources of single-photon pulses such as atomic ensemble inside an optical cavity \cite{halder,thompson}. The SPDC source (the non-linear crystal) must be pumped with a narrowband ($\sim$1 MHz) laser or placed inside a cavity. These kind of cavity-enhanced SPDC sources produce pairs of identical photons with a narrow bandwidth of order of MHz and a spectral brightness of $\sim$1500 photons/s per MHZ bandwidth \cite{neergaard,bao}.

\begin{figure}[t]
\begin{center}
\includegraphics[scale=0.70]{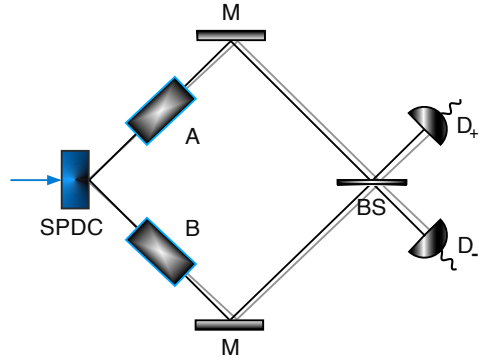}
\end{center}
\caption{Example of an experimental implementation of the entangling protocol. The source of a single-photon pair entangled in the momentum (path) degree of freedom consists of the type I non-linear crystal. \label{implement}}
\end{figure}

In general, successful generation of the entangled state of light depends on the proper setup, where both photons from the SPDC source
recombine at BS$_{1}$ at the same time. In a recent experiment, the coincidence event in the HOM effect occurs with a low
rate of 1500 counts/s \cite{kim}. In fact, it is possible to completely eliminate the coincidence events in the HOM
effect by getting rid of the BS$_{1}$. In place of single-photon sources and BS$_{1}$, one can use a SPDC source generating pairs of
single-photons entangled in momentum (path) degree of freedom \cite{barbieri,rossi}. The state of the photons is given by
$|\phi\rangle_{light}= \frac{1}{\sqrt{2}} (|1,1;0,0\rangle_{A;B} + |0,0;1,1\rangle_{A;B})$, where states $|1,1;0,0\rangle_{A;B}$ and $|0,0;1,1\rangle_{A;B}$ represent two single photons propagating along slightly different paths through upper and lower arm of the interferometer, and interacting with atomic ensembles $A$ and $B$, respectively (see Fig.~\ref{implement}). State $|\phi\rangle_{light}$ represents so-called dual-rail qubit encoding. Moreover, since the SPDC process is a phase and energy matching phenomenon, no phase difference appears between two paths (pairs) $A$ and $B$ \cite{barbieri}. In general, the whole Mach--Zehnder interferometer needs to be phase-stable. In the case of a GHZ state generation, phase locking of a large number of Mach--Zehnder interferometers is very demanding (although possible). Therefore, by replacing single-photon sources and BS$_{1}$ with the SPDC source generating entangled photon pairs, we may simplify experimental realisation of the entangling operation (although the second half of the interferometer after atomic ensembles still requires phase stabilisation). Recently, it has been shown that these kind of entangled pairs of photons can be generated very effectively \cite{rossi}.

Now, assume that an atomic vapour consists of 500 $^{87}$Rb atoms placed in the far off-resonant optical trap (FORT) or magneto-optical trap (MOT). The atomic levels $|g \rangle$, $|e \rangle$, and $|r \rangle$ may correspond to $(5S_{1/2}, F = 1)$, $(5P_{3/2}, F = 2)$ and $43D_{5/2}$ or $58D_{3/2}$, respectively. State  $|s \rangle$ may correspond to the hyperfine state $(5S_{1/2}, F = 2)$, which implies that the transition from $|s\rangle$ to $|r\rangle$ is a two-photon process (see Fig.~\ref{levels}). We have identified state $|e\rangle$ with a short lived state $(5P_{3/2}, F = 2)$, when in fact it must be a long-lived energy level. However, in the case of the MOT, the requirement of a long relaxation time of the state  $|e \rangle$ can be lifted since the trap lasers produce a constant population in the $|e \rangle$ state \cite{li,monroe}. In general, a requirement of state $|e \rangle$ is imposed to simplify experimental realisation of the protocol where usually two-photon excitations are used to obtain Rydberg atoms. The spatial distribution of an atomic cloud is a quasi one-dimensional (cigar shaped) ensemble with probability density given by Eq. (\ref{distribution}). Atomic vapours described with quasi  one-dimensional probability density have been demonstrated experimentally \cite{johnson}.

When a protocol is based on a quantum optical system, its performance is limited by the dark count rate of the photo detectors. The dark count rate of a modern photo detector $\gamma_{dc}$ can be as low as 20 Hz and efficiency reaches $\eta_{D} \approx 30\%$ for wavelengths around 480 nm. The probability of the dark count is $P_{dc} = 1 - \mbox{exp}(-\gamma_{dc} t/p_{success} )$, where $t$ is the time scale of the entangling protocol. In general, the probability of the dark count is negligible for $p_{success} > \gamma_{dc} t$.

Since the length of the atomic ensemble needs to be of the order of several $\mu$m, the most important source of errors is the lack of absorption event. The probability of absorption of a single photon by a cigar shaped atomic ensemble is given by $P_{abs} \cong  1 - e^{-N_{i} \sigma_{0} /A}$, with $N_{i} = N$ the number of atoms in the interaction region, $\sigma_{0} = 3 \lambda^2 \gamma_{0}/(2 \pi \gamma)$ is the on-resonance scattering cross section of a single-photon pulse, where $\gamma_{0}$ is the spontaneous decay rate of the Rydberg state to low-lying levels, and $\gamma$ is the spontaneous decay to other Rydberg states \cite{hammerer,day,saff}. $A = \pi w^2_{0}$ is the area of a single-photon pulse with a waist $w_{0} \approx \pi \lambda$ \cite{walkerabs}. With  $\lambda_{43D} = 485.766$ nm, $\gamma_{0} = 1.1 \times 10^{4}$ Hz and $\gamma = 7.2 \times 10^{4}$ Hz, the probability of absorption for $43D_{5/2}$ state is $P_{abs} \cong 0.69$. For $\lambda_{58D} = 485.081$ nm, $\gamma_{0} = 4.8 \times 10^{3}$ Hz and $\gamma = 2.0 \times 10^{4}$ Hz, the probability of absorption for $58D_{3/2}$ state is $P_{abs} \cong 0.84$. The probability of absorption for both Rydberg states is much too low for reliable operation of the entangling gate, therefore one has to use atomic ensembles with larger number of atoms $N$. A smaller area $A$ does not improve the probability of absorption since it implies a smaller number of atoms $N_{i}$ in the interaction region. In fact, the optimal area $A$ coincides with a size of an atomic ensemble in the transverse directions. Consequently, for a level structure shown in Fig.~\ref{levels} the only solution to a low absorption probability is a higher number of atoms $N$. To render probability of absorption close to unity, one has to use more than 2500 atoms. High fidelity dipole blockade in such a large ensemble is not feasible \cite{saff}. Another potential difficulty may be a rather high atomic density of the sample. Several $\mu$m long atomic vapours consisting of thousand of atoms are hard to prepare.

\begin{figure}[t]
\begin{center}
\includegraphics[scale=0.70]{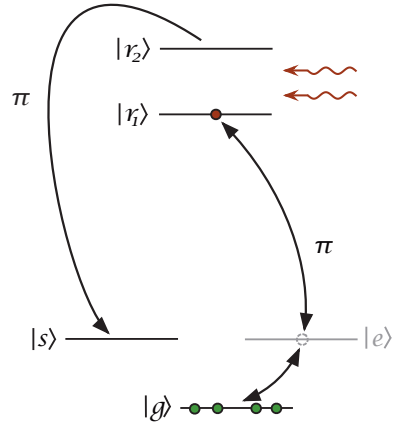}
\end{center}
\caption{Relevant atomic level structure with allowed atomic
transitions. The low-lying state $|g \rangle$ is coupled to the first Rydberg state $|r_{1} \rangle$
through intermediate low-lying level $|e \rangle$ by means of a classical field that implements a $\pi$ pulse. A second $\pi$ pulse realised by a
classical field  is applied to the transition between the second Rydberg level $| r_{2} \rangle$ and the state $| s \rangle$ (it may possibly be a two-photon process). \label{implement_rr}}
\end{figure}

To overcome these difficulties, we propose different experimental implementation based on a level structure shown in Fig.~\ref{implement_rr}.
This level structure implements the entangling operation but in slightly different manner. Here, the dipole blockade is not used to block absorption of a second photon. In this implementation, we employ the dipole blockade mechanism to prepare a single atomic excitation in the Rydberg state $|r_{1} \rangle$ by means of a two-photon process. This is precisely a bit flip operation $X$ described in Sec. 7. However here, the two-photon Rabi frequency of a laser field that realises a $\pi$ pulse between states $|g \rangle$ and $|r_{1} \rangle$ is given by $\Omega = \Omega_{ge}\Omega_{er_{1}}/2\Delta$, where $\Delta$ is a small detuning (not shown in Fig.~\ref{implement_rr}). For $|r_{1} \rangle = 43D_{5/2}$ a bit flip operation can be carried out on a microsecond timescale. Subsequently, two single photons interact with the atomic ensembles: One and only one photon is absorbed by the Rydberg atom in the state $|r_{1} \rangle$. The Rydberg atom is excited to the state $|r_{2} \rangle$. We assume that the probability of a two-photon absorption process is negligible since both photons are on-resonance with a transition between Rydberg states $|r_{1} \rangle$ and $|r_{2} \rangle$ \cite{jacobs,you,nakanishi}. Finally, following a single photo detector click, two $\pi$ pulses realised by classical fields are applied simultaneously to the transitions between the second Rydberg level $| r_{2} \rangle$ and the state $| s \rangle$ (it may possibly be a two-photon process), and between the first Rydberg level $| r_{1} \rangle$ and the state $| g \rangle$. The main strength of this implementation lies in the fact that all $\pi$ pulses that transfer single excitation between Rydberg states and low-lying storage states $| g \rangle$ and $| s \rangle$ are highly reliable operations with fidelity $F_{single} \cong 0.99$. Most importantly, this kind of control over atomic ensembles has been demonstrated experimentally \cite{johnson,singer,deiglmayr}. Consequently, this implementation of the entangling operation requires relatively straightforward experimental extension of known procedures.

Let us now examine if the probability of absorption of a single photon by a single Rydberg atom is high enough for reliable operation of our entangling gate. The two single photons couple to the transition between level $|r \rangle = 45P_{3/2}$ (not shown in Fig.~\ref{implement_rr}) and level $|r_{2} \rangle = 58D_{3/2}$. The reader should note that state $|r_{1} \rangle = 43D_{5/2}$ is only used in a bit flip operation $X$ and then single Rydberg atom is excited by means of a fast microwave pulse to the Rydberg state $|r \rangle = 45P_{3/2}$. For $\lambda_{45P-58D} = 370.783$ $\mu m$, $\gamma_{0} = 4.8 \times 10^{3}$ Hz and $\gamma = 2.0 \times 10^{4}$ Hz, the probability of absorption $P_{abs} \cong  1 - e^{-2 \sigma_{0} /A} \cong 0.90$, where area $A = 0.1 \lambda_{45P-58D}^2$ (this implies a waist of a  single-photon pulse $w_{0} \cong 66$ $\mu$m) \cite{quabis}. High probability of an absorption requires strongly focused light fields with small area $A$ \cite{enk,tey,tey2}. The improved ratio of $\sigma_{0} /A$ and therefore higher probability of an absorption for this experimental implementation is due to the stronger focusing relative to the wavelength of a single-photon pulse. Naturally, the focusing regime is limited by a size of the atomic sample and diffraction limited area of a single-photon pulse \cite{enk,tey,tey2}. To render the probability of absorption close to unity one may apply a mode converter (shaper) to a single photon light field \cite{sondermann}. The probability of absorption depends also on the spontaneous decay rates associated with the Rydberg state. The rich structure of Rydberg levels offers many possible ways for level assignment, therefore one may be able to choose two Rydberg states with higher on-resonance scattering cross section. The overall time scale of the entangling protocol $t$  consists of a time required by the $\pi$ pulse $t_{\pi} = 11.2$ $\mu$s (preparation of a single Rydberg atom in the state $|r_{1} \rangle$) and time of an interaction part of the protocol given by
\begin{equation}
t_{45P-58D} = \frac{\pi}{2 (\sqrt{2} g)} \cong 2.9 \mbox{ ns, with } g = \sqrt{\frac{\sigma_{0} \gamma c}{4 V}}\, ,
\end{equation}
where $g$ is the atom-light coupling constant and $V = A L$ is the interaction volume with $L = 12$ $\mu$m the length of an atomic medium \cite{hammerer}. After successful entanglement preparation the state of atomic ensembles is quickly stored in the long-lived atomic states  $|g \rangle$ and $|s \rangle$ in time significantly shorter than $t_{\pi}$. In summary, the entangling protocol can be carried out on a microsecond time scale.

The two single photons employed in the entangling procedure belong to the far-infrared part of the electromagnetic spectrum. The photo detectors sensitive to this part of the spectrum are under development. The detection range of quantum dot infrared photo detectors such as In(Ga)As quantum ring terahertz photo detector reaches 175 $\mu$m. Another photo detector operating in THz regime is based on hot-electron effect in nanobolometers and used in astrophysics for registration of the cosmic microwave background (CMB) radiation \cite{wei}. A bolometer is a device that measures the energy of incident electromagnetic radiation. Although photo detector based on nanobolometers is characterised with rather complex fabrication and has to work in ultra-cold temperature regime (around 200 mK), it is highly sensitive and capable of detecting single THz photons with quantum efficiency close to 100\% (maintaining at the same time low dark count rate) \cite{wei}. An alternative to the photo detector based on nanobolometers can be given in terms of the atomic-vapour-based high efficiency photon detectors \cite{james,imam}.

The source of single-photon pulses in the far-infrared frequency regime can be based on atomic ensembles or on single ions placed inside an optical cavity. Preferably, the single-photon sources should work on-demand. However, as already mentioned, one may choose different Rydberg levels and implement the entangling gate with single photons from a less extreme part of the electromagnetic spectrum.

The spontaneous emission from the Rydberg state and the black-body transfer (to other Rydberg states) occur with rates of order $10^3$ Hz (or even $10^2$ Hz for higher Rydberg states), and are negligible, since following successful entanglement preparation the state of matter qubit is quickly stored in the long-lived atomic states  $|g \rangle$ and $|s \rangle$. Exact values of these rates are given in Ref.~\cite{day}. The atomic collision rate inside atomic vapour is given by
\begin{equation}
\tau^{-1}_{col}\approx n \sigma_{col}/\sqrt{M/3 k_{B} T},
\end{equation}
with $n$ the number density of atoms, $\sigma_{col}$ the collisional cross section ($\sim$$10^{-14}$ $\mbox{cm}^2$), $M$ the atomic mass, $k_{B}$ Boltzmann's constant, and $T$ the temperature \cite{james}. Assuming a vapour with a number density of atoms of order $10^{12}$  cm$^{-3}$ and a temperature of $\sim$$10^{-3}$ K, the atomic collision rate can be as low as 2 Hz. Moreover, with a sufficiently large energy difference between states $|g\rangle$ and $|s\rangle$ a single collision is not likely to affect the qubit.

A low temperature of an atomic vapour implies negligible Doppler broadening. The Doppler broadening is described by the Gaussian distribution with a standard deviation of $\Delta \lambda = \lambda_{0} \sqrt{k_{B} T/M c^2}$, where $\lambda_{0}$ is the center wavelength of the Doppler profile (wavelength of a transition between states $|r\rangle$ and $|r_{1}\rangle$). For $\lambda_{0}=\lambda_{45P-58D}$, the Doppler broadening is $\Delta \lambda = 0.4 \times 10^{-6}$ $\mu$m. Therefore, the Doppler broadening does not affect fidelity of the entangling protocol.

Considering both the overall and interaction time scales of the protocol, the entangling procedure is mostly affected by the no absorption event, assuming high quantum efficiency and low dark count rate of the THz photo detectors. We assume that the coincidence event rate in the HOM effect and two-photon absorption process are negligible. In the presence of the above noise and decoherence mechanisms, the final state of the system conditional on a single photo detector click is given by
\begin{equation}
\rho_{fin} = (1 - 2\varepsilon) |\psi^{\pm} \rangle \langle \psi^{\pm}| + 2\varepsilon \rho_{noise} + \mathcal{O}(\varepsilon^2)\, ,
\end{equation}
where $|\psi^{\pm}\rangle = \frac{1}{\sqrt{2}}(|sg\rangle \pm i|gs\rangle)$ and $\varepsilon = 1-P_{abs}$, where $P_{abs}$ is the
probability of an absorption of a single photon by a single Rydberg atom. $\rho_{noise}$ denotes the unwanted terms in the state of the two atomic ensembles. It is worth noting that the source efficiency does not affect the fidelity of the final state, it only lowers the success probability.
After taking into account all dominant error mechanisms, the fidelity of the prepared entangled state is given by
\begin{equation}
F = \langle \psi^{\pm} | \rho_{fin} |\psi^{\pm} \rangle \cong 0.90.
\end{equation}
Stronger focusing regime and/or application of a mode converter (shaper) to a single photon light field should render the fidelity of the entangling operation close to current fault-tolerant thresholds of topological codes \cite{raussendorf,kitaev,raussendorf_new,stace}. As already stated, the new entangling protocol is capable of creating cluster states of any degree of connectivity. Since a 3D cluster lattice can be used to implement planar surface codes efficiently, one can exploit topological error-correction capabilities of these codes to perform fault-tolerant quantum computation with our entangling procedure. One can also increase the fidelity of the final states with the use of purification techniques at the price of cluster size reduction when the purification fails \cite{dur,aschauer}.

\subsection{Polarisation-entangled photon pair}
\begin{figure}[t]
\begin{center}
\includegraphics[scale=0.70]{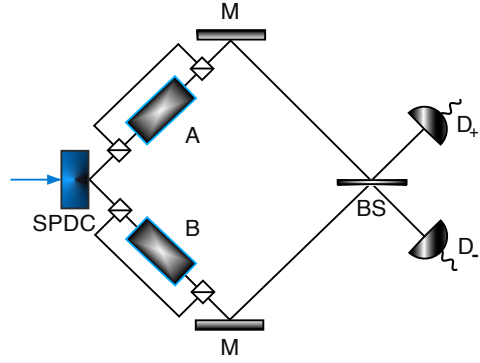}
\end{center}
\caption{Example of an experimental implementation of the entangling protocol exploiting the polarisation-entangled photon pair. The source of a single-photon pair consists of the type I non-linear crystal. A ``square" placed in front of and behind each ensemble depicts the polarisation beam splitter. \label{implement_HV}}
\end{figure}

The dual-rail qubit encoding responsible for the significant experimental simplification of the protocol can be implemented with the polarisation-entangled state of a photon pair $|\phi\rangle_{light} = \frac{1}{\sqrt{2}} (|HV\rangle_{AB} + |VH\rangle_{AB})$, where $H$ and $V$ are the horizontal and vertical polarisation of a photon, respectively (see Fig.~\ref{implement_HV}). Traditionally, polarisation-entangled photon pairs are generated by means of a type I non-linear crystal \cite{barbieri,kwiat}. A pair of photons entangled in a polarisation degree of freedom allows us to simplify the experimental implementation of the protocol even further. In fact, the dipole blockade mechanism is no longer required for a reliable operation of our entangling protocol. Here, the entangling gate works as follows.
Initially, we prepare each ensemble $A$ and $B$ in the collective ground state $|g\rangle \equiv |g_{1},g_{2}, \ldots ,g_{N}\rangle$, where the ground state $|g \rangle$ may correspond to atomic level $(5S_{1/2}, F = 1)$. Subsequently, one of the polarisation-entangled photons, for instance horizontally polarised, interacts with the atomic ensembles: One and only one atom in the ensemble is excited by the photon to the excited state $|e\rangle$ that may correspond to atomic level $(5P_{3/2}, F = 2)$. The absorption of the second photon (the vertically polarised) is completely prohibited since the polarisation beam splitter placed in front of each ensemble prevents vertically polarised photon from interacting with the atomic vapours. Following a balanced beam splitter (BS) interaction and conditioned on a photo detector click, the state of the atomic ensembles is projected onto a pure maximally entangled state. The main advantage of this implementation lies in a fact that the second photon is never absorbed by the atomic ensembles and the probability of absorption of the first photon $P_{abs} \cong  1 - e^{-N_{i} \sigma_{0} /A}$ can practically reach unity since one can exploit optically thick atomic vapours, i.e., highly dense and/or large vapours. Therefore, a single photon should easily couple to atomic medium. Moreover, we are no longer limited by the size of the atomic ensemble, which significantly simplifies the preparation of atomic samples.

The dominant errors and decoherence mechanisms that enter the entangling operation based on the polarisation-entangled photon pair are the following:
\begin{enumerate}
\item the quality of polarisation-entangled photon source
\item the spontaneous emission rate of the exited state $|e\rangle$,
\item the atomic collision rate,
\item the inefficiency and the dark count rate of the photo detectors.
\end{enumerate}
Recently, ultrafast, high quality polarisation-entangled photon sources have been developed with the reported fidelity reaching $F_{source} \cong 0.99$ \cite{rangarajan}. The atomic level $(5P_{3/2}, F = 2)$ corresponding to the excited state $|e\rangle$ is a short-lived level. Therefore, one has to apply an additional laser pulse between the excited state $|e\rangle$ and storage state $|s\rangle$, exploiting stimulated Raman adiabatic passage, to reliably carry out the entangling operation. The atomic collision rate as described in the previous section is insignificant. The dark count rate of a modern photo detector $\gamma_{dc}$ can be negligibly low (20 Hz) with efficiency reaching $\eta_{D} \approx 90\%$ for wavelengths around 780 nm. Considering all the dominant errors and decoherence mechanisms, the implementation of the entangling protocol exploiting polarisation-entangled photon pair can generate entangled states of atomic ensembles with fidelity $F  = F_{source} \cong 0.99$ and success probability $p_{success} = \eta$, where $\eta = \eta_{D} \eta^{2}_{S}$ is the combined detection and source efficiency. It is worth noting that the fidelity of the final state does not depend on $\eta$ and can be as high as the fidelity of the photon source. More importantly, all results associated with the generation of the GHZ and cluster states follow exactly.

\begin{figure}[t]
\begin{center}
\includegraphics[scale=0.70]{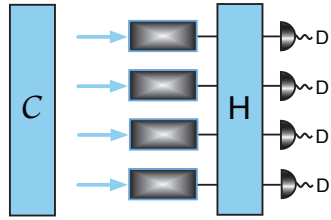}
\end{center}
\caption{A mapping of cluster state on a collection of atomic ensembles. $H$ depicts a multi-port beam splitter that erases which-path information. \label{cluster_map}}
\end{figure}

The new entangling operation presented here allows us to map an entangled state of a photon pair onto two macroscopic atomic ensembles in a heralded fashion. In general, one may imagine an experiment in which a multi-qubit photonic cluster state is mapped onto a collection of atomic ensembles (see Fig.~\ref{cluster_map}).
In this way, a possibly large $Q$-qubit cluster state can be reliably stored in $Q$ atomic vapours.

\section{Summary}
In conclusion, we have reviewed several schemes for probabilistic entanglement generation such as the DLCZ protocol and the double-heralding protocol. We have also presented and studied a new scheme for cluster state generation based on atomic ensembles and the dipole blockade mechanism. The new entangling protocol consists of single-photon sources, ultra-cold atomic ensembles, and regular photo detectors. The protocol generates \emph{in a single step} a GHZ state with success probability $p_{success}\sim\eta^{Q/2}$, where $Q$ is the number of the qubits, and high fidelity $F \cong 0.90$ (or $F \cong 0.99$ when polarisation-entangled implementation is used). Our new entangling gate is more efficient than any previously proposed probabilistic scheme with realistic photo detectors and single-photon sources. Every run of the procedure gives an entangled state of two atomic ensembles with success probability $p_{success} = \eta$, where $\eta$ is the combined detection and source efficiency. The double-heralding protocol produces an entangled state of two matter qubits with the success probability $p_{success} = \eta^2/2$. The protocol proposed by Lim \textit{et al.} \cite{lim_lett} requires on average two repetitions to realise the desired gate operation. Moreover, the successful implementation of this protocol involves detection of a photon pair. The new entangling protocol requires only single photon detection. In general, number-resolution photo detectors are not needed. However, a reliable photon counting detector with low dark count rate would be able to herald any error in the procedure increasing the fidelity close to unity. The GHZ states are locally (up to Hadamard operation) equivalent to star-shaped cluster states. The efficiently generated large GHZ states may serve as building blocks for universal graph states.

We have also reviewed and analysed a scheme implementing any single-qubit operation on the qubit defined as collective states of mesoscopic ensemble. The scheme for single-qubit rotations is based on classical optical pulses and the dipole blockade mechanism. The experimental implementation may be carried out with high fidelity $F_{single} \cong 0.99$ and on the microsecond timescale with current state-of-the-art experimental setups.

The described protocols for single-qubit operations and entangling operation open a possibility of experimental implementation of the
measurement-based quantum computer based on atomic ensembles.  


\addtocontents{toc}{\vspace{2em}} 





\backmatter

\label{Bibliography}
\lhead{\textsc{Bibliography}}  
\bibliographystyle{unsrtnat}  

\begin{thebibliography}{172}
\providecommand{\natexlab}[1]{#1}
\providecommand{\url}[1]{\texttt{#1}}
\expandafter\ifx\csname urlstyle\endcsname\relax
  \providecommand{\doi}[1]{doi: #1}\else
  \providecommand{\doi}{doi: \begingroup \urlstyle{rm}\Url}\fi

\bibitem[{P. Kok and B. W. Lovett}(2010)]{kok_book}
{P. Kok and B. W. Lovett}.
\newblock \emph{{Introduction to Optical Quantum Information Processing}}.
\newblock Cambridge University Press, 2010.

\bibitem[{M. A. Nielsen and I. L. Chuang}(2000)]{nielsen}
{M. A. Nielsen and I. L. Chuang}.
\newblock \emph{{Quantum Computation and Quantum Information}}.
\newblock Cambridge University Press, 2000.

\bibitem[{W. K. Wootters}(1998)]{wootters}
{W. K. Wootters}.
\newblock {Entanglement of formation of an arbitrary state of two qubits}.
\newblock \emph{Phys. Rev. Lett.}, 80:\penalty0 2245, 1998.

\bibitem[{R. Horodecki, P. Horodecki, M. Horodecki, and K.
  Horodecki}(2009)]{horodecki}
{R. Horodecki, P. Horodecki, M. Horodecki, and K. Horodecki}.
\newblock {Quantum entanglement}.
\newblock \emph{Rev. Mod. Phys.}, 81:\penalty0 865, 2009.

\bibitem[{P. Kok}(2010)]{koklecture}
{P. Kok}.
\newblock {Lecture notes on optical quantum computing}.
\newblock \emph{Lecture Notes in Physics}, 787:\penalty0 187, 2010.

\bibitem[{D. Loss and D. P. DiVincenzo}(1998)]{loss}
{D. Loss and D. P. DiVincenzo}.
\newblock {Quantum computation with quantum dots}.
\newblock \emph{Phys. Rev. A}, 57:\penalty0 120, 1998.

\bibitem[{G. Burkard, D. Loss, and D. P. DiVincenzo}(1999)]{burkard}
{G. Burkard, D. Loss, and D. P. DiVincenzo}.
\newblock {Coupled quantum dots as quantum gates}.
\newblock \emph{Phys. Rev. B}, 59:\penalty0 2070, 1999.

\bibitem[{B. E. Kane}(1998)]{kane}
{B. E. Kane}.
\newblock {A silicon-based nuclear spin quantum computer}.
\newblock \emph{Nature}, 393:\penalty0 133, 1998.

\bibitem[{V. Privmana, I. D. Vagnerb, and G. Kventsel}(1998)]{privmana}
{V. Privmana, I. D. Vagnerb, and G. Kventsel}.
\newblock {Quantum computation in quantum-Hall systems}.
\newblock \emph{Phys. Lett. A}, 239:\penalty0 141, 1998.

\bibitem[{D. Jaksch, C. Bruder, J. I. Cirac, C. W. Gardiner, and P.
  Zoller}(1998)]{jaksch98}
{D. Jaksch, C. Bruder, J. I. Cirac, C. W. Gardiner, and P. Zoller}.
\newblock {Cold bosonic atoms in optical lattices}.
\newblock \emph{Phys. Rev. Lett.}, 81:\penalty0 3108, 1998.

\bibitem[{G. K. Brennen, C. M. Caves, P. S. Jessen, and I. H.
  Deutsch}(1999)]{brennen99}
{G. K. Brennen, C. M. Caves, P. S. Jessen, and I. H. Deutsch}.
\newblock {Quantum logic gates in optical lattices}.
\newblock \emph{Phys. Rev. Lett.}, 82:\penalty0 1060, 1999.

\bibitem[{Y. Makhlin, G. Sch\"{o}n, and A. Shnirman}(2001)]{makhlin}
{Y. Makhlin, G. Sch\"{o}n, and A. Shnirman}.
\newblock {Quantum-state engineering with Josephson-junction devices}.
\newblock \emph{Rev. Mod. Phys.}, 73:\penalty0 357, 2001.

\bibitem[{J. M. Martinis, S. Nam, J. Aumentado, and C. Urbina}(2002)]{martinis}
{J. M. Martinis, S. Nam, J. Aumentado, and C. Urbina}.
\newblock {Rabi oscillations in a large Josephson-junction qubit}.
\newblock \emph{Phys. Rev. Lett.}, 89:\penalty0 117901, 2002.

\bibitem[{I. Chiorescu, Y. Nakamura, C. J. P. M. Harmans, and J. E.
  Mooij}(2003)]{chiorescu}
{I. Chiorescu, Y. Nakamura, C. J. P. M. Harmans, and J. E. Mooij}.
\newblock {Coherent quantum dynamics of a superconducting flux qubit}.
\newblock \emph{Science}, 299:\penalty0 1869, 2003.

\bibitem[{M. H. Devoret, A. Wallraff, and J. M. Martinis}(2004)]{devoret}
{M. H. Devoret, A. Wallraff, and J. M. Martinis}.
\newblock {Superconducting qubits: A short review}.
\newblock 2004.
\newblock e-print arXiv:cond-mat/0411174.

\bibitem[{A. Imamo\u{g}lu, D. D. Awschalom, G. Burkard, D. P. DiVincenzo, D.
  Loss, M. Sherwin, and A. Small}(1999)]{imamoglu99}
{A. Imamo\u{g}lu, D. D. Awschalom, G. Burkard, D. P. DiVincenzo, D. Loss, M.
  Sherwin, and A. Small}.
\newblock {Quantum information processing using quantum dot spins and cavity
  QED}.
\newblock \emph{Phys. Rev. Lett.}, 83:\penalty0 4204, 1999.

\bibitem[{J. M. Elzerman, R. Hanson, L. H. Willems Van Bereven, B. Witkamp, L.
  M. Vandersypen, and L. P. Kouwenhoven}(2004)]{elzerman}
{J. M. Elzerman, R. Hanson, L. H. Willems Van Bereven, B. Witkamp, L. M.
  Vandersypen, and L. P. Kouwenhoven}.
\newblock {Single-shot read-out of an individual electron spin in a quantum
  dot}.
\newblock \emph{Nature}, 430:\penalty0 431, 2004.

\bibitem[{E. Knill, R. Laflamme, and G. J. Milburn}(2001)]{knill}
{E. Knill, R. Laflamme, and G. J. Milburn}.
\newblock {A scheme for efficient quantum computation with linear optics}.
\newblock \emph{Nature}, 409:\penalty0 46, 2001.

\bibitem[{P. Kok, W. J. Munro, K. Nemoto, T. C. Ralph, J. P. Dowling, and G. J.
  Milburn}(2007)]{koklinear}
{P. Kok, W. J. Munro, K. Nemoto, T. C. Ralph, J. P. Dowling, and G. J.
  Milburn}.
\newblock {Linear optical quantum computing with photonic qubits}.
\newblock \emph{Rev. Mod. Phys.}, 79:\penalty0 135, 2007.

\bibitem[{L.-M. Duan, M. D. Lukin, J. I. Cirac, and P. Zoller}(2001)]{duan}
{L.-M. Duan, M. D. Lukin, J. I. Cirac, and P. Zoller}.
\newblock {Long-distance quantum communication with atomic ensembles and linear
  optics}.
\newblock \emph{Nature}, 414:\penalty0 413, 2001.

\bibitem[{P. W. Shor}(1996)]{shor96}
{P. W. Shor}.
\newblock {Fault-tolerant quantum computation}.
\newblock In \emph{{FOCS '96 Proceedings of the 37th Annual Symposium on
  Foundations of Computer Science}}, pages 56–--65, 1996.

\bibitem[{L. M. K. Vandersypen, M. Steffen, G. Breyta, C. S. Yannoni, M. H.
  Sherwood, and I. L. Chuang}(2001)]{vandersypen}
{L. M. K. Vandersypen, M. Steffen, G. Breyta, C. S. Yannoni, M. H. Sherwood,
  and I. L. Chuang}.
\newblock {Experimental realization of Shor's quantum factoring algorithm using
  nuclear magnetic resonance}.
\newblock \emph{Nature}, 414:\penalty0 883, 2001.

\bibitem[{I. L. Chuang, N. Gershenfeld, and M. Kubinec}(1998)]{chuang1}
{I. L. Chuang, N. Gershenfeld, and M. Kubinec}.
\newblock {Experimental implementation of fast quantum searching}.
\newblock \emph{Phys. Rev. Lett.}, 80:\penalty0 3408, 1998.

\bibitem[{I. L. Chuang, L. M. K. Vandersypen, X. Zhou, D. W. Leung, and S.
  Lloyd}(1998)]{chuang2}
{I. L. Chuang, L. M. K. Vandersypen, X. Zhou, D. W. Leung, and S. Lloyd}.
\newblock {Experimental realization of a quantum algorithm}.
\newblock \emph{Nature}, 393:\penalty0 143, 1998.

\bibitem[{V. Danos, E. D'Hondt, E. Kashefi, and P. Panangaden}(2007)]{danos}
{V. Danos, E. D'Hondt, E. Kashefi, and P. Panangaden}.
\newblock {Distributed measurement-based quantum computation}.
\newblock \emph{Electronic Notes in Theoretical Computer Science},
  170:\penalty0 73, 2007.

\bibitem[{R. Raussendorf, D. E. Browne, and H. J. Briegel}(2003)]{rauss}
{R. Raussendorf, D. E. Browne, and H. J. Briegel}.
\newblock {Measurement-based quantum computation on cluster states}.
\newblock \emph{Phys. Rev. A}, 68:\penalty0 022312, 2003.

\bibitem[{M. Hein, J. Eisert, and H. J. Briegel}(2004)]{hein}
{M. Hein, J. Eisert, and H. J. Briegel}.
\newblock {Multiparty entanglement in graph states}.
\newblock \emph{Phys. Rev. A}, 69:\penalty0 062311, 2004.

\bibitem[{Y. L. Lim, S. D. Barrett, A. Beige, P. Kok, and L. C.
  Kwek}(2006)]{lim}
{Y. L. Lim, S. D. Barrett, A. Beige, P. Kok, and L. C. Kwek}.
\newblock {Repeat-until-success quantum computing using stationary and flying
  qubits}.
\newblock \emph{Phys. Rev. A}, 73:\penalty0 012304, 2006.

\bibitem[{Y. L. Lim, A. Beige, and L. C. Kwek}(2005)]{lim_lett}
{Y. L. Lim, A. Beige, and L. C. Kwek}.
\newblock {Repeat-until-success linear optics distributed quantum computing}.
\newblock \emph{Phys. Rev. Lett.}, 95:\penalty0 030505, 2005.

\bibitem[{J. I. Cirac, A. K. Ekert, S. F. Huelga, and C.
  Macchiavello}(1999)]{cirac}
{J. I. Cirac, A. K. Ekert, S. F. Huelga, and C. Macchiavello}.
\newblock {Distributed quantum computation over noisy channels}.
\newblock \emph{Phys. Rev. A}, 59:\penalty0 4249, 1999.

\bibitem[{R. Van Meter, K. Nemoto, and W. J. Munro}(2007)]{meter}
{R. Van Meter, K. Nemoto, and W. J. Munro}.
\newblock {Communication links for distributed quantum computation}.
\newblock \emph{IEEE Transactions on Computers}, 56:\penalty0 1643, 2007.

\bibitem[{L. K. Grover}(1997)]{grover}
{L. K. Grover}.
\newblock {Quantum telecomputation}.
\newblock \emph{e-print arXiv:quant-ph/9704012}, 1997.

\bibitem[{J. Eisert, K. Jacobs, P. Papadopoulos, and M. B.
  Plenio}(2000)]{eisert}
{J. Eisert, K. Jacobs, P. Papadopoulos, and M. B. Plenio}.
\newblock {Optimal local implementation of nonlocal quantum gates}.
\newblock \emph{Phys. Rev. A}, 62:\penalty0 052317, 2000.

\bibitem[{V. S. Denchev and G. Pandurangan}(2008)]{denchev}
{V. S. Denchev and G. Pandurangan}.
\newblock {Distributed quantum computing: A new frontier in distributed systems
  or science fiction?}
\newblock \emph{ACM SIGACT News}, 39:\penalty0 77, 2008.

\bibitem[{H. Buhrman and H. R\"{o}hrig}(2003)]{buhrman}
{H. Buhrman and H. R\"{o}hrig}.
\newblock {Distributed quantum computing}.
\newblock \emph{Mathematical Foundations of Computer Science-Lecture Notes in
  Computer Science}, 2747:\penalty0 1, 2003.

\bibitem[{K. Hammerer, A. S. S{\o}rensen, and E. S. Polzik}(2010)]{hammerer}
{K. Hammerer, A. S. S{\o}rensen, and E. S. Polzik}.
\newblock {Quantum interface between light and atomic ensembles}.
\newblock \emph{Rev. Mod. Phys.}, 82:\penalty0 1041, 2010.

\bibitem[{S. L. Braunstein, C. A. Fuchs, and H. J. Kimble}(2000)]{braunstein}
{S. L. Braunstein, C. A. Fuchs, and H. J. Kimble}.
\newblock {Criteria for continuous-variable quantum teleportation}.
\newblock \emph{Journal of Modern Optics}, 47:\penalty0 267, 2000.

\bibitem[{S. Massar and S. Popescu}(1995)]{massar}
{S. Massar and S. Popescu}.
\newblock {Optimal extraction of information from finite quantum ensembles}.
\newblock \emph{Phys. Rev. Lett.}, 74:\penalty0 1259, 1995.

\bibitem[{M. A. Nielsen}(2004)]{nielsen_microcluster}
{M. A. Nielsen}.
\newblock {Optical quantum computation using cluster states}.
\newblock \emph{Phys. Rev. Lett.}, 93:\penalty0 040503, 2004.

\bibitem[{H. J. Briegel, D. E. Browne, W. D\"{u}r, R. Raussendorf, and M. Van
  den Nest}(2009)]{briegel09}
{H. J. Briegel, D. E. Browne, W. D\"{u}r, R. Raussendorf, and M. Van den Nest}.
\newblock {Measurement-based quantum computation}.
\newblock \emph{Nature Physics}, pages 19--26, 2009.

\bibitem[{H. J. Briegel and R. Raussendorf}(2001)]{briegel01}
{H. J. Briegel and R. Raussendorf}.
\newblock {Persistent entanglement in arrays of interacting particles}.
\newblock \emph{Phys. Rev. Lett.}, 86:\penalty0 910--913, 2001.

\bibitem[{D. Jaksch, H. J. Briegel, J. I. Cirac, C. W. Gardiner, and P.
  Zoller}(1999)]{jaksch99}
{D. Jaksch, H. J. Briegel, J. I. Cirac, C. W. Gardiner, and P. Zoller}.
\newblock {Entanglement of atoms via cold controlled collisions}.
\newblock \emph{Phys. Rev. Lett.}, 82:\penalty0 1975--1978, 1999.

\bibitem[{J. Anders, M. Hajdu\v{s}ek, D. Markham, and V.
  Vedral}(2008)]{anders08}
{J. Anders, M. Hajdu\v{s}ek, D. Markham, and V. Vedral}.
\newblock {How much of one-way computation is \textit{just} thermodynamics?}
\newblock \emph{Foundations of Physics}, 38:\penalty0 506--522, 2008.

\bibitem[{S. L. Braunstein and C. M. Caves}(1994)]{braunstein94}
{S. L. Braunstein and C. M. Caves}.
\newblock {Statistical distance and the geometry of quantum states}.
\newblock \emph{{Phys. Rev. Lett.}}, 72:\penalty0 3439--3443, 1994.

\bibitem[{U. Leonhardt}(1997)]{leonhardt}
{U. Leonhardt}.
\newblock \emph{{Measuring the Quantum State of Light}}.
\newblock Cambridge University Press, 1997.

\bibitem[{S. L. Braunstein and P. van Loock}(2005)]{braunstein05}
{S. L. Braunstein and P. van Loock}.
\newblock {Quantum information with continuous variables}.
\newblock \emph{{Rev. Mod. Phys.}}, 77:\penalty0 513, 2005.

\bibitem[{V. M. Kendon, K. Nemoto, and W. J. Munro}(2010)]{kendon}
{V. M. Kendon, K. Nemoto, and W. J. Munro}.
\newblock {Quantum analogue computing}.
\newblock \emph{{Phil. Trans. R. Soc. A}}, 368:\penalty0 3609, 2010.

\bibitem[{S. Lloyd and S. L. Braunstein}(1999)]{lloyd}
{S. Lloyd and S. L. Braunstein}.
\newblock {Quantum computation over continuous variables}.
\newblock \emph{{Phys. Rev. Lett.}}, 82:\penalty0 1784, 1999.

\bibitem[{S. L. Braunstein and H. J. Kimble}(1998)]{braunstein98}
{S. L. Braunstein and H. J. Kimble}.
\newblock {Teleportation of continuous quantum variables}.
\newblock \emph{{Phys. Rev. Lett.}}, 80:\penalty0 869, 1998.

\bibitem[{S. L. Braunstein}(1998{\natexlab{a}})]{braunstein_errora}
{S. L. Braunstein}.
\newblock {Error correction for continuous quantum variables}.
\newblock \emph{{Phys. Rev. Lett.}}, 80:\penalty0 4084, 1998{\natexlab{a}}.

\bibitem[{S. L. Braunstein}(1998{\natexlab{b}})]{braunstein_errorb}
{S. L. Braunstein}.
\newblock {Quantum error correction for communication with linear optics}.
\newblock \emph{{Nature}}, 394:\penalty0 47, 1998{\natexlab{b}}.

\bibitem[{A. Furusawa, J. L. S{\o}rensen, S. L. Braunstein, C. A. Fuchs, H. J.
  Kimble, and E. S. Polzik}(1998)]{furusawa}
{A. Furusawa, J. L. S{\o}rensen, S. L. Braunstein, C. A. Fuchs, H. J. Kimble,
  and E. S. Polzik}.
\newblock {Unconditional Quantum Teleportation}.
\newblock \emph{{Science}}, 282:\penalty0 706, 1998.

\bibitem[{T. C. Ralph}(1999)]{ralph}
{T. C. Ralph}.
\newblock {Continuous variable quantum cryptography}.
\newblock \emph{{Phys. Rev. A}}, 61:\penalty0 010303(R), 1999.

\bibitem[{M. Hillery}(2000)]{hillery}
{M. Hillery}.
\newblock {Quantum cryptography with squeezed states}.
\newblock \emph{{Phys. Rev. A}}, 61:\penalty0 022309, 2000.

\bibitem[{D. Gottesman and J. Preskill}(2001)]{gottesman}
{D. Gottesman and J. Preskill}.
\newblock {Secure quantum key distribution using squeezed states}.
\newblock \emph{{Phys. Rev. A}}, 63:\penalty0 022309, 2001.

\bibitem[{M. Fleischhauer, A. Imamo\u{g}lu, and J. P. Marangos}(2005)]{fleisch}
{M. Fleischhauer, A. Imamo\u{g}lu, and J. P. Marangos}.
\newblock {Electromagnetically induced transparency: Optics in coherent media}.
\newblock \emph{Rev. Mod. Phys.}, 77:\penalty0 633, 2005.

\bibitem[{M. D. Lukin}(2003)]{lukin}
{M. D. Lukin}.
\newblock {\textit{Colloquium}: Trapping and manipulating photon states in
  atomic ensembles}.
\newblock \emph{Rev. Mod. Phys.}, 75:\penalty0 457, 2003.

\bibitem[{S. D. Barrett, P. P. Rohde, and T. M. Stace}(2010)]{barrett}
{S. D. Barrett, P. P. Rohde, and T. M. Stace}.
\newblock {Scalable quantum computing with atomic ensembles}.
\newblock \emph{New J. Phys.}, 12:\penalty0 093032, 2010.

\bibitem[{S. Manz, T. Fernholz, J. Schmiedmayer, and J.-W. Pan}(2007)]{manz}
{S. Manz, T. Fernholz, J. Schmiedmayer, and J.-W. Pan}.
\newblock {Collisional decoherence during writing and reading quantum states}.
\newblock \emph{Phys. Rev. A}, 75:\penalty0 040101(R), 2007.

\bibitem[{I. Novikova, A. V. Gorshkov, D. F. Phillips, A. S. S{\o}rensen, M. D.
  Lukin, and R. L. Walsworth}(2007)]{novikova}
{I. Novikova, A. V. Gorshkov, D. F. Phillips, A. S. S{\o}rensen, M. D. Lukin,
  and R. L. Walsworth}.
\newblock {Optimal control of light pulse storage and retrieval}.
\newblock \emph{Phys. Rev. Lett.}, 98:\penalty0 243602, 2007.

\bibitem[{R. Pugatch, M. Shuker, O. Firstenberg, A. Ron, and N.
  Davidson}(2007)]{pugatch}
{R. Pugatch, M. Shuker, O. Firstenberg, A. Ron, and N. Davidson}.
\newblock {Topological stability of stored optical vortices}.
\newblock \emph{Phys. Rev. Lett.}, 98:\penalty0 203601, 2007.

\bibitem[{T. Hong, A. V. Gorshkov, D. Patterson, A. S. Zibrov, J. M. Doyle, M.
  D. Lukin, and M. G. Prentiss}(2009)]{thong}
{T. Hong, A. V. Gorshkov, D. Patterson, A. S. Zibrov, J. M. Doyle, M. D. Lukin,
  and M. G. Prentiss}.
\newblock {Realization of coherent optically dense media via buffer-gas
  cooling}.
\newblock \emph{Phys. Rev. A}, 79:\penalty0 013806, 2009.

\bibitem[{C. M. Caves}(1981)]{caves}
{C. M. Caves}.
\newblock {Quantum-mechanical noise in an interferometer}.
\newblock \emph{{Phys. Rev. D}}, 23:\penalty0 1693, 1981.

\bibitem[{C. W. Helstrom}(1967)]{helstrom67}
{C. W. Helstrom}.
\newblock {Minimum mean-squared error of estimates in quantum statistics}.
\newblock \emph{{Phys. Letters}}, 25A:\penalty0 101--102, 1967.

\bibitem[{C. W. Helstrom}(1976)]{helstrom76}
{C. W. Helstrom}.
\newblock \emph{{Quantum Detection and Estimation Theory}}.
\newblock Academic Press, New York, 1976.

\bibitem[{C. W. Gardiner and P. Zoller}(2010)]{gardiner04}
{C. W. Gardiner and P. Zoller}.
\newblock \emph{{Quantum Noise}}.
\newblock Springer-Verlag, {3rd} edition, 2010.

\bibitem[{M. J. Holland and K. Burnett}(1993)]{holland93}
{M. J. Holland and K. Burnett}.
\newblock {Interferometric detection of optical phase shifts at the Heisenberg
  limit}.
\newblock \emph{{Phys. Rev. Lett.}}, 71:\penalty0 1355--1358, 1993.

\bibitem[{J. Beltr\'{a}n and A. Luis}(2005)]{luis05}
{J. Beltr\'{a}n and A. Luis}.
\newblock {Breaking the Heisenberg limit with inefficient detectors}.
\newblock \emph{{Phys. Rev. A}}, 72:\penalty0 045801, 2005.

\bibitem[{S. Boixo, S. T. Flammia, C. M. Caves, and J. M.
  Geremia}(2007)]{boixo}
{S. Boixo, S. T. Flammia, C. M. Caves, and J. M. Geremia}.
\newblock {Generalized limits for single-parameter quantum estimation}.
\newblock \emph{{Phys. Rev. Lett.}}, 98:\penalty0 090401, 2007.

\bibitem[{S. Roy and S. L. Braunstein}(2008)]{Roy08}
{S. Roy and S. L. Braunstein}.
\newblock {Exponentially enhanced quantum metrology}.
\newblock \emph{{Phys. Rev. Lett.}}, 100:\penalty0 220501, 2008.

\bibitem[{V. Giovannetti, S. Lloyd, and L. Maccone}(2006)]{Giovannetti06}
{V. Giovannetti, S. Lloyd, and L. Maccone}.
\newblock {Quantum metrology}.
\newblock \emph{{Phys. Rev. Lett.}}, 96:\penalty0 010401, 2006.

\bibitem[{G. Breit and E. Wigner}(1936)]{breit}
{G. Breit and E. Wigner}.
\newblock {Capture of slow neutrons}.
\newblock \emph{{Phys. Rev.}}, 49:\penalty0 519–--531, 1936.

\bibitem[{J. Uffink}(1993)]{uffink93}
{J. Uffink}.
\newblock {The rate of evolution of a quantum state}.
\newblock \emph{{Am. J. Phys.}}, 61:\penalty0 935--936, 1993.

\bibitem[{A. Bhattacharyya}(1943)]{bhatta43}
{A. Bhattacharyya}.
\newblock {On a measure of divergence between two statistical populations
  defined by probability distributions}.
\newblock \emph{{Bull. Calcutta Math. Soc.}}, 35:\penalty0 99--109, 1943.

\bibitem[{W. K. Wootters}(1981)]{wootters81}
{W. K. Wootters}.
\newblock {Statistical distance and Hilbert space}.
\newblock \emph{{Phys. Rev. D}}, 23:\penalty0 357–--362, 1981.

\bibitem[{J. Anandan and Y. Aharonov}(1990)]{anandan}
{J. Anandan and Y. Aharonov}.
\newblock {Geometry of quantum evolution}.
\newblock \emph{{Phys. Rev. Lett.}}, 65:\penalty0 1697--1700, 1990.

\bibitem[{S. L. Braunstein, C. M. Caves, and G. J.
  Milburn}(1996)]{braunstein96}
{S. L. Braunstein, C. M. Caves, and G. J. Milburn}.
\newblock {Generalized uncertainty relations: Theory, examples, and Lorentz
  invariance}.
\newblock \emph{{Ann. Phys.}}, 247:\penalty0 135--173, 1996.

\bibitem[{W. Heitler}(1954)]{heitler54}
{W. Heitler}.
\newblock \emph{{The Quantum Theory of Radiation}}.
\newblock Clarendon Press, Oxford, {3rd} edition, 1954.

\bibitem[{L. Mandelstam and I. Tamm}(1945)]{mandelstam45}
{L. Mandelstam and I. Tamm}.
\newblock {The uncertainty relation between energy and time in non-relativistic
  quantum mechanics}.
\newblock \emph{{J. Phys. (USSR)}}, 9:\penalty0 249–--254, 1945.

\bibitem[{A. K. Pati and S. L. Braunstein}(2003)]{pati}
{A. K. Pati and S. L. Braunstein}.
\newblock {Deutsch-Jozsa algorithm for continuous variables}.
\newblock In \emph{{Quantum Information with Continuous Variables}}. {Kluwer
  Academic Publisher}, 2003.

\bibitem[{M. R. A. Adcock, P. H{\o}yer, and B. C. Sanders}(2009)]{sanders09}
{M. R. A. Adcock, P. H{\o}yer, and B. C. Sanders}.
\newblock {Limitations on continuous variable quantum algorithms with Fourier
  transforms}.
\newblock \emph{{New J. Phys.}}, 11:\penalty0 103035, 2009.

\bibitem[{D. Deutsch and R. Jozsa}(1992)]{deutsch}
{D. Deutsch and R. Jozsa}.
\newblock {Rapid solution of problems by quantum computation}.
\newblock In \emph{{Proceedings of the Royal Society of London}}, volume A439,
  pages 553–--558, 1992.

\bibitem[{R. Cleve, A. Ekert, C. Macchiavello, and M. Mosca}(1998)]{cleve}
{R. Cleve, A. Ekert, C. Macchiavello, and M. Mosca}.
\newblock {Quantum algorithms revisited}.
\newblock In \emph{{Proceedings of the Royal Society of London}}, volume A454,
  pages 339--354, 1998.
\newblock e-print arXiv:quant-ph/9708016.

\bibitem[{R. Bl\"{u}mel}(2009)]{blumel}
{R. Bl\"{u}mel}.
\newblock \emph{{Foundations of Quantum Mechanics: from Photons to Quantum
  Computers}}.
\newblock Jones and Bartlett Publishers, Inc, 2009.

\bibitem[{C. W. Helstrom}(1969)]{helstrom69}
{C. W. Helstrom}.
\newblock {Quantum detection and estimation theory}.
\newblock \emph{{Journal of Statistical Physics}}, 1:\penalty0 231--252, 1969.

\bibitem[{M. Zwierz, C. A. P\'{e}rez-Delgado, and P. Kok}(2010)]{zwierz10}
{M. Zwierz, C. A. P\'{e}rez-Delgado, and P. Kok}.
\newblock {General optimality of the Heisenberg limit for quantum metrology}.
\newblock \emph{{Phys. Rev. Lett.}}, 105:\penalty0 180402, 2010.

\bibitem[{T. Tilma, S. Hamaji, W. J. Munro, and K. Nemoto}(2010)]{tilma10}
{T. Tilma, S. Hamaji, W. J. Munro, and K. Nemoto}.
\newblock {Entanglement is not a critical resource for quantum metrology}.
\newblock \emph{{Phys. Rev. A}}, 81:\penalty0 022108, 2010.

\bibitem[{O. Pinel, J. Fade, N. Treps, and C. Fabre}(2010)]{pinel10}
{O. Pinel, J. Fade, N. Treps, and C. Fabre}.
\newblock {General Cram\'{e}r-Rao bound for parameter estimation using Gaussian
  multimode quantum resources}.
\newblock 2010.
\newblock e-print arXiv:1008.0844.

\bibitem[{S. E. Harris, J. E. Field, and A. Imamo\u{g}lu}(1990)]{harris}
{S. E. Harris, J. E. Field, and A. Imamo\u{g}lu}.
\newblock {Nonlinear optical processes using electromagnetically induced
  transparency}.
\newblock \emph{Phys. Rev. Lett.}, 64:\penalty0 1107, 1990.

\bibitem[{S. E. Harris}(1997)]{harris_phys}
{S. E. Harris}.
\newblock {Electromagnetically induced transparency}.
\newblock \emph{Phys. Today}, 50:\penalty0 36, 1997.

\bibitem[{R. G. Beausoleil, W. J. Munro, and T. P. Spiller}(2004)]{munro}
{R. G. Beausoleil, W. J. Munro, and T. P. Spiller}.
\newblock {Applications of coherent population transfer to quantum information
  processing}.
\newblock \emph{Topic Review in JMO}, 51:\penalty0 1559, 2004.

\bibitem[{E. Arimondo}(1997)]{arimondo}
{E. Arimondo}.
\newblock {Coherent population trapping in laser spectroscopy}.
\newblock \emph{Progress in Optics}, 35:\penalty0 259, 1997.

\bibitem[{K.-J. Boller, A. Imamo\u{g}lu, and S. E. Harris}(1991)]{boller}
{K.-J. Boller, A. Imamo\u{g}lu, and S. E. Harris}.
\newblock {Observation of electromagnetically induced transparency}.
\newblock \emph{Phys. Rev. Lett.}, 66:\penalty0 2593, 1991.

\bibitem[{M. Fleischhauer and M. D. Lukin}(2000)]{fleisch_dark}
{M. Fleischhauer and M. D. Lukin}.
\newblock {Dark-state polaritons in electromagnetically induced transparency}.
\newblock \emph{Phys. Rev. Lett.}, 84:\penalty0 5094, 2000.

\bibitem[{S. E. Harris, J. E. Field, and A. Kasapi}(1992)]{harris_PRA}
{S. E. Harris, J. E. Field, and A. Kasapi}.
\newblock {Dispersive properties of electromagnetically induced transparency}.
\newblock \emph{Phys. Rev. A}, 46:\penalty0 R29, 1992.

\bibitem[{C. Liu, Z. Dutton, C. H. Behroozi, and L. V. Hau}(2001)]{liu}
{C. Liu, Z. Dutton, C. H. Behroozi, and L. V. Hau}.
\newblock {Observation of coherent optical information storage in an atomic
  medium using halted light pulses}.
\newblock \emph{Nature}, 409:\penalty0 490, 2001.

\bibitem[{D. F. Phillips, A. Fleischhauer, A. Mair, R. L. Walsworth, and M. D.
  Lukin}(2001)]{phillips}
{D. F. Phillips, A. Fleischhauer, A. Mair, R. L. Walsworth, and M. D. Lukin}.
\newblock {Storage of light in atomic vapor}.
\newblock \emph{Phys. Rev. Lett.}, 86:\penalty0 783, 2001.

\bibitem[{A. V. Turukhin, V. S. Sudarshanam, M. S. Shahriar, J. A. Musser, B.
  S. Ham, and P. R. Hemmer}(2001)]{turukhin}
{A. V. Turukhin, V. S. Sudarshanam, M. S. Shahriar, J. A. Musser, B. S. Ham,
  and P. R. Hemmer}.
\newblock {Observation of ultraslow and stored light pulses in a solid}.
\newblock \emph{Phys. Rev. Lett.}, 88:\penalty0 023602, 2001.

\bibitem[{L. V. Hau, S. E. Harris, Z. Dutton, and C. H. Behroozi}(1999)]{hau}
{L. V. Hau, S. E. Harris, Z. Dutton, and C. H. Behroozi}.
\newblock {Light speed reduction to 17 metres per second in an ultracold atomic
  gas}.
\newblock \emph{Nature}, 397:\penalty0 594, 1999.

\bibitem[{M. D. Lukin, S. F. Yelin, and M. Fleischhauer}(2000)]{lukin_PRL}
{M. D. Lukin, S. F. Yelin, and M. Fleischhauer}.
\newblock {Entanglement of atomic ensembles by trapping correlated photon
  states}.
\newblock \emph{Phys. Rev. Lett.}, 84:\penalty0 4232, 2000.

\bibitem[{M. Fleischhauer and M. D. Lukin}(2002)]{fleisch_darkPRA}
{M. Fleischhauer and M. D. Lukin}.
\newblock {Quantum memory for photons: Dark-state polaritons}.
\newblock \emph{Phys. Rev. A}, 65:\penalty0 022314, 2002.

\bibitem[{C. Mewes and M. Fleischhauer}(2002)]{mewes}
{C. Mewes and M. Fleischhauer}.
\newblock {Two-photon linewidth of light "stopping" via electromagnetically
  induced transparency}.
\newblock \emph{Phys. Rev. A}, 66:\penalty0 033820, 2002.

\bibitem[{A. Kuzmich, W. P. Bowen, A. D. Boozer, A. Boca, C. W. Chou, L.-M.
  Duan, and H. J. Kimble}(2003)]{kuzmich}
{A. Kuzmich, W. P. Bowen, A. D. Boozer, A. Boca, C. W. Chou, L.-M. Duan, and H.
  J. Kimble}.
\newblock {Generation of nonclassical photon pairs for scalable quantum
  communication with atomic ensembles}.
\newblock \emph{Nature}, 423:\penalty0 731, 2003.

\bibitem[{C. H. van der Wal, M. D. Eisaman, A. Andr\'{e}, R. L. Walsworth, D.
  F. Phillips, A. S. Zibrov, and M. D. Lukin}(2003)]{wal}
{C. H. van der Wal, M. D. Eisaman, A. Andr\'{e}, R. L. Walsworth, D. F.
  Phillips, A. S. Zibrov, and M. D. Lukin}.
\newblock {Atomic memory for correlated photon states}.
\newblock \emph{Science}, 301:\penalty0 196, 2003.

\bibitem[{J. Laurat, K. S. Choi, H. Deng, C. W. Chou, and H. J.
  Kimble}(2007)]{laurat}
{J. Laurat, K. S. Choi, H. Deng, C. W. Chou, and H. J. Kimble}.
\newblock {Heralded entanglement between atomic ensembles: Preparation,
  decoherence, and scaling}.
\newblock \emph{Phys. Rev. Lett.}, 99:\penalty0 180504, 2007.

\bibitem[{D. Felinto, C. W. Chou, H. de Riedmatten, S. V. Polyakov, and H. J.
  Kimble}(2005)]{felinto}
{D. Felinto, C. W. Chou, H. de Riedmatten, S. V. Polyakov, and H. J. Kimble}.
\newblock {Control of decoherence in the generation of photon pairs from atomic
  ensembles}.
\newblock \emph{Phys. Rev. A}, 72:\penalty0 053809, 2005.

\bibitem[{T. Chaneli\`{e}re, D. N. Matsukevich, S. D. Jenkins, S.-Y. Lan, T. A.
  B. Kennedy, and A. Kuzmich}(2005)]{chaneliere}
{T. Chaneli\`{e}re, D. N. Matsukevich, S. D. Jenkins, S.-Y. Lan, T. A. B.
  Kennedy, and A. Kuzmich}.
\newblock {Storage and retrieval of single photons transmitted between remote
  quantum memories}.
\newblock \emph{Nature}, 438:\penalty0 833, 2005.

\bibitem[{M. D. Eisaman, A. Andr\'{e}, F. Massou, M. Fleischhauer, A. S.
  Zibrov, and M. D. Lukin}(2005)]{eisaman}
{M. D. Eisaman, A. Andr\'{e}, F. Massou, M. Fleischhauer, A. S. Zibrov, and M.
  D. Lukin}.
\newblock {Electromagnetically induced transparency with tunable single-photon
  pulses}.
\newblock \emph{Nature}, 438:\penalty0 837, 2005.

\bibitem[{K. S. Choi, H. Deng, J. Laurat, and H. J. Kimble}(2008)]{choi}
{K. S. Choi, H. Deng, J. Laurat, and H. J. Kimble}.
\newblock {Mapping photonic entanglement into and out of a quantum memory}.
\newblock \emph{Nature}, 452:\penalty0 67, 2008.

\bibitem[{D. F. V. James and P. G. Kwiat}(2002)]{james}
{D. F. V. James and P. G. Kwiat}.
\newblock {Atomic-vapor-based high efficiency optical detectors with photon
  number resolution}.
\newblock \emph{Phys. Rev. Lett.}, 89:\penalty0 183601, 2002.

\bibitem[{A. Imamo\u{g}lu}(2002)]{imam}
{A. Imamo\u{g}lu}.
\newblock {High efficiency photon counting using stored light}.
\newblock \emph{Phys. Rev. Lett.}, 89:\penalty0 163602, 2002.

\bibitem[{T. F. Gallagher}(1994)]{gallagher}
{T. F. Gallagher}.
\newblock \emph{{Rydberg Atoms}}.
\newblock Cambridge University Press, 1994.

\bibitem[{W. Li, I. Mourachko, M. W. Noel, and T. F. Gallagher}(2003)]{li}
{W. Li, I. Mourachko, M. W. Noel, and T. F. Gallagher}.
\newblock {Millimeter-wave spectroscopy of cold Rb Rydberg atoms in a
  magneto-optical trap: Quantum defects of the ns, np, and nd series}.
\newblock \emph{Phys. Rev. A}, 67:\penalty0 052502, 2003.

\bibitem[{C. Gerry and P. Knight}(2005)]{gerry}
{C. Gerry and P. Knight}.
\newblock \emph{{Introductory Quantum Optics}}.
\newblock Cambridge University Press, 2005.

\bibitem[{T. A. Johnson, E. Urban, T. Henage, L. Isenhower, D. D. Yavuz, T. G.
  Walker, and M. Saffman}(2008)]{johnson}
{T. A. Johnson, E. Urban, T. Henage, L. Isenhower, D. D. Yavuz, T. G. Walker,
  and M. Saffman}.
\newblock {Rabi oscillations between ground and Rydberg states with
  dipole-dipole atomic interactions}.
\newblock \emph{Phys. Rev. Lett.}, 100:\penalty0 113003, 2008.

\bibitem[{C. S. E. van Ditzhuijzen, A. F. Koenderink, J. V. Hern\'{a}ndez, F.
  Robicheaux, L. D. Noordam, and H. B. van Linden van den Heuvell}(2008)]{van}
{C. S. E. van Ditzhuijzen, A. F. Koenderink, J. V. Hern\'{a}ndez, F.
  Robicheaux, L. D. Noordam, and H. B. van Linden van den Heuvell}.
\newblock {Spatially resolved observation of dipole-dipole interaction between
  Rydberg atoms}.
\newblock \emph{Phys. Rev. Lett.}, 100:\penalty0 243201, 2008.

\bibitem[{T. G. Walker and M. Saffman}(2008)]{walkerbloc}
{T. G. Walker and M. Saffman}.
\newblock {Consequences of Zeeman degeneracy for the van der Waals blockade
  between Rydberg atoms}.
\newblock \emph{Phys. Rev. A}, 77:\penalty0 032723, 2008.

\bibitem[{T. G. Walker and M. Saffman}(2005)]{walker}
{T. G. Walker and M. Saffman}.
\newblock {Zeros of Rydberg-Rydberg F\"{o}ster interactions}.
\newblock \emph{J. Phys. B}, 38:\penalty0 S309, 2005.

\bibitem[{M. Saffman, T. G. Walker, and K. M{\o}lmer}(2010)]{saffman10}
{M. Saffman, T. G. Walker, and K. M{\o}lmer}.
\newblock {Quantum information with Rydberg atoms}.
\newblock \emph{Rev. Mod. Phys.}, 82:\penalty0 2313, 2010.

\bibitem[{R. C. Stoneman, M. D. Adams, and T. F. Gallagher}(1987)]{stoneman}
{R. C. Stoneman, M. D. Adams, and T. F. Gallagher}.
\newblock {Resonant-collision spectroscopy of Rydberg atoms}.
\newblock \emph{Phys. Rev. Lett.}, 58:\penalty0 1324, 1987.

\bibitem[{D. Jaksch, J. I. Cirac, P. Zoller, S. L. Rolston, R. C\^{o}t\'{e},
  and M. D. Lukin}(2000)]{jaksch}
{D. Jaksch, J. I. Cirac, P. Zoller, S. L. Rolston, R. C\^{o}t\'{e}, and M. D.
  Lukin}.
\newblock {Fast quantum gates for neutral atoms}.
\newblock \emph{Phys. Rev. Lett.}, 85:\penalty0 2208, 2000.

\bibitem[{M. D. Lukin, M. Fleischhauer, R. C\^{o}t\'{e}, L.-M. Duan, D. Jaksch,
  J. I. Cirac, and P. Zoller}(2001)]{lukingate}
{M. D. Lukin, M. Fleischhauer, R. C\^{o}t\'{e}, L.-M. Duan, D. Jaksch, J. I.
  Cirac, and P. Zoller}.
\newblock {Dipole blockade and quantum information processing in mesoscopic
  atomic ensembles}.
\newblock \emph{Phys. Rev. Lett.}, 87:\penalty0 037901, 2001.

\bibitem[{E. Brion, K. M{\o}lmer, and M. Saffman}(2007)]{brion}
{E. Brion, K. M{\o}lmer, and M. Saffman}.
\newblock {Quantum computing with collective ensembles of multilevel systems}.
\newblock \emph{Phys. Rev. Lett.}, 99:\penalty0 260501, 2007.

\bibitem[{I. Friedler, D. Petrosyan, M. Fleischhauer, and G.
  Kurizki}(2005)]{friedler}
{I. Friedler, D. Petrosyan, M. Fleischhauer, and G. Kurizki}.
\newblock {Long-range interactions and entanglement of slow single-photon
  pulses}.
\newblock \emph{Phys. Rev. A}, 72:\penalty0 043803, 2005.

\bibitem[{A. K. Mohapatra, T. R. Jackson, and C. S. Adams}(2007)]{mohapatra}
{A. K. Mohapatra, T. R. Jackson, and C. S. Adams}.
\newblock {Coherent optical detection of highly excited Rydberg states using
  electromagnetically induced transparency}.
\newblock \emph{Phys. Rev. Lett.}, 98:\penalty0 113003, 2007.

\bibitem[{D. Petrosyan and M. Fleischhauer}(2008)]{petrosyan}
{D. Petrosyan and M. Fleischhauer}.
\newblock {Quantum information processing with single photons and atomic
  ensembles in microwave coplanar waveguide resonators}.
\newblock \emph{Phys. Rev. Lett.}, 100:\penalty0 170501, 2008.

\bibitem[{M. Zwierz and P. Kok}(2009)]{zwierz}
{M. Zwierz and P. Kok}.
\newblock {High-efficiency cluster-state generation with atomic ensembles via
  the dipole-blockade mechanism}.
\newblock \emph{Phys. Rev. A}, 79:\penalty0 022304, 2009.

\bibitem[{S. D. Barrett and P. Kok}(2005)]{kok}
{S. D. Barrett and P. Kok}.
\newblock {Efficient high-fidelity quantum computation using matter qubits and
  linear optics}.
\newblock \emph{Phys. Rev. A}, 71:\penalty0 060310(R), 2005.

\bibitem[{S. J. van Enk, N. L\"{u}tkenhaus, and H. J. Kimble}(2007)]{vanEnk}
{S. J. van Enk, N. L\"{u}tkenhaus, and H. J. Kimble}.
\newblock {Experimental procedures for entanglement verification}.
\newblock \emph{Phys. Rev. A}, 75:\penalty0 052318, 2007.

\bibitem[{C. W. Chou, S. V. Polyakov, A. Kuzmich, and H. J.
  Kimble}(2004)]{chou}
{C. W. Chou, S. V. Polyakov, A. Kuzmich, and H. J. Kimble}.
\newblock {Single-photon generation from stored excitation in an atomic
  ensemble}.
\newblock \emph{Phys. Rev. Lett.}, 92:\penalty0 213601, 2004.

\bibitem[{C. W. Chou, H. de Riedmatten, D. Felinto, S. V. Polyakov, S. J. van
  Enk, and H. J. Kimble}(2005)]{chou_nature}
{C. W. Chou, H. de Riedmatten, D. Felinto, S. V. Polyakov, S. J. van Enk, and
  H. J. Kimble}.
\newblock {Measurement-induced entanglement for excitation stored in remote
  atomic ensembles}.
\newblock \emph{Nature}, 438:\penalty0 828, 2005.

\bibitem[{C. W. Chou, J. Laurat, H. Deng, K. S. Choi, H. de Riedmatten, D.
  Felinto, and H. J. Kimble}(2007)]{chou_science}
{C. W. Chou, J. Laurat, H. Deng, K. S. Choi, H. de Riedmatten, D. Felinto, and
  H. J. Kimble}.
\newblock {Functional quantum nodes for entanglement distribution over scalable
  quantum networks}.
\newblock \emph{Science}, 316:\penalty0 1316, 2007.

\bibitem[{Z.-S. Yuan, Y.-A. Chen, B. Zhao, S. Chen, J. Schmiedmayer, and J.-W.
  Pan}(2008)]{yuan}
{Z.-S. Yuan, Y.-A. Chen, B. Zhao, S. Chen, J. Schmiedmayer, and J.-W. Pan}.
\newblock {Experimental demonstration of a BDCZ quantum repeater node}.
\newblock \emph{Nature}, 454:\penalty0 1098, 2008.

\bibitem[{Y.-A. Chen, S. Chen, Z.-S. Yuan, B. Zhao, C.-S. Chuu, J.
  Schmiedmayer, and J.-W. Pan}(2008)]{chen}
{Y.-A. Chen, S. Chen, Z.-S. Yuan, B. Zhao, C.-S. Chuu, J. Schmiedmayer, and
  J.-W. Pan}.
\newblock {Memory-built-in quantum teleportation with photonic and atomic
  qubits}.
\newblock \emph{Nature Physics}, 4:\penalty0 103, 2008.

\bibitem[{S. C. Benjamin}(2005)]{benjamin}
{S. C. Benjamin}.
\newblock {Comment on "Efficient high-fidelity quantum computation using matter
  qubits and linear optics"}.
\newblock \emph{Phys. Rev. A}, 72:\penalty0 056302, 2005.

\bibitem[{S. C. Benjamin, D. E. Browne, J. Fitzsimons, and J. J. L.
  Morton}(2006)]{benjamin_broker}
{S. C. Benjamin, D. E. Browne, J. Fitzsimons, and J. J. L. Morton}.
\newblock {Brokered graph-state quantum computation}.
\newblock \emph{New J. Phys.}, 8:\penalty0 141, 2006.

\bibitem[{D. L. Moehring, P. Maunz, S. Olmschenk, K. C. Younge, D. N.
  Matsukevich, L.-M. Duan, and C. Monroe}(2007)]{moehring}
{D. L. Moehring, P. Maunz, S. Olmschenk, K. C. Younge, D. N. Matsukevich, L.-M.
  Duan, and C. Monroe}.
\newblock {Entanglement of single-atom quantum bits at a distance}.
\newblock \emph{Nature}, 449:\penalty0 68, 2007.

\bibitem[{D. N. Matsukevich, P. Maunz, D. L. Moehring, S. Olmschenk, and C.
  Monroe}(2008)]{matsukevich}
{D. N. Matsukevich, P. Maunz, D. L. Moehring, S. Olmschenk, and C. Monroe}.
\newblock {Bell inequality violation with two remote atomic qubits}.
\newblock \emph{Phys. Rev. Lett.}, 100:\penalty0 150404, 2008.

\bibitem[{M. Saffman and T. G. Walker}(2002)]{saffman}
{M. Saffman and T. G. Walker}.
\newblock {Creating single-atom and single-photon sources from entangled atomic
  ensembles}.
\newblock \emph{Phys. Rev. A}, 66:\penalty0 065403, 2002.

\bibitem[{J. O. Day, E. Brekke, and T. G. Walker}(2008)]{day}
{J. O. Day, E. Brekke, and T. G. Walker}.
\newblock {Dynamics of low-density ultracold Rydberg gases}.
\newblock \emph{Phys. Rev. A}, 77:\penalty0 052712, 2008.

\bibitem[{C. K. Hong, Z. Y. Ou, and L. Mendel}(1987)]{hong}
{C. K. Hong, Z. Y. Ou, and L. Mendel}.
\newblock {Measurement of subpicosecond time intervals between two photons by
  interference}.
\newblock \emph{Phys. Rev. Lett.}, 59:\penalty0 2044, 1987.

\bibitem[{R. Ghosh and L. Mendel}(1987)]{ghosh}
{R. Ghosh and L. Mendel}.
\newblock {Observation of nonclassical effects in the interference of two
  photons}.
\newblock \emph{Phys. Rev. Lett.}, 59:\penalty0 1903, 1987.

\bibitem[{K. Kieling, D. Gross, and J. Eisert}(2007)]{kieling}
{K. Kieling, D. Gross, and J. Eisert}.
\newblock {Cluster state preparation using gates operating at arbitrary success
  probabilities}.
\newblock \emph{New J. Phys.}, 9:\penalty0 200, 2007.

\bibitem[{M. Halder, A. Beveratos, R. T. Thew, C. Jorel, H. Zbinden, and N.
  Gisin}(2008)]{halder}
{M. Halder, A. Beveratos, R. T. Thew, C. Jorel, H. Zbinden, and N. Gisin}.
\newblock {High coherence photon pair source for quantum communication}.
\newblock \emph{New J. Phys.}, 10:\penalty0 023027, 2008.

\bibitem[{J. K. Thompson, J. Simon, H. Loh, and V. Vuletic}(2006)]{thompson}
{J. K. Thompson, J. Simon, H. Loh, and V. Vuletic}.
\newblock {A high-brightness source of narrowband, identical-photon pairs}.
\newblock \emph{Science}, 313:\penalty0 74, 2006.

\bibitem[{J. S. Neergaard-Nielsen, B. M. Nielsen, H. Takahashi, A. I. Vistnes,
  and E. S. Polzik}(2007)]{neergaard}
{J. S. Neergaard-Nielsen, B. M. Nielsen, H. Takahashi, A. I. Vistnes, and E. S.
  Polzik}.
\newblock {High purity bright single photon source}.
\newblock \emph{Opt. Express}, 15:\penalty0 7940, 2007.

\bibitem[{X.-H. Bao, Y. Qian, J. Yang, H. Zhang, Z.-B. Chen, T. Yang, and J.-W.
  Pan}(2008)]{bao}
{X.-H. Bao, Y. Qian, J. Yang, H. Zhang, Z.-B. Chen, T. Yang, and J.-W. Pan}.
\newblock {Generation of narrow-band polarization-entangled photon pairs for
  atomic quantum memories}.
\newblock \emph{Phys. Rev. Lett.}, 101:\penalty0 190501, 2008.

\bibitem[{H. Kim, O. Kwon, W. Kim, and T. Kim}(2006)]{kim}
{H. Kim, O. Kwon, W. Kim, and T. Kim}.
\newblock {Spatial two-photon interference in a Hong-Ou-Mandel interferometer}.
\newblock \emph{Phys. Rev. A}, 73:\penalty0 023820, 2006.

\bibitem[{M. Barbieri, C. Cinelli, P. Mataloni, and F. De
  Martini}(2005)]{barbieri}
{M. Barbieri, C. Cinelli, P. Mataloni, and F. De Martini}.
\newblock {Polarization-momentum hyperentangled states: Realization and
  characterization}.
\newblock \emph{Phys. Rev. A}, 72:\penalty0 052110, 2005.

\bibitem[{A. Rossi, G. Vallone, A. Chiuri, F. De Martini, and P.
  Mataloni}(2009)]{rossi}
{A. Rossi, G. Vallone, A. Chiuri, F. De Martini, and P. Mataloni}.
\newblock {Multipath entanglement of two photons}.
\newblock \emph{Phys. Rev. Lett.}, 102:\penalty0 153902, 2009.

\bibitem[{C. Monroe, W. Swann, H. Robinson, and C. Wieman}(1990)]{monroe}
{C. Monroe, W. Swann, H. Robinson, and C. Wieman}.
\newblock {Very cold trapped atoms in a vapor cell}.
\newblock \emph{Phys. Rev. Lett.}, 65:\penalty0 1571, 1990.

\bibitem[{M. Saffman}(2009)]{saff}
{M. Saffman}.
\newblock {Private communication}.
\newblock 2009.

\bibitem[{M. Saffman and T. G. Walker}(2005)]{walkerabs}
{M. Saffman and T. G. Walker}.
\newblock {Entangling single- and N -atom qubits for fast quantum state
  detection and transmission}.
\newblock \emph{Phys. Rev. A}, 72:\penalty0 042302, 2005.

\bibitem[{B. C. Jacobs and J. D. Franson}(2009)]{jacobs}
{B. C. Jacobs and J. D. Franson}.
\newblock {All-optical switching using the quantum Zeno effect and two-photon
  absorption}.
\newblock \emph{Phys. Rev. A}, 79:\penalty0 063830, 2009.

\bibitem[{H. You, S. M. Hendrickson, and J. D. Franson}(2008)]{you}
{H. You, S. M. Hendrickson, and J. D. Franson}.
\newblock {Analysis of enhanced two-photon absorption in tapered optical
  fibers}.
\newblock \emph{Phys. Rev. A}, 78:\penalty0 053803, 2008.

\bibitem[{T. Nakanishi, H. Kobayashi, K. Sugiyama, and M.
  Kitano}(2009)]{nakanishi}
{T. Nakanishi, H. Kobayashi, K. Sugiyama, and M. Kitano}.
\newblock {Full quantum analysis of two-photon absorption using two-photon
  wavefunction: Comparison with one-photon absorption}.
\newblock \emph{J. Phys. Soc. Jpn.}, 78:\penalty0 104401, 2009.

\bibitem[{K. Singer, M. Reetz-Lamour, T. Amthor, L. G. Marcassa, and M.
  Weidem\"{u}ller}(2004)]{singer}
{K. Singer, M. Reetz-Lamour, T. Amthor, L. G. Marcassa, and M.
  Weidem\"{u}ller}.
\newblock {Suppression of excitation and spectral broadening induced by
  interactions in a cold gas of Rydberg atoms}.
\newblock \emph{Phys. Rev. Lett.}, 93:\penalty0 163001, 2004.

\bibitem[{J. Deiglmayra, M. Reetz-Lamour, T. Amthora, S. Westermanna, A. L. de
  Oliveira, and M. Weidem\"{u}llera}(2006)]{deiglmayr}
{J. Deiglmayra, M. Reetz-Lamour, T. Amthora, S. Westermanna, A. L. de Oliveira,
  and M. Weidem\"{u}llera}.
\newblock {Coherent excitation of Rydberg atoms in an ultracold gas}.
\newblock \emph{Optics Communications}, 264:\penalty0 293, 2006.

\bibitem[{S. Quabis, R. Dorn, M. Eberler, O. Gl\"{o}ckl, and G.
  Leuchs}(2001)]{quabis}
{S. Quabis, R. Dorn, M. Eberler, O. Gl\"{o}ckl, and G. Leuchs}.
\newblock {The focus of light –- theoretical calculation and experimental
  tomographic reconstruction}.
\newblock \emph{Appl. Phys. B}, 72:\penalty0 109, 2001.

\bibitem[{S. J. van Enk}(2004)]{enk}
{S. J. van Enk}.
\newblock {Atoms, dipole waves, and strongly focused light beams}.
\newblock \emph{Phys. Rev. A}, 69:\penalty0 043813, 2004.

\bibitem[{M. K. Tey, Z. Chen, S. A. Aljunid, B. Chng, F. Huber, G. Maslennikov,
  and C. Kurtsiefer}(2008)]{tey}
{M. K. Tey, Z. Chen, S. A. Aljunid, B. Chng, F. Huber, G. Maslennikov, and C.
  Kurtsiefer}.
\newblock {Strong interaction between light and a single trapped atom without
  the need for a cavity}.
\newblock \emph{Nature Physics}, 4:\penalty0 924, 2008.

\bibitem[{M. K. Tey, G. Maslennikov, T. C. H. Liew, S. A. Aljunid, F. Huber, B.
  Chng, Z. Chen, V. Scarani, and C. Kurtsiefer}(2009)]{tey2}
{M. K. Tey, G. Maslennikov, T. C. H. Liew, S. A. Aljunid, F. Huber, B. Chng, Z.
  Chen, V. Scarani, and C. Kurtsiefer}.
\newblock {Interfacing light and single atoms with a lens}.
\newblock \emph{New J. Phys.}, 11:\penalty0 043011, 2009.

\bibitem[{M. Sondermann, R. Maiwald, H. Konermann, N. Lindlein, U. Peschel, and
  G. Leuchs}(2007)]{sondermann}
{M. Sondermann, R. Maiwald, H. Konermann, N. Lindlein, U. Peschel, and G.
  Leuchs}.
\newblock {Design of a mode converter for efficient light-atom coupling in free
  space}.
\newblock \emph{Appl. Phys. B}, 89:\penalty0 489, 2007.

\bibitem[{J. Wei, D. Olaya, B. S. Karasik, S. V. Pereverzev, A. V. Sergeev, and
  M. E. Gershenson}(2008)]{wei}
{J. Wei, D. Olaya, B. S. Karasik, S. V. Pereverzev, A. V. Sergeev, and M. E.
  Gershenson}.
\newblock {Ultrasensitive hot-electron nanobolometers for terahertz
  astrophysics}.
\newblock \emph{Nature Nanotechnology}, 3:\penalty0 496, 2008.

\bibitem[{R. Raussendorf, J. Harrington, and K. Goyal}(2006)]{raussendorf}
{R. Raussendorf, J. Harrington, and K. Goyal}.
\newblock {A fault-tolerant one-way quantum computer}.
\newblock \emph{Ann. Phys.}, 321:\penalty0 2242, 2006.

\bibitem[{A. Y. Kitaev}(2003)]{kitaev}
{A. Y. Kitaev}.
\newblock {Fault-tolerant quantum computation by anyons}.
\newblock \emph{Ann. Phys.}, 303:\penalty0 2, 2003.

\bibitem[{R. Raussendorf, J. Harrington, and K. Goyal}(2007)]{raussendorf_new}
{R. Raussendorf, J. Harrington, and K. Goyal}.
\newblock {Topological fault-tolerance in cluster state quantum computation}.
\newblock \emph{New J. Phys.}, 9:\penalty0 199, 2007.

\bibitem[{T. M. Stace, S. D. Barrett, and A. C. Doherty}(2009)]{stace}
{T. M. Stace, S. D. Barrett, and A. C. Doherty}.
\newblock {Thresholds for topological codes in the presence of loss}.
\newblock \emph{Phys. Rev. Lett.}, 102:\penalty0 200501, 2009.

\bibitem[{W. D\"{u}r, H. Aschauer, and H. J. Briegel}(2003)]{dur}
{W. D\"{u}r, H. Aschauer, and H. J. Briegel}.
\newblock {Multiparticle entanglement purification for graph states}.
\newblock \emph{Phys. Rev. Lett.}, 91:\penalty0 107903, 2003.

\bibitem[{H. Aschauer, W. D\"{u}r, and H. J. Briegel}(2005)]{aschauer}
{H. Aschauer, W. D\"{u}r, and H. J. Briegel}.
\newblock {Multiparticle entanglement purification for two-colorable graph
  states}.
\newblock \emph{Phys. Rev. A}, 71:\penalty0 012319, 2005.

\bibitem[{P. G. Kwiat, E. Waks, A. G. White, I. Appelbaum, and P. H.
  Eberhard}(1999)]{kwiat}
{P. G. Kwiat, E. Waks, A. G. White, I. Appelbaum, and P. H. Eberhard}.
\newblock {Ultrabright source of polarization-entangled photons}.
\newblock \emph{Phys. Rev. A}, 60:\penalty0 R773--R776, 1999.

\bibitem[{R. Rangarajan, M. Goggin, and P. G. Kwiat}(2009)]{rangarajan}
{R. Rangarajan, M. Goggin, and P. G. Kwiat}.
\newblock {Optimizing type-I polarization-entangled photons}.
\newblock \emph{Optics Express}, 17:\penalty0 18920--18933, 2009.

\end{thebibliography}

\end{document}